\documentclass[a4paper,preprintnumbers,floatfix,twocolumn,aps,prb,unsortedaddress,superscriptaddress]{revtex4-2}

\usepackage[paperwidth=210mm,paperheight=297mm,centering,hmargin=1.7cm,vmargin=2.5cm]{geometry}

\usepackage[para]{threeparttable} 
\usepackage{amsmath}
\usepackage{graphicx} 
\usepackage{booktabs}
\usepackage{subcaption}
\usepackage{color}
\usepackage{url}
\usepackage{hyperref}
\usepackage{miller}

\usepackage{array}
\newcolumntype{'}{>{\global\let\currentrowstyle\relax}}
\newcolumntype{^}{>{\currentrowstyle}}
\newcommand{\rowstyle}[1]{\gdef\currentrowstyle{#1}%
  #1\ignorespaces
}

\usepackage[utf8]{inputenc}

\begin{document}

\title{Gaussian approximation potentials for body-centered-cubic transition metals}

\author{J. Byggmästar}
\thanks{Corresponding author}
\email{jesper.byggmastar@helsinki.fi}
\affiliation{Department of Physics, P.O. Box 43, FI-00014 University of Helsinki, Finland}
\author{K. Nordlund}
\affiliation{Department of Physics, P.O. Box 43, FI-00014 University of Helsinki, Finland}
\author{F. Djurabekova}
\affiliation{Department of Physics, P.O. Box 43, FI-00014 University of Helsinki, Finland}
\affiliation{Helsinki Institute of Physics, Helsinki, Finland}

\date{\today}

\begin{abstract}
{We develop a set of machine-learning interatomic potentials for elemental V, Nb, Mo, Ta, and W using the Gaussian approximation potential framework. The potentials show good accuracy and transferability for elastic, thermal, liquid, defect, and surface properties. All potentials are augmented with accurate repulsive potentials, making them applicable to radiation damage simulations involving high-energy collisions. We study melting and liquid properties in detail and use the potentials to provide melting curves up to 400 GPa for all five elements.}
\end{abstract}

\maketitle

\section{Introduction}
\label{sec:intro}

The use of machine learning approaches to construct interatomic potentials has rapidly gained popularity in the last decade. Different classes of machine-learning potentials have been developed with different underlying machine learning architectures and descriptors for representing the local environments of atoms~\cite{behler_perspective_2016,bartok_representing_2013}. Well-established frameworks exist for interatomic potentials using artificial neural networks~\cite{behler_generalized_2007,ssmith_ani-1_2017,zhang_deep_2018,schutt_schnet_2018}, Gaussian process regression and other kernel methods~\cite{bartok_gaussian_2010,glielmo_efficient_2018,vandermause_--fly_2020}, and linear regression~\cite{thompson_spectral_2015,shapeev_moment_2016}. Although the field is still relatively new, it has already reached a level of maturity that high-quality machine learning potentials are now routinely trained for a variety of materials and molecules~\cite{deringer_machine_2019,mueller_machine_2020,chen_accurate_2017,wood_data-driven_2019,zuo_performance_2020}.

Metals and metal alloys are typically modelled using embedded atom method
(EAM) potentials in classical molecular dynamics simulations~\cite{daw_embedded-atom_1984}.
For the nonmagnetic body-centered transition metals V, Nb, Mo, Ta, and W, several parametrisations have been developed that reproduce
a variety of material properties with reasonable accuracy, while also being computationally extremely efficient~\cite{ackland_improved_1987,derlet_multiscale_2007,chen_development_2019,chen_interatomic_2020}.
Nevertheless, certain properties are often difficult to reproduce by traditional interatomic potentials with fixed functional forms.
Examples include surface energies and the energetics and structures of vacancy clusters, self-interstitial clusters, and dislocations~\cite{byggmastar_machine-learning_2019,bonny_many-body_2014}.
Developing more accurate interatomic potentials for these elements is therefore well-motivated.
Machine-learning potentials provide a useful complement to traditional analytical potentials and expensive electronic-structure calculations, lying somewhere between the two in both computational cost and accuracy.

The aim of this article is to develop robust and accurate potentials for V, Nb, Mo, Ta, and W (the W potential has been developed previously~\cite{byggmastar_machine-learning_2019}, but included also here to allow for comprehensive comparison of the five elements). These elements belong to the family of refractory metals, characterised by high melting points and good resistance to deformation and heat, which make them attractive for a number of applications. For example, W is the top candidate for the parts most exposed to heat and irradiation in fusion reactors~\cite{rieth_recent_2013,stork_materials_2014} and Mo is a candidate for diagnostic mirrors in fusion test reactors~\cite{litnovsky_diagnostic_2007}. All five elements are also commonly used in various high-strength alloys~\cite{liu_nanostructured_2013,buckman_new_2000,sheftel_niobiumbase_1993,moskalyk_processing_2003}.

The rest of the article is structured as follows. In Sec.~\ref{sec:training} we describe the training details and strategy. In Sec.~\ref{sec:results} we extensively benchmark the potentials, and use them to simulate melting curves of all five elements. Finally, we discuss the results and provide a brief outlook in Sec.~\ref{sec:concl}.

\section{Training}
\label{sec:training}

We use the Gaussian approximation potential framework~\cite{bartok_gaussian_2010,bartok_gaussian_2015} to train the potentials. Our training strategy closely follows the methods described in detail in our previous work~\cite{byggmastar_machine-learning_2019}. The total energy of $N$ atoms is given by
\begin{align}
 \begin{split}
    E_\mathrm{tot} &= \sum_{i, j>i}^N V_\mathrm{pair}(r_{ij}) + \delta_\mathrm{2b}^2 \sum_i^{N_\mathrm{pairs}} \sum_s^{M_\mathrm{2b}} \alpha_{s, \mathrm{2b}} K_\mathrm{2b} (\textbf{q}_{i, \mathrm{2b}}, \textbf{q}_{s, \mathrm{2b}}) \\
    &+ \delta_\mathrm{mb}^2 \sum_i^N \sum_s^{M_\mathrm{mb}} \alpha_{s, \mathrm{mb}} K_\mathrm{mb} (\textbf{q}_{i, \mathrm{mb}}, \textbf{q}_{s, \mathrm{mb}}),
  \end{split}
\end{align}
where $V_\mathrm{pair}$ is a purely repulsive pair potential in the form of a screened Coulomb potential. We fit $V_\mathrm{pair}$ separately for each atomic pair to all-electron DFT data from Ref.~\cite{nordlund_repulsive_1997}. The last two terms make up the machine-learning contributions. The second term carries out Gaussian process regression with a two-body (interatomic distance) descriptor $\textbf{q}_\mathrm{2b}$ and the squared exponential kernel $K_\mathrm{2b}$. The last term contains the smooth overlap of atomic positions (SOAP) kernel~\cite{bartok_representing_2013} for many-body interactions. We use $n_\mathrm{max} = l_\mathrm{max} = 8$ for the spherical harmonics expansion in SOAP. The prefactors of the machine-learning terms are $\delta^2_\mathrm{2b}=10^2$ eV and $\delta_\mathrm{mb}^2=2^2$ eV. $M_\mathrm{2b}$ and $M_\mathrm{mb}$ are the (sparsified) number of training environments ($M_\mathrm{2b}=20$, $M_\mathrm{mb}=4000$). $\alpha_\mathrm{2b}$ and $\alpha_\mathrm{mb}$ are the regression coefficients that are optimized during training. For more details about the GAP framework, we refer to Refs.~\cite{bartok_gaussian_2015,bartok_representing_2013}.

The training database for each element consists of the same structures as in our previous W GAP~\cite{byggmastar_machine-learning_2019}, but rescaled to the correct lattice spacing. The training structures include elastically distorted unit cells of bcc, high-temperature bcc crystals, structures containing vacancies and self-interstitial atoms, surfaces (both flat and disordered), and liquids. These structures ensure machine-learning of elastic, thermal, and defect properties, as well as surface energetics, melting, and the structure of the liquid phase. To improve transferability and cover a variety of crystal symmetries, we also include several other crystal structures (fcc, hcp, simple cubic, diamond, A15, and C15) as elastically distorted unit cells. The A15 structure is the $\beta$ phase of tungsten, and C15 crystals are possible radiation-induced self-interstitial defect clusters in bcc~\cite{marinica_irradiation-induced_2012}. The training database additionally includes the dimer curve sampled down to distances of significant repulsion, and bcc crystals containing short interatomic bonds. These are crucial to ensure a smooth connection to the analytical screened Coulomb potential~\cite{byggmastar_machine-learning_2019}. We also made the following small additions. As W is considerably heavier and elastically stiffer than all the other elements, we extended the volume range of the elastically distorted unit cells from $\pm30$\% around the equilibrium volume for W to about $\pm50$\% for V, Nb, Mo, and Ta. For the same reasons, we also prepared new high-temperature bcc crystals for each element separately, using MD simulations with preliminary potentials trained to all other data. Furthermore, we added structures with the \hkl(110), \hkl(100) and \hkl(112) unstable stacking fault surfaces ($\gamma$ surfaces) to ensure some transferability to plastic deformation. Our W GAP in Ref.~\cite{byggmastar_machine-learning_2019} was not trained to $\gamma$ surfaces, but for all other properties it is virtually identical to the W GAP presented here.

The cutoff for the interaction range is 5 Å for W and Mo (which have almost identical lattice constants). For the other elements, we rescale the cutoff distance according to the lattice constants, so that the range includes the same numbers of neighbors in bcc for all elements (5.2 Å for Ta and Nb, and 4.7 Å for V). A key detail of the GAP framework is the use of heavy regularization in the regression fit to avoid overfitting and to only reproduce the training data to a desired accuracy. The regularization errors are chosen according to the assumed uncertainty of ideal GAP predictions, i.e. the convergence accuracy of the DFT data combined with the approximation of a finite interaction range of the GAP. As default regularization errors when training the GAPs, we used $\sigma_0^E=1$ meV/atom, $\sigma_0^F=0.04$ eV/Å, and $\sigma_0^S=0.04$ eV for energies, forces, and virial stresses. Larger errors were assumed for liquid training structures ($\sigma=10\sigma_0$), structures containing short interatomic bonds ($\sigma=10\sigma_0$), and for the strongly distorted bcc unit cells and $\gamma$ surfaces ($\sigma=2\sigma_0$).

Energies, forces, and for the distorted unit cells, stresses, of all training structures are calculated using \textsc{vasp}~\cite{kresse_ab_1993,kresse_ab_1994,kresse_efficiency_1996,kresse_efficient_1996} with the PBE GGA exchange-correlation functional~\cite{perdew_generalized_1996}. Hard projector-augmented wave (PAW)~\cite{blochl_projector_1994,kresse_ultrasoft_1999} potentials were used (\texttt{W\_sv}, \texttt{Mo\_sv}, \texttt{Ta\_pv}, \texttt{Nb\_sv}, \texttt{V\_sv}), with 14 valence electrons for W and Mo, 11 for Ta, and 13 for Nb and V. The cutoff of the plane-wave expansion was 500 eV for all elements. A 0.1 eV smearing was applied using the first-order Methfessel-Paxton method~\cite{methfessel_high-precision_1989}. $k$-points were sampled on Monkhorst-Pack grids~\cite{monkhorst_special_1976} with a maximum spacing of 0.15 Å$^{-1}$.

The GAPs were trained using the code \textsc{quip}~\cite{quip}. \textsc{lammps}~\cite{plimpton_fast_1995} was used for all molecular dynamics simulations. Phonon dispersions and nudged elastic band calculations were obtained using the Atomic Simulation Environment (ASE) framework~\cite{larsen_atomic_2017} and quasi-harmonic approximation calculations using \textsc{phonopy}~\cite{togo_first_2015}.

\section{Validation}
\label{sec:results}

\subsection{Bulk properties}

\begin{figure}
    \centering
    \includegraphics[width=\linewidth]{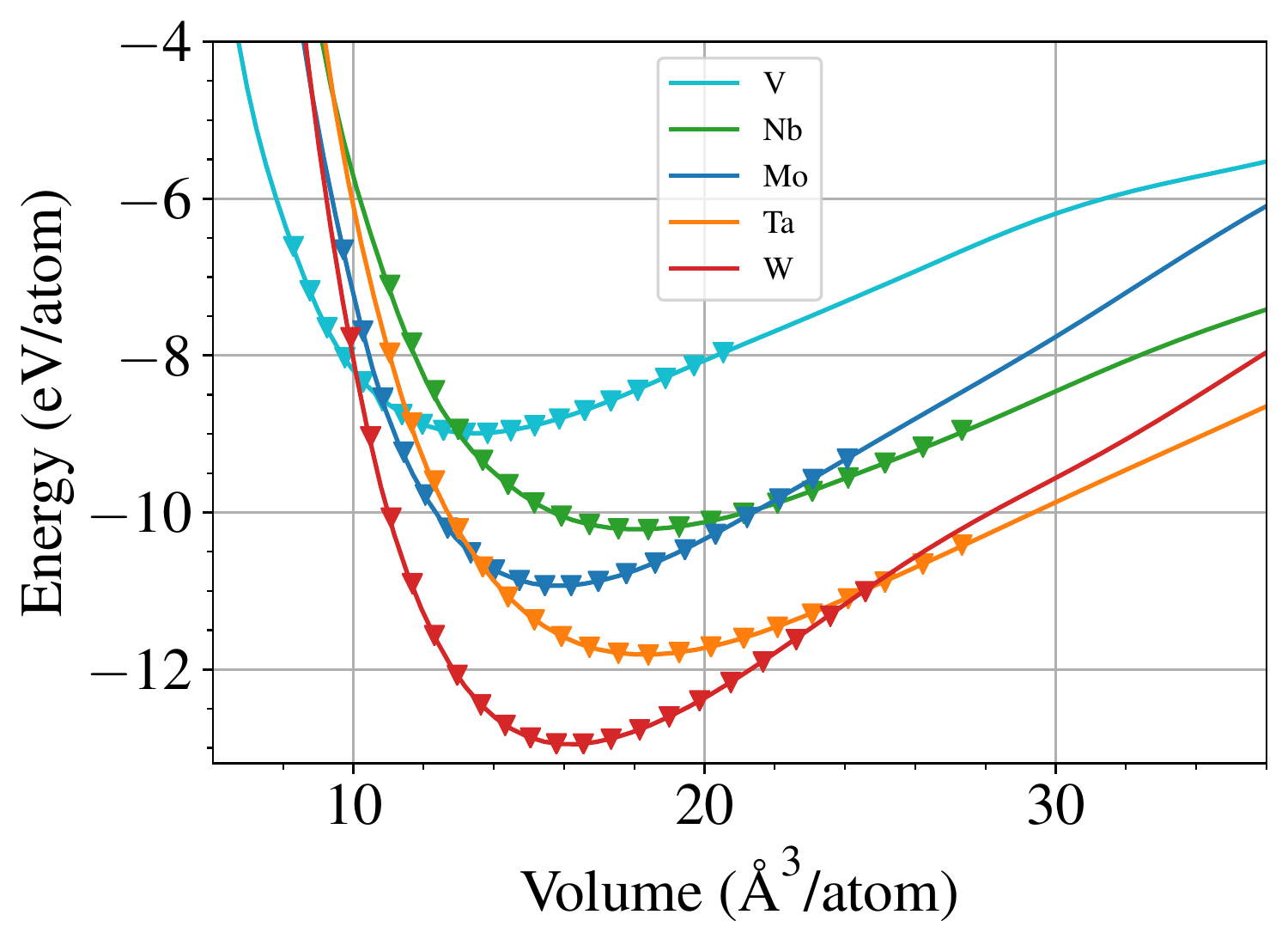}
    \caption{Energy-volume curves for the bcc phase of the different elements. The lines are predictions by GAP and the data points are from DFT.}
    \label{fig:bulkvol}
\end{figure}

Table~\ref{tab:bulk} lists the energies, lattice constants, and elastic constants of the bcc ground state for all five elements, compared between GAP (bold values), DFT, and experimental data (italic values). The GAPs reproduce the DFT data well. There are, however, some noteworthy discrepancies between DFT (and thus also GAP) and experiments, in particular the elastic constants of V and $C_{44}$ of Nb. Figure~\ref{fig:bulkvol} shows energy-volume relations with DFT data as data points and GAP data as solid lines. The training data contains only randomly strained bcc crystals covering at most volumes in the $\pm 50$\% range around the equilibrium. That the GAPs show physically reasonable extrapolations well beyond this range is a promising indicator of good transferability. Transferability to high pressures is ensured by the external repulsive potential, which will be further demonstrated when discussing the high-pressure phase diagram in Sec.~\ref{sec:liquid}.

\begin{table*}
 \centering
 \caption{Energy per atom of the bcc phase, $E_\mathrm{bcc}$, cohesive energy of bcc, $E_\mathrm{coh}$, lattice constant, $a$, and elastic constants, $C_{ij}$. Bold values are GAP, plain font are DFT, and italic values are experimental data from Ref.~\cite{rumble_crc_2019}.}
 \label{tab:bulk}
  \begin{tabular}{'l^l^l^l^l^l}
   \toprule
   & V & Nb & Mo & Ta & W \\
   \midrule
    \rowstyle{\bfseries}
    $E_\mathrm{bcc}$ \textnormal{(eV/atom)} & $-$8.992 & $-$10.216 & $-$10.937 & $-$11.813 & $-$12.956  \\  
                               & $-8.992$ & $-10.216$ & $-10.936$ & $-11.812$ & $-12.957$ \\  
                               & & & & & \\  
    \rowstyle{\bfseries}
    $E_\mathrm{coh}$ \textnormal{(eV/atom)} & $-$5.384 & $-$7.004 & $-$6.288 & $-$8.114 & $-$8.39 \\  
                               & $-5.384$ & $-7.003$ & $-6.288$ & $-8.113$ & $-8.386$  \\  
    \rowstyle{\itshape}
                               & $-$5.329 & $-$7.523 & $-$6.821 & $-$8.105 & $-$8.803 \\ 
    \rowstyle{\bfseries}
    $a$ \textnormal{(Å)} & 2.997 & 3.308 & 3.163 & 3.321 & \textbf{3.185} \\
            & 2.997 & 3.307 & 3.163 & 3.319 & 3.185  \\  
    \rowstyle{\itshape}
            & 3.024 & 3.300 & 3.147 & 3.303 & 3.165 \\ 
    \rowstyle{\bfseries}
    $C_{11}$ \textnormal{(GPa)} & 271 & 243 & 472 & 267 & 524  \\
                   & 269 & 237 & 468 & 266 & 521 \\
    \rowstyle{\itshape}
                   & 229 & 247 & 464 & 260 & 522 \\
    \rowstyle{\bfseries}
    $C_{12}$ \textnormal{(GPa)} & 145 & 137 & 163 & 161 & 200 \\
                   & 146 & 138 & 155 & 161 & 195 \\
    \rowstyle{\itshape}
                   & 119 & 135 & 159 & 154 & 204 \\
    \rowstyle{\bfseries}
    $C_{44}$ \textnormal{(GPa)} & 23 & 13 & 105 & 77 & 148 \\
                   & 22 & 11 & 100 & 77 & 147 \\
    \rowstyle{\itshape}
                   & 43 & 29 & 109 & 83 & 161 \\
   \bottomrule
 \end{tabular}
\end{table*}

\begin{table}
 \centering
 \caption{Thermal properties at room temperature calculated within the quasi-harmonic approximation with the GAP (bold values) and DFT (plain font), compared with experimental data (italic). $\alpha_L$: linear thermal expansion coefficient in $10^{-6}$ K$^{-1}$, $C_p$: heat capacity at constant pressure in J mol$^{-1}$ K$^{-1}$, and $\gamma$: thermodynamical Gr\"uneisen parameter. GAP values in parentheses are from MD simulations. Experimental data for thermal expansion and heat capacities are from Ref.~\cite{rumble_crc_2019}, and Gr\"uneisen parameters from Ref.~\cite{white_heat_1988}.}
 \label{tab:thermal}
  \begin{tabular}{'l^l^l^l^l^l}
   \toprule
   & V & Nb & Mo & Ta & W \\
   \midrule
    \rowstyle{\bfseries}
    $\alpha_L$ & 12.1 (10.0) & 8.9 (8.5) & 5.6 (6.3) & 7.4 (7.7) & 5.1 (5.2) \\
                                    & 10.8 & 8.0 & 5.8 & 8.6 & 4.9  \\  
    \rowstyle{\itshape}
                                    & 8.4 & 7.3 & 4.8 & 6.3 & 4.5 \\ 
    \rowstyle{\bfseries}
    $C_p$    & 23.90 & 24.14 & 23.50 & 24.65 & 23.98 \\
                                    & 23.81 & 24.03 & 23.50 & 24.78 & 23.95  \\  
    \rowstyle{\itshape}
                                    & 24.89 & 24.60 & 24.06 & 25.36 & 24.27 \\ 
    \rowstyle{\bfseries}
    $\gamma$ & 2.2 & 2.0 & 1.8 & 1.9 & 1.9 \\
             & 2.0 & 1.8 & 1.8 & 2.2 & 1.8 \\
    \rowstyle{\itshape}
             & 1.5 & 1.6 & 1.6 & 1.6 & 1.6 \\
   \bottomrule
 \end{tabular}
\end{table}

\begin{figure*}
    \centering
    \includegraphics[width=0.32\linewidth]{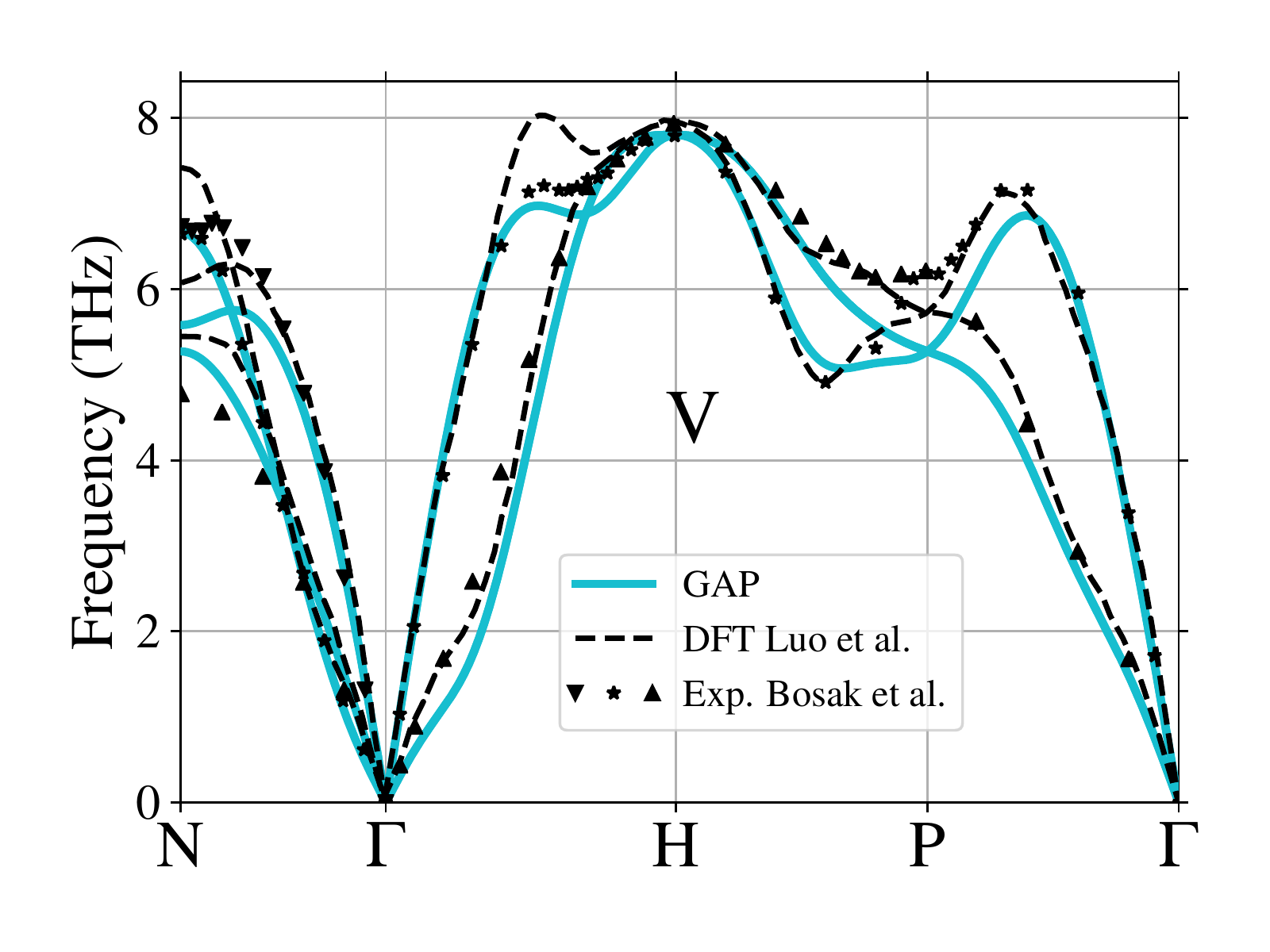}
    \includegraphics[width=0.32\linewidth]{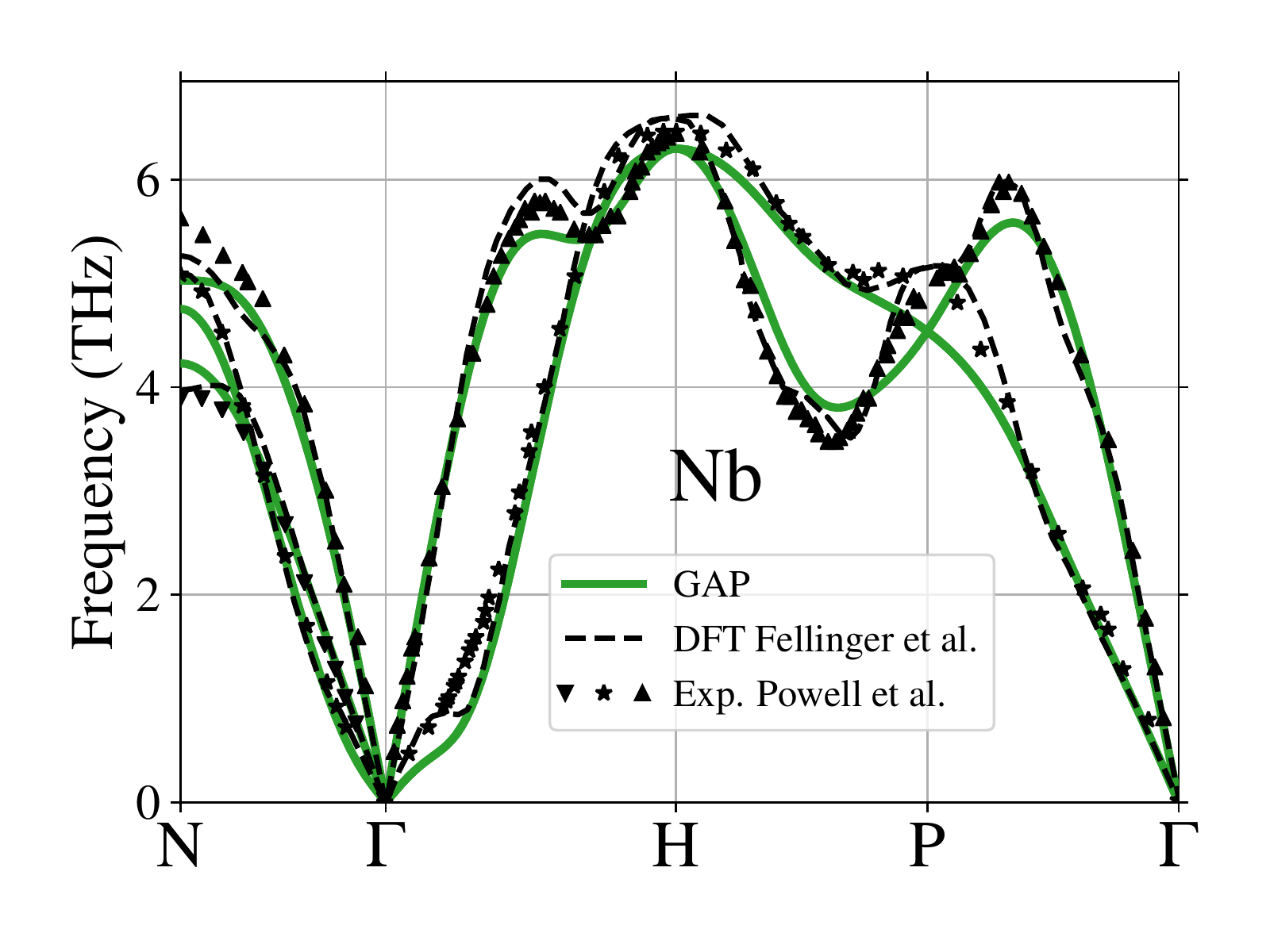}
    \includegraphics[width=0.32\linewidth]{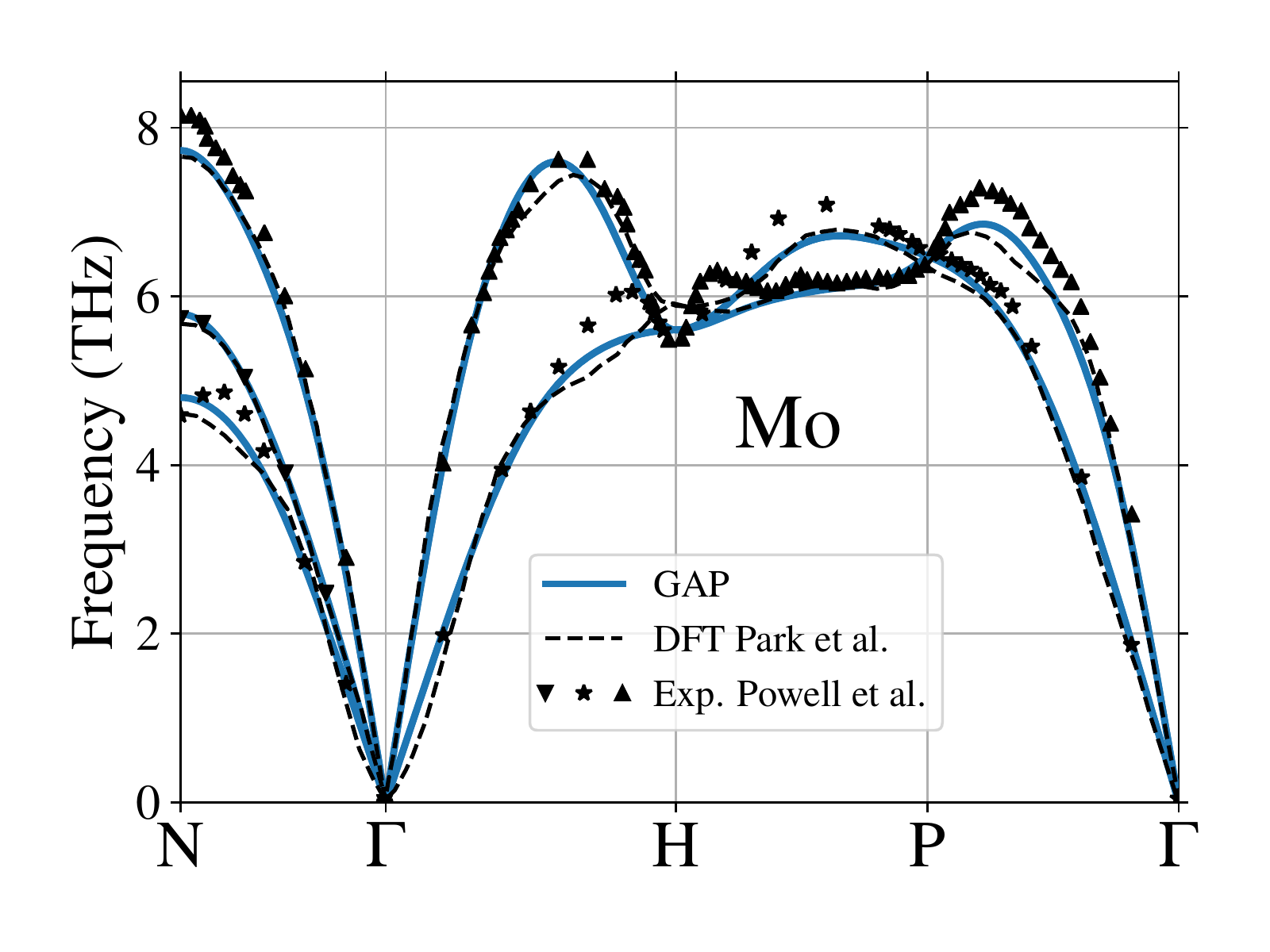}
    \includegraphics[width=0.32\linewidth]{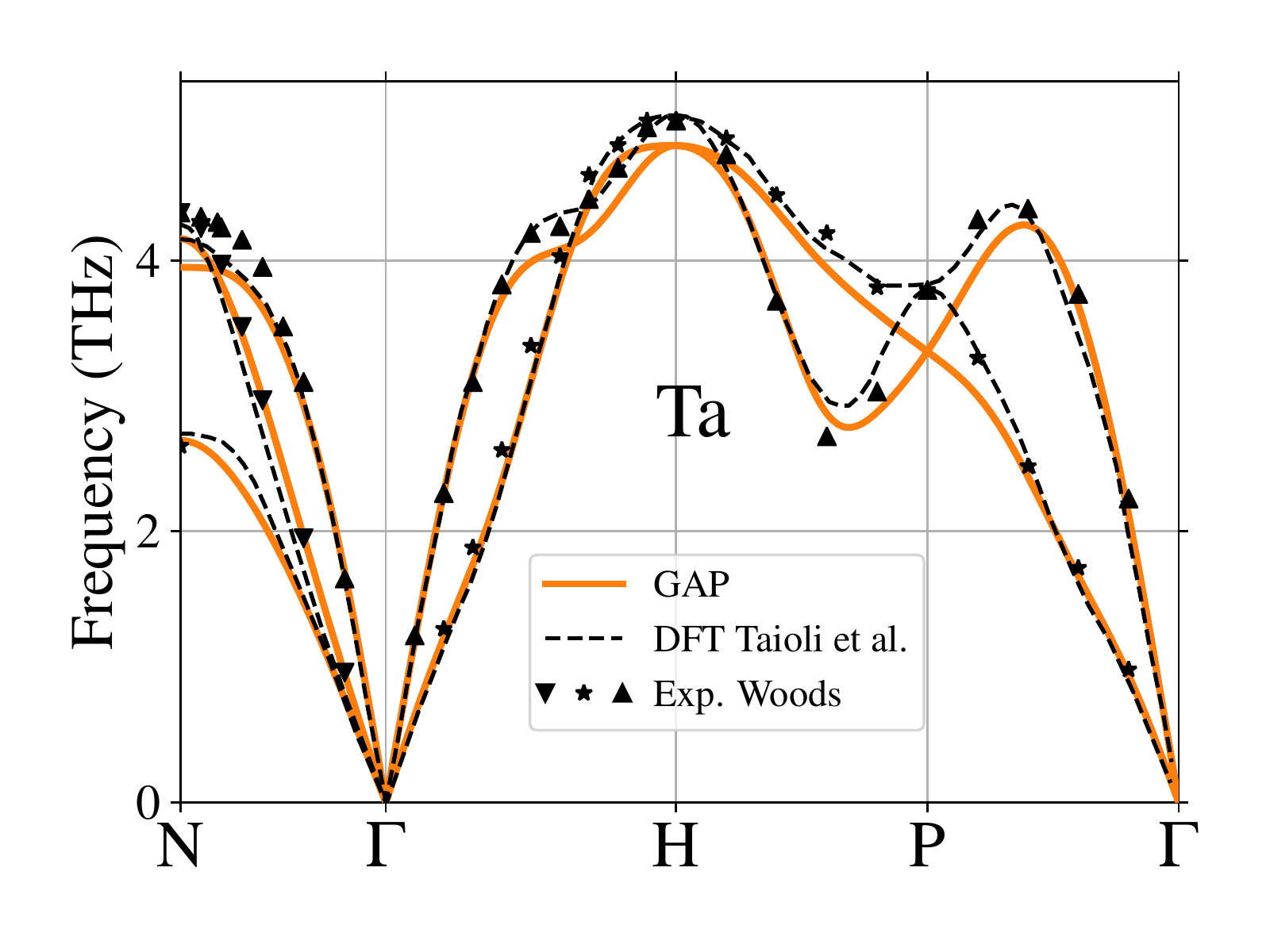}
    \includegraphics[width=0.32\linewidth]{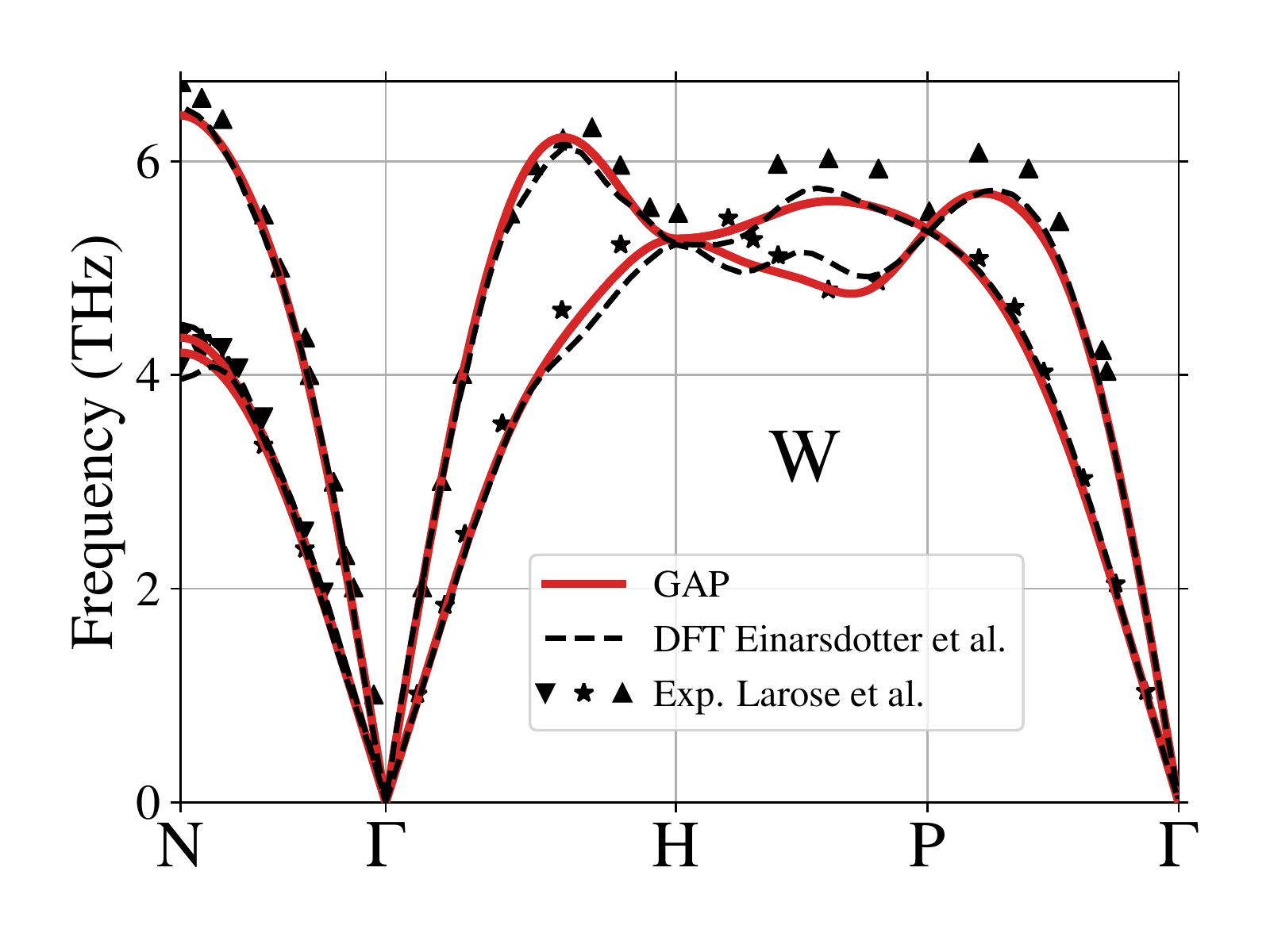}
    \caption{Phonon dispersions of the elements, compared between GAP, DFT results from the literature~\cite{luo_unusual_2007,fellinger_force-matched_2010,park_ab_2012,taioli_melting_2007,einarsdotter_phonon_1997}, and experimental data~\cite{bosak_lattice_2008-1,powell_lattice_1968,woods_lattice_1964,larose_lattice_1976}.}
    \label{fig:phonon}
\end{figure*}

Phonon dispersion plots are shown in Fig.~\ref{fig:phonon}, compared with DFT data and experimental measurements. DFT results are reproduced from the literature~\cite{luo_unusual_2007,fellinger_force-matched_2010,park_ab_2012,taioli_melting_2007,einarsdotter_phonon_1997}. Note that a direct comparison between GAP and the DFT data must therefore be made with care, as the DFT results were obtained with different settings and are not always directly consistent with our training data. All elements show similar and overall good quantitative agreement with experimental data. The consistent training of all potentials is apparent in that the small discrepancies with experiments are very similar in the GAPs. For the three group 5 metals (V, Nb, Ta), all GAPs underestimate the frequency at the P point slightly, while the group 6 elements (Mo, W) fail to capture the exact trends between the H and P points.

We used the quasi-harmonic approximation (QHA) to estimate thermal properties of the elements in both GAP and DFT. Table~\ref{tab:thermal} lists the results at 300 K: the linear thermal expansion coefficient, the heat capacity, and the thermodynamical Gr\"uneisen parameter $\gamma = V \alpha_V B / C_V$. DFT consistently overestimates the experimental thermal expansion and Gr\"uneisen parameter, and the GAPs further slightly overestimates the DFT results (except for Ta). The heat capacities at 300 K are in good agreement between GAP, DFT, and experiments for all elements. The thermal expansion coefficients were also simulated at 300 K in MD with the GAPs (given in parentheses in tab.~\ref{tab:thermal}), which gives an indication of the reliability of the QHA results. The QHA includes zero-point lattice vibrational energies but neglects the true anharmonic effects predicted by the potentials, and is therefore accurate at low temperatures. The MD results are roughly consistent with the QHA for the heavier elements, but shows a 20\% difference for the thermal expansion of V.

\subsection{Defects}

\begin{table}
 \centering
 \caption{Formation energies, $E_\mathrm{f}$, relaxation volumes, $\Omega_\mathrm{rel.}$ (in units of the atomic volume), and migration energies, $E_\mathrm{mig.}$, of single vacancies calculated with GAP (bold values) and DFT. DFT formation energies and relaxation volumes are obtained in this work and the migration energies are from Ref.~\cite{ma_effect_2019}. Experimental data (italic) are from Ref.~\cite{ullmaier_atomic_1991}.}
 \label{tab:vacancy}
  \begin{tabular}{'l^l^l^l^l^l}
   \toprule
   & V & Nb & Mo & Ta & W \\
   \midrule
   \rowstyle{\bfseries}
    $E_\mathrm{f}$ \textnormal{(eV)}     & 2.56 & 2.85 & 2.84 & 3.09 & 3.32 \\
                            & 2.49 & 2.77 & 2.83 & 2.95 & 3.36 \\
    \rowstyle{\itshape}
                            & 2.1--2.2 & 2.6--3.07 & 3.0--3.24 & 2.2--3.1 & 3.51--4.1 \\
    \rowstyle{\bfseries}
    $\Omega_\mathrm{rel.}$  & $-$0.45 & $-$0.39 & $-$0.33 & $-$0.47 & $-$0.31 \\
                            & $-0.47$ & $-0.40$ & $-0.38$ & $-0.44$ & $-0.36$ \\
                            & \\
    \rowstyle{\bfseries}
    $E_\mathrm{mig.}$ \textnormal{(eV)}  & 0.39 & 0.42 & 1.28 & 0.62 & 1.71 \\
                            & 0.65 & 0.65 & 1.24 & 0.76 & 1.73 \\
    \rowstyle{\itshape}
                            & 0.5 & 0.55 & 1.35--1.62 & 0.7 & 1.70--2.02 \\
   \bottomrule
  \end{tabular}
\end{table}

\begin{table}
    \centering
    \caption{Formation energies of self-interstitial atoms in different configurations and relaxation volumes of the \hkl<111> intersitital calculated with GAP (bold values) and DFT. DFT data are from Refs.~\cite{ma_universality_2019,ma_symmetry-broken_2019} unless otherwise indicated.}
    \begin{tabular}{'l^l^l^l^l^l}
     \toprule
     & V & Nb & Mo & Ta & W \\
     \midrule
     \rowstyle{\bfseries}
      \hkl<11\xi> & & & 7.47 & & 10.32 \\
                  & & & 7.40 & & 10.25 \\
                  \rowstyle{\bfseries}
      \hkl<111>   & 2.80 & 4.01 & 7.56 & 4.84 & 10.35 \\
                  & 2.41, 2.75* & 3.95 & 7.48 & 4.77 & 10.29 \\
                  \rowstyle{\bfseries}
      \hkl<110>   & 3.06 & 4.20 & 7.61 & 5.47 & 10.58 \\
                  & 2.68 & 4.20 & 7.58 & 5.48 & 10.58 \\
                  \rowstyle{\bfseries}
      \hkl<100>   & 3.30 & 4.63 & 8.99 & 5.99 & 12.23 \\
                  & 2.83 & 4.50 & 8.89 & 5.89 & 12.20 \\
                  \rowstyle{\bfseries}
      \textnormal{Octa}        & 3.37 & 4.76 & 9.00 & 6.06 & 12.33 \\
                  & 2.90 & 4.62 & 8.92 & 5.95 & 12.27 \\
                  \rowstyle{\bfseries}
      \textnormal{Tetra}       & 3.32 & 4.66 & 8.44 & 5.98 & 11.77 \\
                  & 2.90 & 4.42 & 8.36 & 5.77 & 11.72 \\
                  \rowstyle{\bfseries}
      $\Omega_\mathrm{rel.}$ & 1.40 & 1.48 & 1.58 & 1.35 & 1.85 \\
                             & 1.47 & 1.55 & 1.54 & 1.52 & 1.71 \\
     \bottomrule
     * This work.
    \end{tabular}
    \label{tab:SIA}
\end{table}

\begin{figure}
    \centering
    \includegraphics[width=\linewidth]{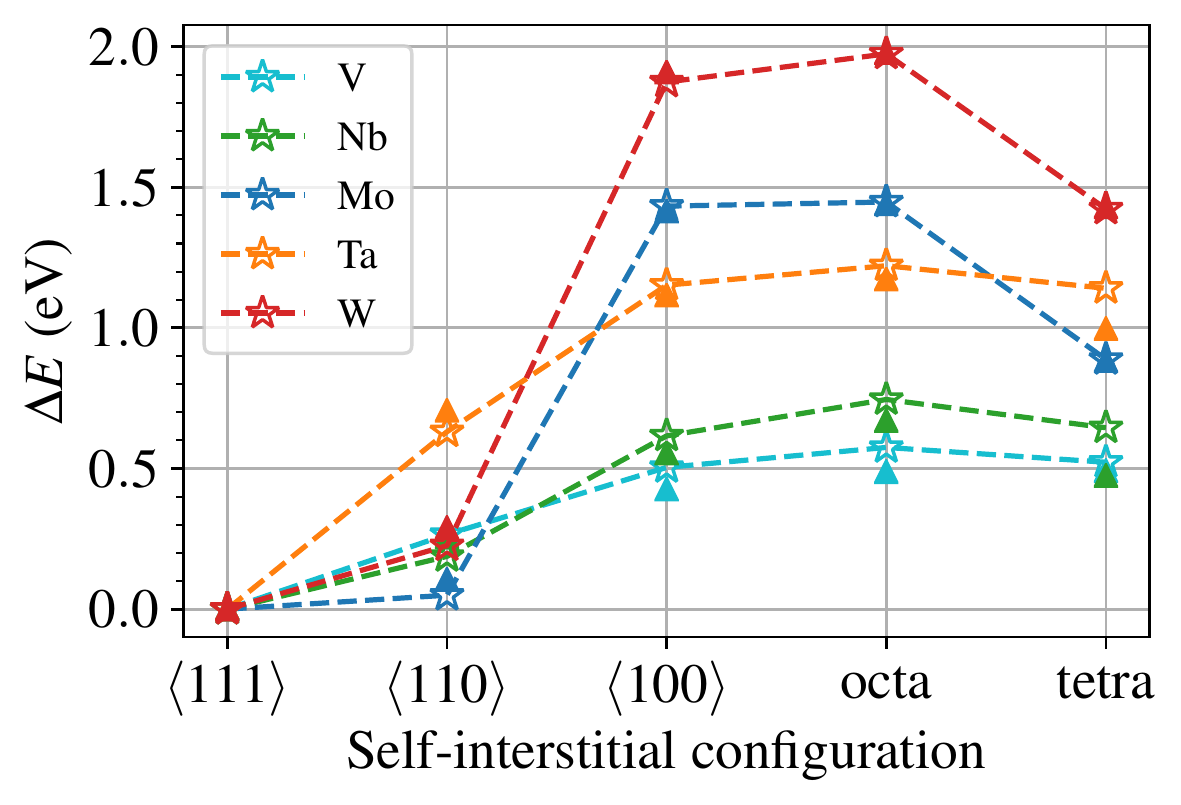}
    \caption{Relative formation energies of self-interstitial atoms, given as differences to the \hkl<111> configuration. Stars are GAP and solid triangles DFT data from Ref.~\cite{ma_universality_2019}.}
    \label{fig:sia}
\end{figure}

\begin{figure}
    \centering
    \includegraphics[width=\linewidth]{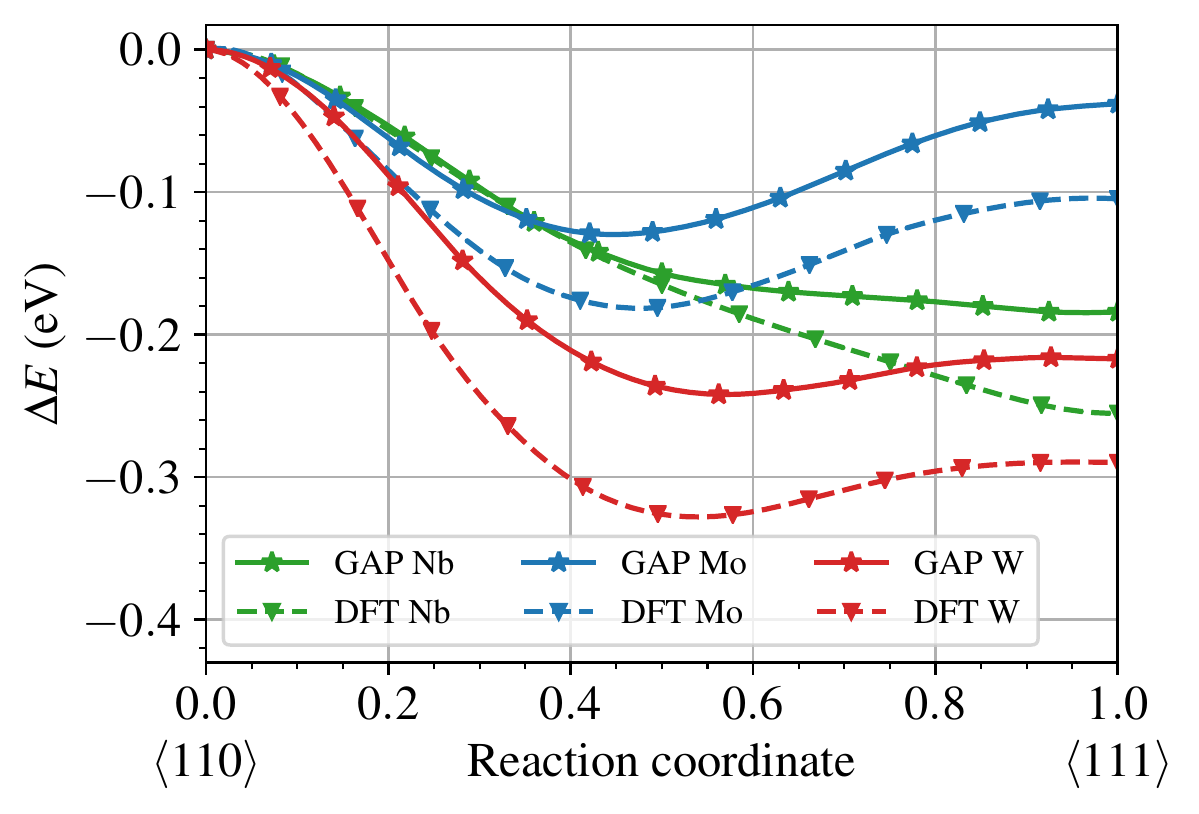}
    \caption{Nudged elastic band calculations of a self-interstitial atom rotating from the \hkl<110> dumbbell to \hkl<111>. DFT data are from Ref.~\cite{ma_symmetry-broken_2019}. GAP reproduces the global minimum \hkl<11\xi> configuration in Mo and W. Nb (along with V and Ta) show no minimum along the path.}
    \label{fig:neb110-111}
\end{figure}

\begin{figure}
    \centering
    \begin{subfigure}{\linewidth}
        \centering
        \includegraphics[width=\linewidth]{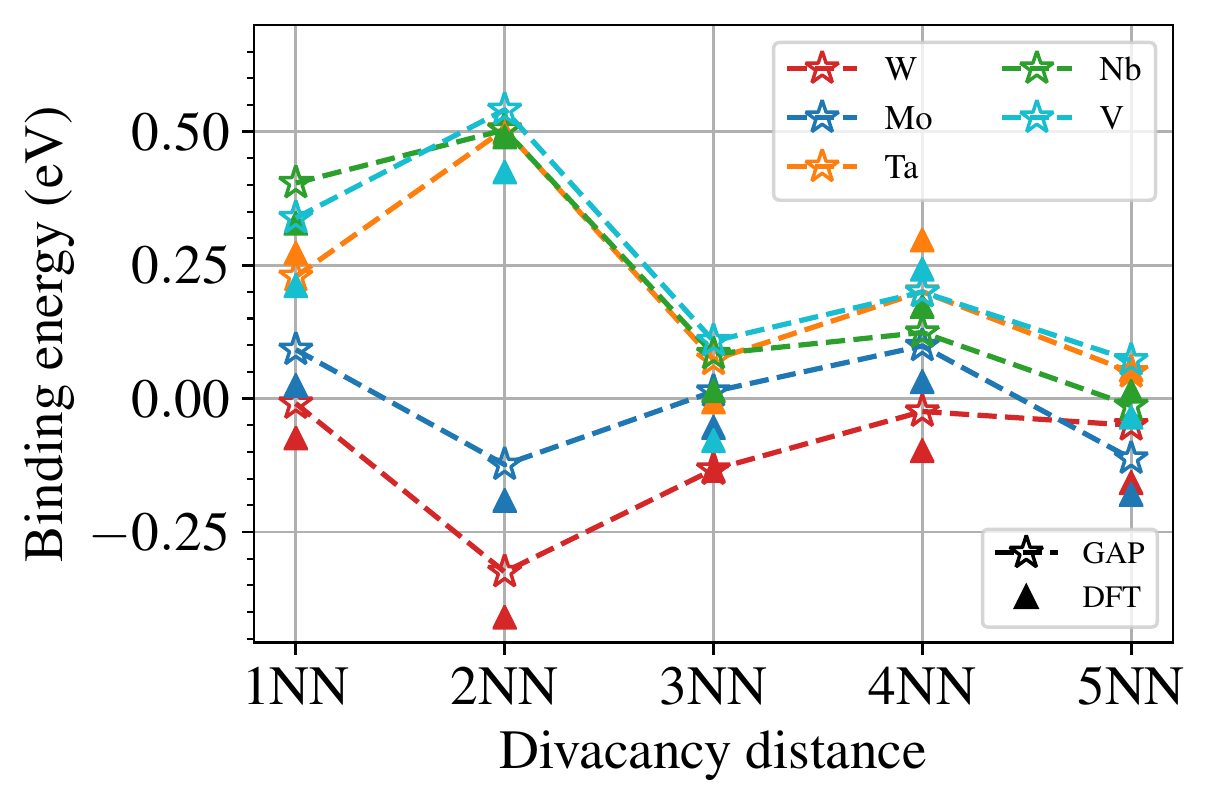}
        \caption{}
    \end{subfigure}
    \begin{subfigure}{\linewidth}
        \centering
        \includegraphics[width=\linewidth]{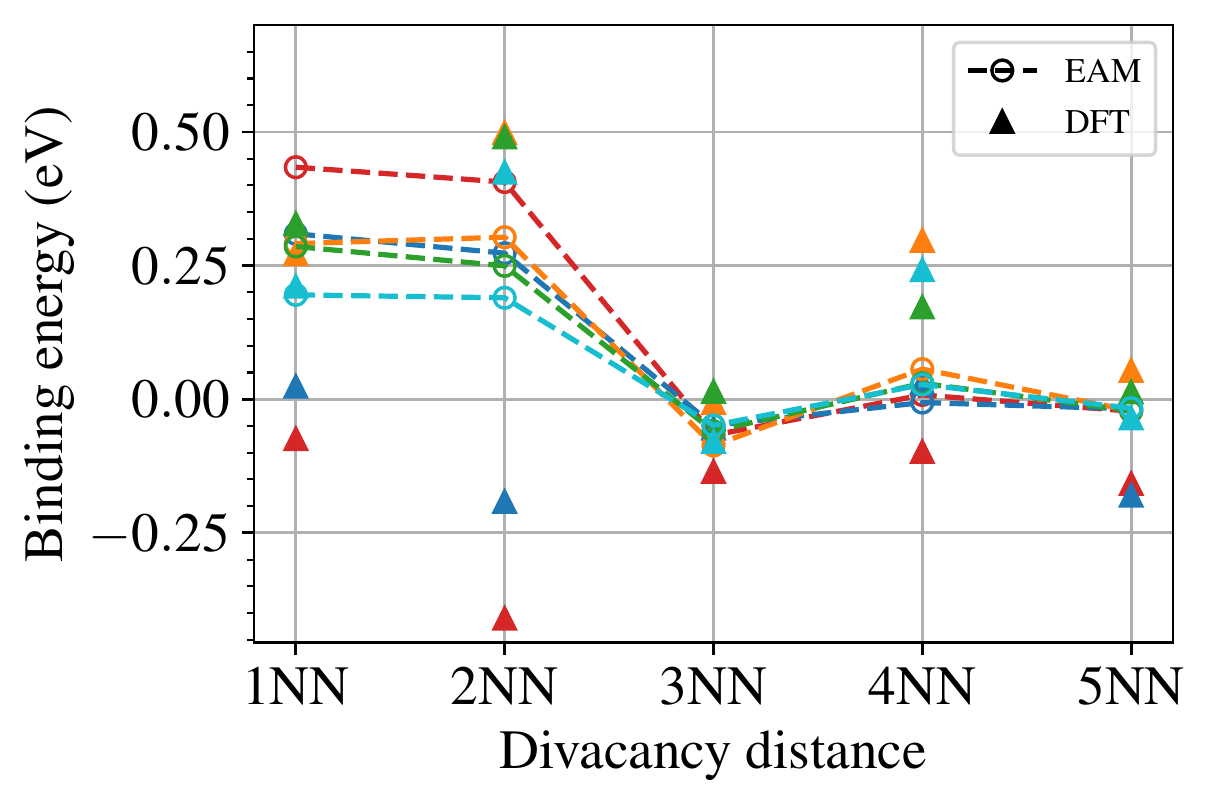}
        \caption{}
    \end{subfigure}
    \caption{Binding energies of divacancies at different nearest-neighbor (NN) separations in the different elements. (a) Comparison between GAP and DFT. (b) Comparison between the EAM potentials from Ref.~\cite{ackland_improved_1987} and DFT.}
    \label{fig:divac}
\end{figure}

\begin{figure}
    \centering
    \includegraphics[width=\linewidth]{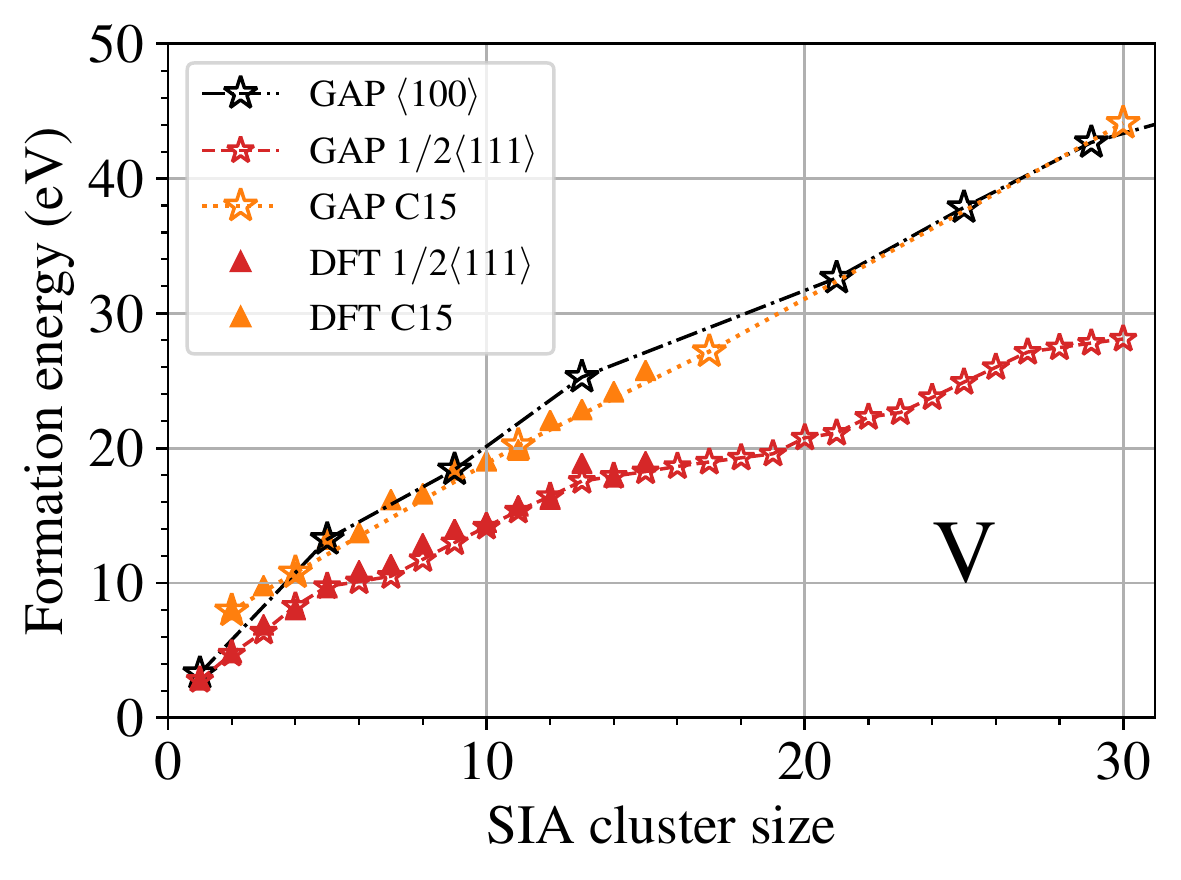}
    \caption{Formation energy of self-interstitial clusters in V compared between the GAP and DFT data from Ref.~\cite{alexander_ab_2016}. The GAP is only trained to single and di-interstitial structures.}
    \label{fig:sia_clusters_V}
\end{figure}

The training database includes structures containing vacancies and self-interstitial atoms in various configurations. Tables~\ref{tab:vacancy} and \ref{tab:SIA} list formation energies of single vacancies and interstitials, compared with DFT data. The vacancy formation energies were calculated in systems of $3 \times 3 \times 3$ unit cells. The vacancy formation energies by GAP are consistent with DFT to within a few percent. The largest discrepancy is seen for Ta, where GAP overestimates the vacancy formation energy by 0.14 eV. Note that for the 53-atom training structures, this corresponds only to a 2.6 meV/atom error, which is the level of accuracy that can be expected from the GAPs for crystalline systems. Slightly larger discrepancies between GAP and DFT and experiments are revealed for the vacancy migration energies. In V, Nb, and Ta, GAP clearly underestimate the DFT vacancy migration energies, but actually agrees better with the experimental values than DFT. The migration energies in Mo and W are, however, well reproduced by the GAP. Again, it is interesting to note the consistent accuracy and predictions of the GAPs between the group 5 and group 6 elements, although it remains unclear why only the GAPs for the latter successfully reproduce the vacancy migration energies.

Formation energies of self-interstitials in the common high-symmetry configurations are well reproduced by the GAPs, as seen in Tab.~\ref{tab:SIA}. The formation energies in the GAPs are calculated in noncubic boxes of 421 atoms. The DFT calculations in Ref.~\cite{ma_universality_2019} used smaller boxes, but made corrections for the elastic interactions across the periodic boundaries, and should therefore be consistent with our GAP results. There is a systematic 0.4 eV offset in the formation energies in V between GAP and the DFT results from Ref.~\cite{ma_universality_2019}, which are also significantly lower than other DFT results~\cite{han_interatomic_2003}. To check that the systematic offset is not an artefact of the GAP, we calculated the formation energy for the \hkl<111> interstitial in DFT using a noncubic box of 121 atoms. Our DFT formation energy is 2.75 eV, which is close to the 2.80 eV predicted by the GAP.

The GAP correctly predicts the $\hkl<11\xi>$ to be the most stable configuration in Mo and W~\cite{ma_symmetry-broken_2019}, while the straight \hkl<111> interstitial is lowest in energy for the other three elements. In V, we found using the GAP that the \hkl<110> interstitial easily rotates to a dumbbell close to the \hkl<210> direction, which the GAP predicts to be 0.05 eV lower in energy. We confirmed that in DFT the \hkl<210> is indeed lower in energy than the \hkl<110> dumbbell, by 0.12 eV with a formation energy of 2.9 eV, and is therefore the second-most stable self-interstitial configuration in V (only 0.15 eV higher in energy than \hkl<111>).

To better highlight the relative energies of self-interstitials and differences between the elements, Fig.~\ref{fig:sia} shows the differences in energy to the \hkl<111> configuration. The difference in energy between the \hkl<111> and the \hkl<110> interstitial is consistently slightly underestimated by the GAPs. This is further evident in Fig.~\ref{fig:neb110-111}, which shows the energy landscape for rotation from \hkl<110> to \hkl<111>, obtained from nudged elastic band calculations and compared with DFT data from Ref.~\cite{ma_symmetry-broken_2019}. Fig.~\ref{fig:neb110-111} also illustrates the global minimum at $\hkl<11\xi>$ in Mo and W, while Nb (along with V and Ta) show no minimum in the $\hkl<110> \rightarrow \hkl<111>$ rotation path.

The balance between the relaxation volumes of vacancies and interstitial atoms determines the macroscopic swelling of a material during irradiation. Therefore, Tabs.~\ref{tab:vacancy} and \ref{tab:SIA} also list the relaxation volumes of single vacancies and the \hkl<111> self-interstitial configuration. Overall, the GAPs are consistent with DFT. Irradiation further results in clusters of defects. To ensure that the GAPs are applicable to radiation damage studies, it is crucial to test properties beyond single vacancies and interstitials. Fig.~\ref{fig:divac} shows the binding energies of divacancies at different nearest-neighbor (NN) separations compared between GAP and DFT (Fig.~\ref{fig:divac}a) and EAM and DFT (Fig.~\ref{fig:divac}b). The divacancy interaction is strikingly different between the group 5 and group 6 elements. Divacancies show only weak interactions in W and Mo, with the 2NN divacancy even being strongly repulsive. In contrast, 2NN divacancies in V, Nb, and Ta are the most favorable configuration with binding energies close to 0.5 eV. The GAPs successfully capture this group-specific trend, which is a noteworthy improvement over traditional EAM potentials that predict strong and almost equal binding for 1NN and 2NN divacancies in all elements, as is clear from Fig.~\ref{fig:divac}b (see also the Supplemental Material~\cite{supplemental} for results from other interatomic potentials).

Previously~\cite{byggmastar_machine-learning_2019}, we found that the W GAP shows good transferability to larger self-interstitial clusters, despite being trained to only single and di-interstitial configurations. With the same training structures, it is reasonable to assume that the remaining GAPs show similar transferability. Nevertheless, we tested this assumption. DFT data for self-interstitial clusters in V are available from Ref.~\cite{alexander_ab_2016}. Figure~\ref{fig:sia_clusters_V} shows formation energies of interstitial-type $1/2\hkl<111>$ and \hkl<100> dislocation loops, and the C15 Laves phase clusters~\cite{marinica_irradiation-induced_2012} in V. DFT data for \hkl<100> loops is not available, and Fig.~\ref{fig:sia_clusters_V} only shows GAP data for symmetric \hkl<100> loops (asymmetric loops collapse during relaxation). The formation energies for $1/2\hkl<111>$ loops and C15 clusters obtained by the GAP closely match the DFT energies in the available size range, and shows reasonable extrapolation to cluster sizes beyond the DFT range. Hence, we conclude that the GAPs are well-suited for simulations of radiation damage with point defects and defect clusters of arbitrary sizes.

\subsection{Surfaces}

\begin{figure}
    \centering
        \begin{subfigure}{\linewidth}
        \centering
        \includegraphics[width=\linewidth]{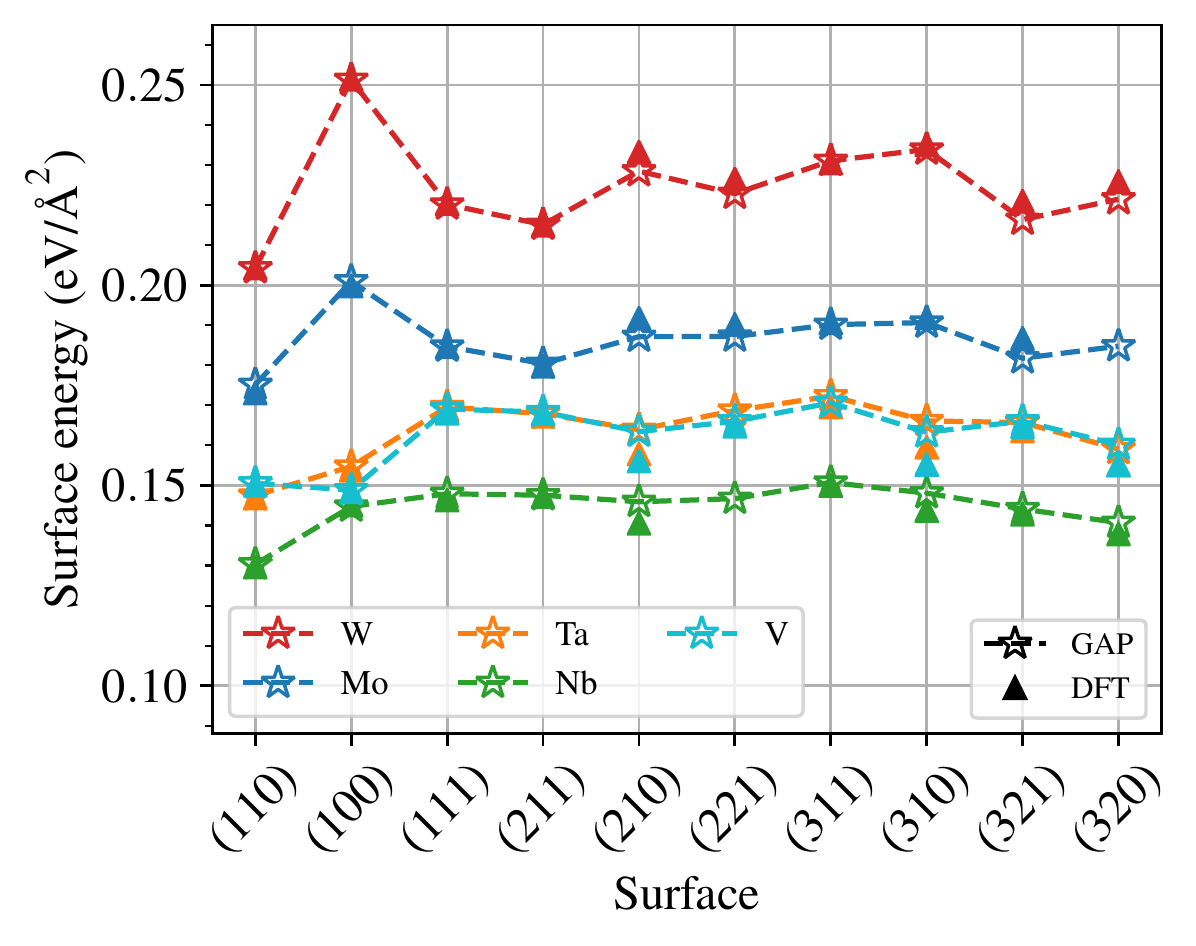}
        \caption{}
    \end{subfigure}
    \begin{subfigure}{\linewidth}
        \centering
        \includegraphics[width=\linewidth]{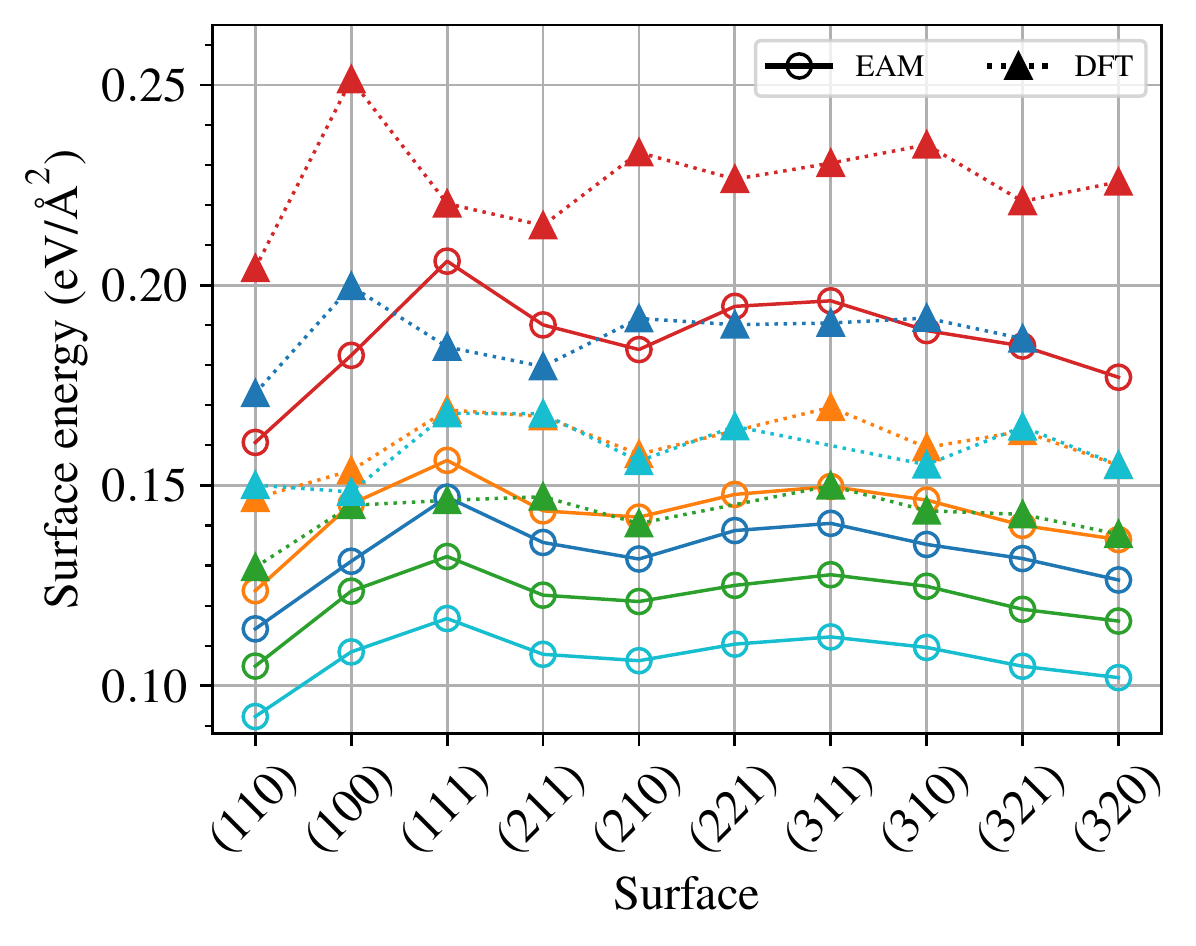}
        \caption{}
    \end{subfigure}
    \caption{Surface energies of ten different surfaces. (a) Comparison between GAP (stars) and our DFT results (solid triangles). The first four surfaces were included in the training data. (b) Comparison between the EAM potentials from Ref.~\cite{ackland_improved_1987} and DFT.}
    \label{fig:esurf}
\end{figure}

Surface energies for the ten lowest-index surfaces are shown in Fig.~\ref{fig:esurf}a, compared between GAP and our DFT results. The training database includes only the \hkl(100), \hkl(110), \hkl(111), and \hkl(211) surfaces, but GAP shows good transferability to other surfaces too. The largest discrepancies between GAP and DFT are seen for the \hkl(210) and \hkl(310) surfaces, although the errors are only about 5 meV/Å$^2$ (translating to around 10 meV/atom).

Reproducing surface energies is often difficult for analytical potentials. Fig.~\ref{fig:esurf}b shows results using EAM potentials~\cite{ackland_improved_1987}. In all elements, the EAM strongly underestimates the surface energies, and always predicts the \hkl(100) surface to be lower in energy than the \hkl(111) surface, which is true for the group 5 elements but not true for Mo and W. In the Supplemental Material~\cite{supplemental}, we show results obtained using a number of other interatomic potentials. These results show that although some traditional potentials do reproduce a reasonable average surface energy, the order of stability of different surfaces is only reproduced by machine-learning-based potentials.

The close-packed \hkl(110) surface is the most stable surface in all elements except V, where DFT interestingly shows that the \hkl(100) surface is slightly lower in energy at zero Kelvin. There is evidence and some controversy whether the V\hkl(100) surface is ferromagnetic~\cite{mathon_magnetism_1988}. However, investigating surface magnetism and its effects on surface stability is both beyond the scope of this work and beyond the capabilities of GAP, and will not be considered here.

\subsection{Repulsive potential}

\begin{figure}
    \centering
    \includegraphics[width=\linewidth]{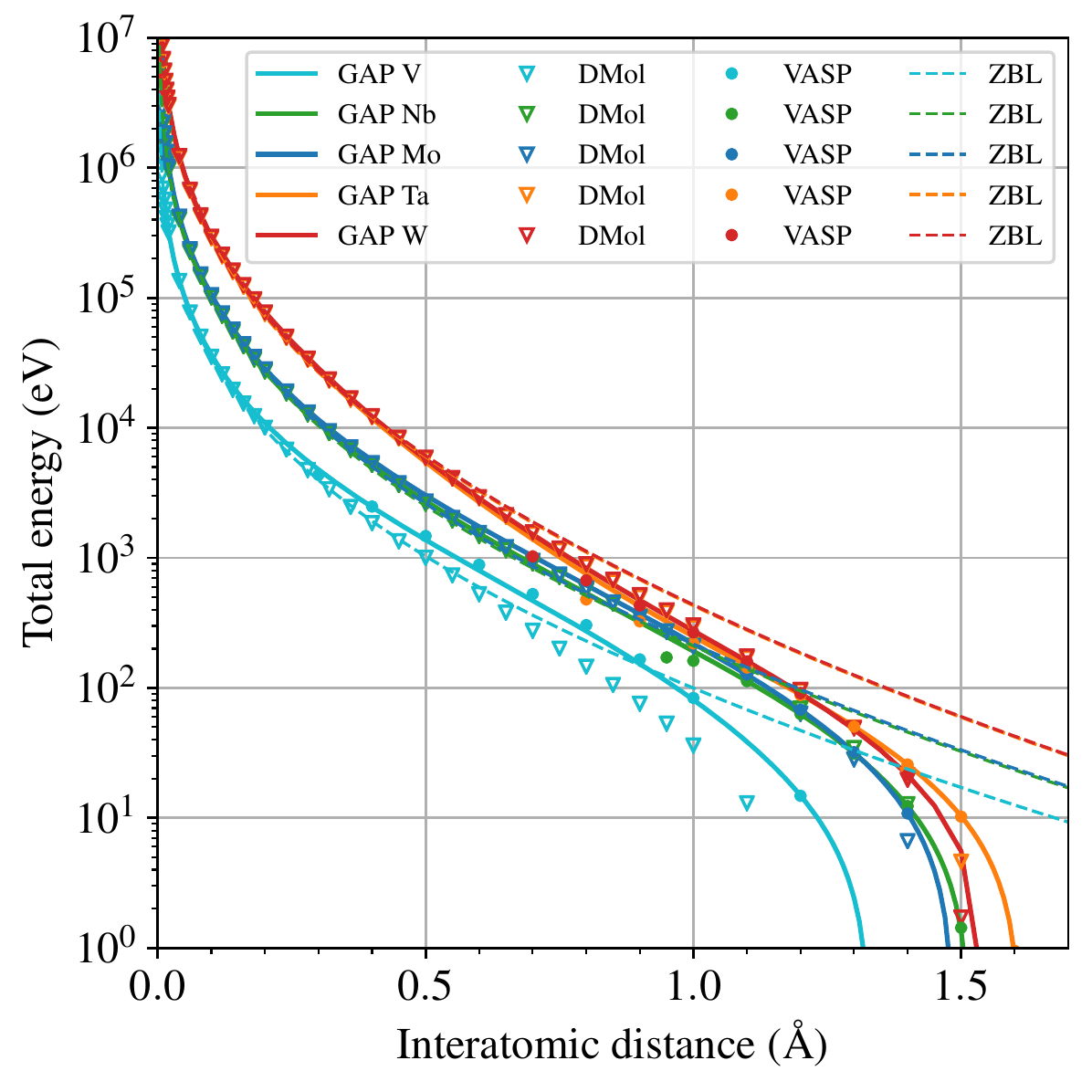}
    \caption{Repulsive parts of the dimer curves compared between GAP, data from all-electron DFT using DMol~\cite{nordlund_repulsive_1997}, \textsc{vasp}, and the universal ZBL potential~\cite{ziegler_stopping_1985}.}
    \label{fig:reppots}
\end{figure}

\begin{figure}
    \centering
    \includegraphics[width=\linewidth]{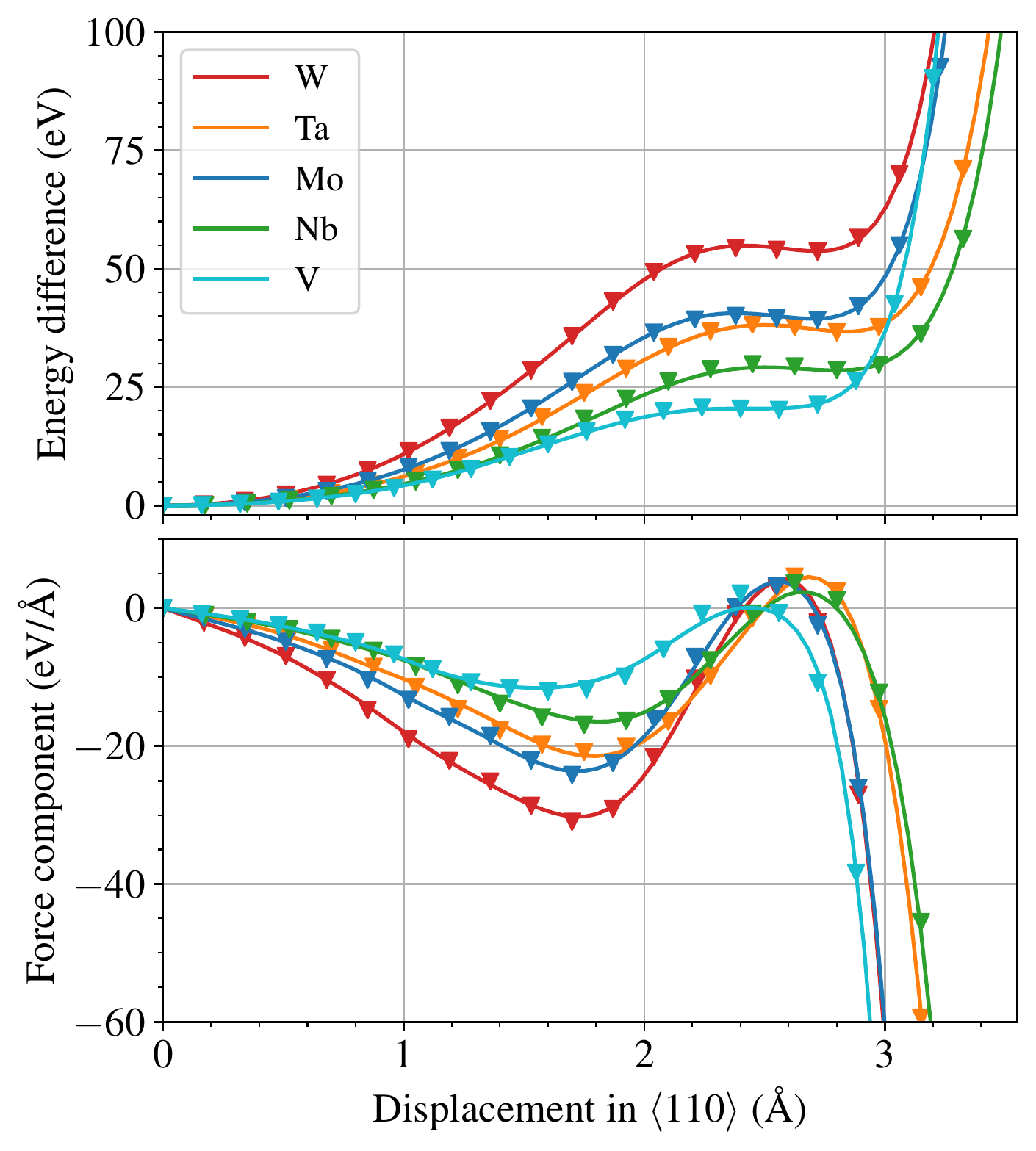}
    \caption{Energy difference (top) and force (bottom) for an atom statically displaced from its lattice position along the \hkl<110> direction in bcc. Lines are GAP data and points are DFT.}
    \label{fig:mbreppot}
\end{figure}

\begin{figure}
    \centering
    \includegraphics[width=\linewidth]{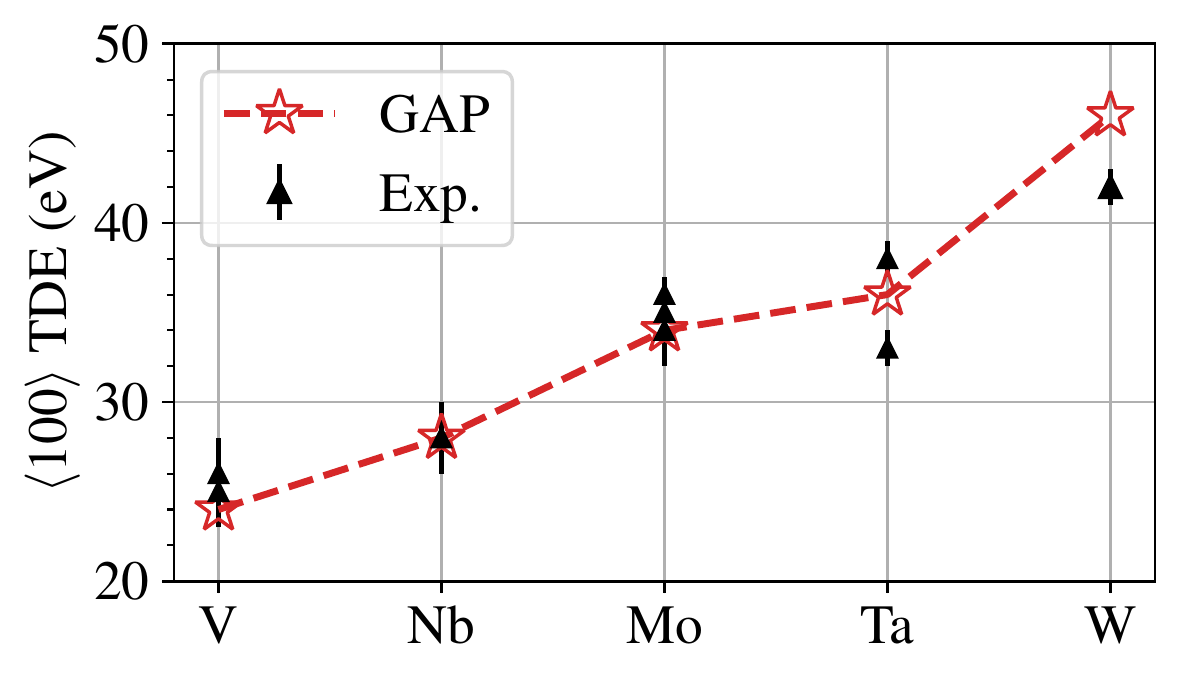}
    \caption{Threshold displacement energies in the \hkl<100> direction obtained with GAP compared to experimental data: V~\cite{miller_defect_1974,jung_damage_1975}, Nb~\cite{jung_damage_1975}, Mo~\cite{maury_anisotropy_1975,jung_damage_1975,biget_study_1974}, Ta~\cite{biget_near-threshold_1979}, W~\cite{maury_frenkel_1978}. The uncertainty of the GAP values is $\pm1$ eV.}
    \label{fig:TDE}
\end{figure}

All GAPs are augmented with screened Coulomb potentials fitted to the all-electron DFT DMol data~\cite{nordlund_repulsive_1997}, with a smooth machine-learned connection to the near-equilibrium energies. A smooth connection is ensured by including structures with short interatomic distances in the training database, and then only machine-learning the difference between the screened Coulomb potential and the \textsc{vasp} training data~\cite{byggmastar_machine-learning_2019}. These structures include simple dimers and bcc crystals with randomly placed interstitial atoms that are close, but not too close, to the nearest atom. The closest allowed distance in the training structures is chosen by comparing \textsc{vasp} data with the all-electron DMol data, to see where \textsc{vasp} becomes unreliable (due to frozen core electrons) and starts diverging from DMol. Figure~\ref{fig:reppots} shows the repulsive part of the dimer curves in GAP, DFT-DMol, DFT-\textsc{vasp}, and the commonly used universal ZBL screened Coulomb potential~\cite{ziegler_stopping_1985}. For all elements except V, there is good agreement between \textsc{vasp} and DMol in the 1--1.4 Å range. Consequently, GAP also closely overlaps with both and follows DMol at $<1$ Å when \textsc{vasp} becomes unreliable, and vice versa at $>1.4$ Å. In V, there is some difference between \textsc{vasp} and DMol, and consequently a similar difference between GAP and DMol as GAP is trained to \textsc{vasp} data for distances above 1.1 Å.

We test the accuracy of repulsion inside a crystal by displacing an atom along the \hkl<110> direction and tracking the change in energy as well as the total force on the displaced atom. The results are shown in Fig.~\ref{fig:mbreppot}, with GAP data as lines and DFT data as points. All GAPs closely follow the DFT data, and show only small discrepancies in the forces. The \hkl<110> direction is chosen as it involves many-body interactions with the nearest neighbors along the displacement path (producing local maxima and minima in Fig.~\ref{fig:mbreppot}), before colliding head-on with the \hkl<110> neighbor.

To dynamically test the repulsive parts of the GAPs and the creation of defects, we simulate the minimum threshold displacement energies (TDEs) according to the methods described in Ref.~\cite{nordlund_molecular_2006}. We use a 2 eV increment in the trial kinetic energies, yielding an uncertainty of $\pm 1$ eV in the TDEs. The lowest TDE is known to be in the \hkl<100> direction in bcc metals, which the W GAP reproduces~\cite{byggmastar_machine-learning_2019}. Experimental data is available for all five elements from low-temperature (typically 4 K) electron irradiation experiments~\cite{miller_defect_1974,jung_damage_1975,maury_anisotropy_1975,biget_study_1974,biget_near-threshold_1979,maury_frenkel_1978}. Previously we found that simulations of the TDEs at 0 K and 40 K yielded within the statistical uncertainties identical results in W~\cite{byggmastar_machine-learning_2019}. We therefore use 0 K in the simulations, which is then directly comparable to the experiments. However, at 0 K the TDE in exactly the \hkl<100> direction is sometimes significantly higher than for a few degrees away from \hkl<100>. Additionally, in the experimental data there is always a significant uncertainty in the direction due to spreading of the electron beam. We therefore simulate six different directions within 10 degrees from \hkl<100> and report the minimum TDEs. Figure~\ref{fig:TDE} shows the results compared with the experimental measurements. The predictions by the GAPs are in excellent agreement with experiments, with only a small overestimation of the single experimental point in W. This indicates the GAPs can be used to obtain accurate threshold displacement energies also for other directions that are unattainable from experimental measurements.

\subsection{Melting and liquid properties}
\label{sec:liquid}

\begin{table}
    \centering
    \caption{Liquid properties compared between GAP and experiments. Melting temperature ($T_\mathrm{melt}$), density of the liquid phase at the melting temperature ($\rho_\mathrm{l}$), and enthalpy of fusion ($\Delta H_\mathrm{f}$). Experimental data are from Ref.~\cite{rumble_crc_2019} unless otherwise indicated.}
    \begin{tabular}{'l^l^l^l^l^l}
     \toprule
     & V & Nb & Mo & Ta & W \\
     \midrule
     \rowstyle{\bfseries}
     $T_\mathrm{melt}$ \textnormal{(K)} & 2130 & 2550 & 2750 & 3010 & 3540 \\
     \rowstyle{\itshape}
                           & 2183 & 2750 & 2895 & 3290 & 3687 \\
     $T_\mathrm{melt}$ \textnormal{error} & $-$2.4\% & $-$7.3\% & $-$5.0\% & $-$8.5\% & $-$4.0\% \\
     \rowstyle{\bfseries}
     $\rho_\mathrm{l}$ \textnormal{(g/cm$^3$)} & 5.68 & 7.62 & 8.92 & 14.62 & 16.49 \\
     \rowstyle{\itshape}
                                  & 5.5  & 7.62* & 9.33 & 15, 14.4* & 17.6, 16.2* \\
     \rowstyle{\bfseries}
     $\Delta H_\mathrm{f}$ \textnormal{(eV/atom)} & 0.24 & 0.29 & 0.38 & 0.30 & 0.50 \\
     \rowstyle{\itshape}
                                     & 0.22 & 0.31 & 0.39 & 0.38 & 0.54 \\
     \bottomrule
     * Ref.~\cite{vinet_surface_1993}
    \end{tabular}
    \label{tab:melting}
\end{table}

Calculating various properties of the liquid phase provides a stringent test of how well the GAPs describe arbitrary low-symmetry local atomic geometries. The melting point at zero pressure, the density of the liquid phase at the melting temperature, and the heat of fusion (latent heat) for all five elements are listed in Tab.~\ref{tab:melting}. The data predicted by the GAPs are compared with experimental data~\cite{rumble_crc_2019,vinet_surface_1993}. The melting temperatures are simulated using the liquid-crystalline interface method with systems of 1372 atoms. The heat of fusion is calculated as the average energy difference between completely molten and completely crystalline systems simulated at the melting temperature.

The GAPs provide melting temperatures in close agreement with experimental measurements, although they are consistently underestimated by 2--9\%. It is likely that this systematic underestimation is inherited from the underlying DFT training data, given that similar observations have been made previously using DFT and attributed to the PBE functional~\cite{jinnouchi_on-the-fly_2019}. The experimental liquid densities and heats of fusion are quantitatively well reproduced by the GAPs.

\begin{figure}
    \centering
    \includegraphics[width=\linewidth]{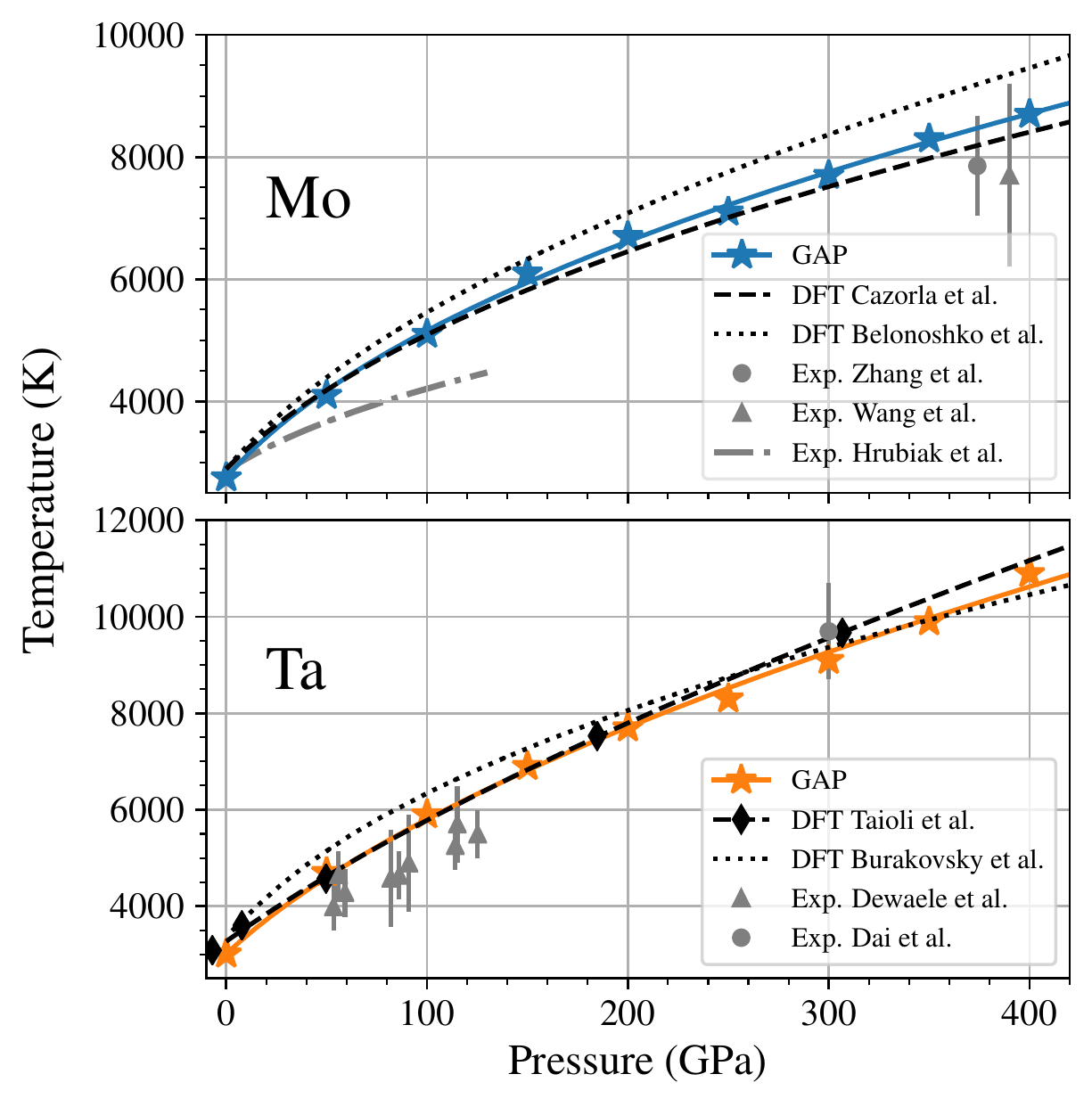}
    \caption{Melting curves of Mo and Ta compared between GAP, DFT results~\cite{cazorla_ab_2007,belonoshko_molybdenum_2008,taioli_melting_2007,burakovsky_high-pressurehigh-temperature_2010}, and experimental measurements~\cite{zhang_temperature_2008,wang_x-ray_2015,hrubiak_microstructures_2017,dewaele_high_2010,dai_hugoniot_2009}.}
    \label{fig:melt_Mo-Ta}
\end{figure}

\begin{table}
    \centering
    \caption{Fitting parameters of the $T_\mathrm{melt}(P) = T_0 \left( 1 + P / \eta \right)^\zeta$ melting curves in Fig.~\ref{fig:melt_gap}.}
    \begin{tabular}{llllll}
        \toprule
        & V & Nb & Mo & Ta & W \\
        \midrule
        $\eta$ (GPa) & 56.89 & 47.81 & 31.62 & 40.88 & 32.48  \\
        $\zeta$ & 0.52 & 0.58 & 0.44 & 0.53 & 0.47 \\
        \bottomrule
    \end{tabular}
    \label{tab:tmelt_fits}
\end{table}

\begin{figure}
    \centering
    \includegraphics[width=\linewidth]{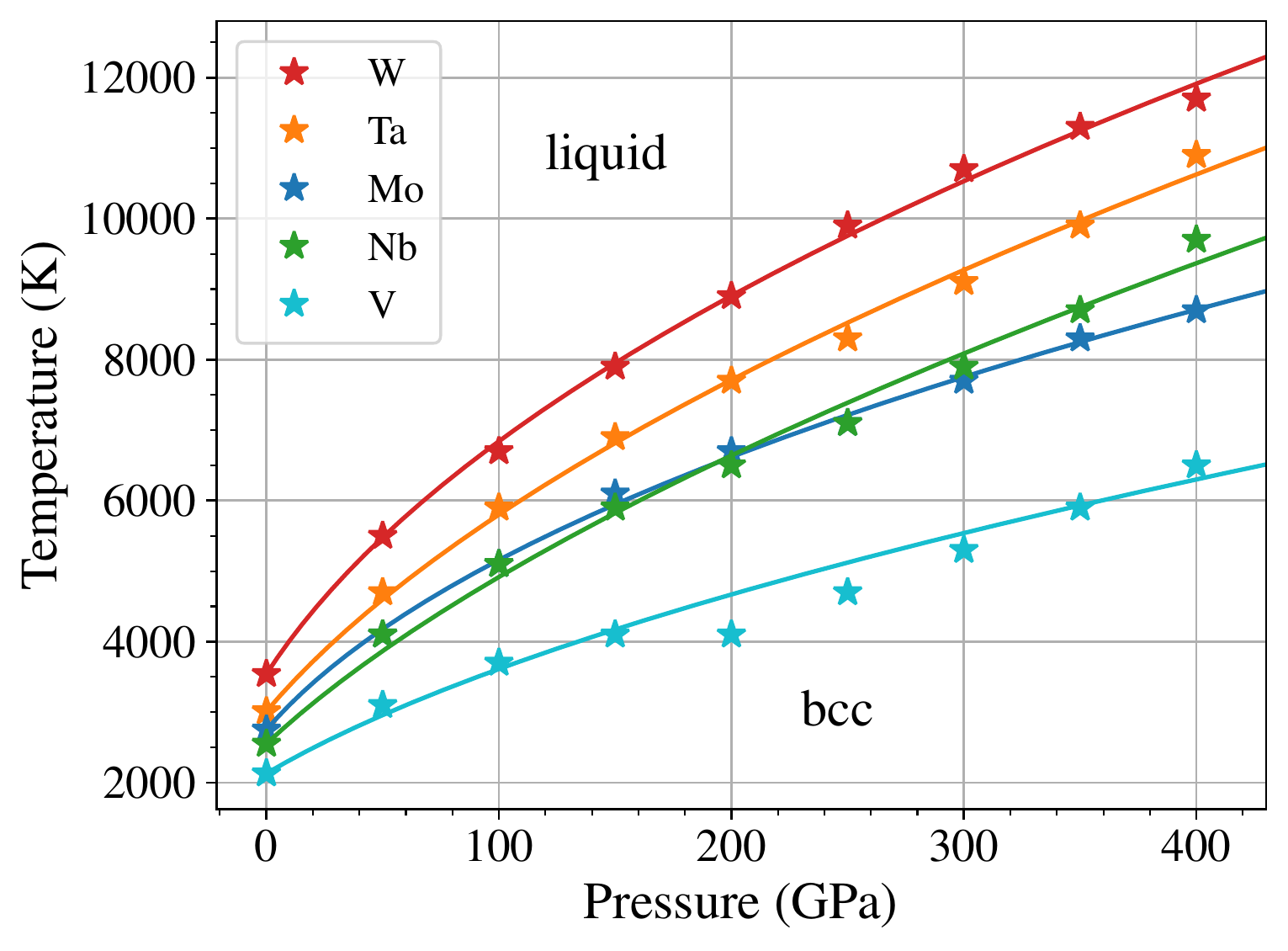}
    \caption{Melting curves simulated with the GAPs. The parameters of the fitted curves are listed in Tab.~\ref{tab:tmelt_fits}.}
    \label{fig:melt_gap}
\end{figure}

There has been a considerable interest in studying the high-pressure phase diagram of transition metals. This has partly been motivated by extreme discrepancies between experimental diamond anvil cell (DAC) measurements, shock-melting measurements, and theoretical estimates using DFT~\cite{errandonea_systematics_2001,cazorla_ab_2007,taioli_melting_2007,errandonea_melting_2019}. Obtaining reliable melting temperatures in DFT is nontrivial due to limitations in system size and simulation time. Nevertheless, successful methods have been developed to overcome this, for example by calculating free-energy contributions from DFT to correct results from classical potentials~\cite{alfe_complementary_2002,cazorla_ab_2007}, or by exploiting conditions of superheating in the so-called $Z$ method~\cite{belonoshko_melting_2006}. Using the GAPs we can use system sizes and time scales beyond reach of DFT, and directly simulate the melting temperatures at a desired pressure in MD simulations. We use the same liquid-solid system (1372 atoms) as in the zero-pressure melting simulations, and determine the melting temperatures in $NPT$ simulations for pressures in the range 0--400 GPa at intervals of 50 GPa.

Among the five elements, Mo and Ta have received most attention, and high-pressure melting curves have been calculated in a number of DFT studies and compared with experimental data~\cite{cazorla_ab_2007,belonoshko_molybdenum_2008,taioli_melting_2007,burakovsky_high-pressurehigh-temperature_2010,zhang_temperature_2008,wang_x-ray_2015,hrubiak_microstructures_2017,dewaele_high_2010,dai_hugoniot_2009}. As a validation of our GAPs, we therefore first focus on the melting curves of Mo and Ta. Fig.~\ref{fig:melt_Mo-Ta} shows the results, compared with DFT results and experimental data. In both Mo and Ta, the GAP curves fall between the two DFT curves and agree well with the experimental shock-melting data points at high pressures. This is encouraging for two reasons. First, the liquid structures in the GAP training databases only cover pressures up to around 100 GPa, yet the GAP shows good transferability up to at least 400 GPa. This can partly be attributed to the accurate repulsive part of our GAPs, which become increasingly important at extreme pressures. Second, the GAP results can be seen as further validation of the above-mentioned DFT methods for obtaining melting curves, given that the GAPs are trained to tightly converged DFT data and the melting curves are obtained in systems of relatively large size. Figure~\ref{fig:melt_Mo-Ta} shows two DFT curves each for Mo and Ta, where one is obtained using the $Z$ method (Belonoshko et al.~\cite{belonoshko_molybdenum_2008} and Burakovsky et al.~\cite{burakovsky_high-pressurehigh-temperature_2010}) and the other by the thermodynamic correction of a reference classical potential (Cazorla et al.~\cite{cazorla_ab_2007} and Taioli et al.~\cite{taioli_melting_2007}). The GAP curves are close to both, although it can be noted that they are in somewhat closer agreement with the latter DFT method. Compared to experiments, GAP and DFT is consistent with the shock-melting points at 300--400 GPa, but overestimates the DAC data at lower pressure, especially in Mo.

Having validated the accuracy of GAP for high-pressure melting, we simulate the melting curves for all five elements. The results are shown in Fig.~\ref{fig:melt_gap}. The data points are fitted to the commonly used Simon-Glatzel equation~\cite{simon_bemerkungen_1929} $T_\mathrm{melt}(P) = T_0 \left( 1 + P / \eta \right)^\zeta$, where $T_0$ is the melting temperature at zero pressure and $\eta$ and $\zeta$ are fitting parameters. The fitting parameters are listed in Tab~\ref{tab:tmelt_fits}. The melting curves at nonzero pressures follow the order of the zero-pressure melting points, with only Nb and Mo showing a crossover. V shows a peculiar trend in the 100--250 GPa range. It is unclear whether this is an artefact of the GAP. Melting curves for V have also been obtained in very recent studies~\cite{errandonea_melting_2019,zhang_melting_2019}. Errandonea et al.~\cite{errandonea_melting_2019} presented both DFT and experimental results, and obtained good agreement between the two. However, the melting curve obtained in the DFT study of Zhang et al.~\cite{zhang_melting_2019} is significantly lower. The GAP curve agrees well with the latter, and hence underestimates the experimental data from Ref.~\cite{errandonea_melting_2019}.

We note that the phase diagram of some of the elements may contain other crystalline phases than bcc. At least V has been seen to stabilise a rhombohedral phase in a certain pressure-temperature region~\cite{errandonea_melting_2019}. Solid--solid phase transitions have also been suggested to occur in Mo and Ta~\cite{hixson_acoustic_1989,belonoshko_molybdenum_2008,burakovsky_high-pressurehigh-temperature_2010}, although further theoretical investigations refute this~\cite{cazorla_constraints_2012,haskins_polymorphism_2012} and some experimental studies report no evidence of phases other than bcc~\cite{hrubiak_microstructures_2017,dewaele_high_2010}. Although the GAPs are trained to some crystal structures other than bcc, it is difficult to assess their reliability in describing arbitrary crystal symmetries at high pressures. Hence, we make no attempt to investigate possible solid-solid phase transitions in the high-pressure phase diagram.

\section{Discussion and outlook}
\label{sec:concl}

We have developed machine-learning interatomic potentials for five nonmagnetic bcc metals, and demonstrated their performance for a wide range of properties. The potentials show good accuracy for the targeted properties, which include basic elastic and thermal properties, energetics of defects and surfaces, and properties of the liquid phase. They also show good transferability to properties that are not directly covered by the structures in the training database. Examples include formation energies of self-interstitial clusters and melting at extreme pressures. Both are crucial properties in simulations of radiation damage. During the heat spike of an energetic radiation-induced collision cascade, the cascade core is an extremely hot liquid-like region with large pressure gradients. The extent and morphology of the defects surviving the cascade is affected by the efficiency of atomic mixing and recrystallization of the hot cascade core, and the energetics of defect clusters~\cite{bjorkas_assessment_2009,byggmastar_dynamical_2020}, which based on our results should be well described by our GAPs.

We attribute the transferability of our GAPs to a diverse DFT training database that in many ways put physical constraints on the machine-learning predictions. The analytical screened Coulomb potential for the short-range repulsion combined with short-range structures in the training database ensure that the GAPs produce physically reasonable predictions for systems containing short interatomic distances. Furthermore, our training database includes liquid-like structures at various densities. Fitting the forces of liquid structures has long been a successful strategy for developing robust and transferable analytical potentials, in particular embedded atom method potentials~\cite{ercolessi_interatomic_1994,mendelev_development_2003}. A key difference between fitting machine-learning potentials and analytical potentials is, however, that for the latter one can rely on the physical constraints and transferability provided by the analytical functions. In a machine-learning potential, physical constraints must be imposed by the training database. When training liquids, we found that only including physically relevant liquid densities is not enough. This sometimes resulted in a potential that predicted clearly artificial low-density structures to be even lower in energy than the ground state bcc phase. Our liquid structures in the training database therefore cover a wide range of densities, from relatively dense liquids to unphysically low densities. Both extremes are significantly higher in energy than the zero-pressure equilibrium liquid (by up to several eV/atom), and therefore effectively constrain the predictions by the GAP.

In the community of machine-learning potentials, major efforts are focused on strategies for automating the training process~\cite{bernstein_novo_2019,vandermause_--fly_2020,deringer_data-driven_2018,zhang_active_2019}. This is motivated by the hope of bypassing the need for time-consuming expert human input required to fit and benchmark a robust interatomic potential. Our strategy is very much based on informed human decisions when constructing the training database. Nevertheless, our work also demonstrates a natural and alternative way of minimising the time invested on fitting potentials. In this article, we almost completely rely on our previously developed database of structures for W~\cite{byggmastar_machine-learning_2019} to train a set of similar potentials for a family of similar metals, with little human effort.

The training structures and potential files are freely available as Supplemental Material~\cite{supplemental} from Ref.~\cite{gitlab_gap}. We anticipate that the training structures may be valuable for training potentials using other machine-learning frameworks with little effort, for which our results can serve as a useful benchmark for the performance of the GAP framework. Our set of training structures and potentials also serve as starting points for the development of potentials for bcc alloys.

\section*{Acknowledgements}

This work has been carried out within the framework of the EUROfusion Consortium and has received funding from the Euratom research and training programme 2014-2018 and 2019-2020 under grant agreement No 633053. The views and opinions expressed herein do not necessarily reflect those of the European Commission. Grants of computer capacity from CSC - IT Center for Science, Finland, as well as from the Finnish Grid and Cloud Infrastructure (persistent identifier urn:nbn:fi:research-infras-2016072533) are gratefully acknowledged.

\bibliography{mybib}

\begin{thebibliography}{104}%
\makeatletter
\providecommand \@ifxundefined [1]{%
 \@ifx{#1\undefined}
}%
\providecommand \@ifnum [1]{%
 \ifnum #1\expandafter \@firstoftwo
 \else \expandafter \@secondoftwo
 \fi
}%
\providecommand \@ifx [1]{%
 \ifx #1\expandafter \@firstoftwo
 \else \expandafter \@secondoftwo
 \fi
}%
\providecommand \natexlab [1]{#1}%
\providecommand \enquote  [1]{``#1''}%
\providecommand \bibnamefont  [1]{#1}%
\providecommand \bibfnamefont [1]{#1}%
\providecommand \citenamefont [1]{#1}%
\providecommand \href@noop [0]{\@secondoftwo}%
\providecommand \href [0]{\begingroup \@sanitize@url \@href}%
\providecommand \@href[1]{\@@startlink{#1}\@@href}%
\providecommand \@@href[1]{\endgroup#1\@@endlink}%
\providecommand \@sanitize@url [0]{\catcode `\\12\catcode `\$12\catcode
  `\&12\catcode `\#12\catcode `\^12\catcode `\_12\catcode `\%12\relax}%
\providecommand \@@startlink[1]{}%
\providecommand \@@endlink[0]{}%
\providecommand \url  [0]{\begingroup\@sanitize@url \@url }%
\providecommand \@url [1]{\endgroup\@href {#1}{\urlprefix }}%
\providecommand \urlprefix  [0]{URL }%
\providecommand \Eprint [0]{\href }%
\providecommand \doibase [0]{https://doi.org/}%
\providecommand \selectlanguage [0]{\@gobble}%
\providecommand \bibinfo  [0]{\@secondoftwo}%
\providecommand \bibfield  [0]{\@secondoftwo}%
\providecommand \translation [1]{[#1]}%
\providecommand \BibitemOpen [0]{}%
\providecommand \bibitemStop [0]{}%
\providecommand \bibitemNoStop [0]{.\EOS\space}%
\providecommand \EOS [0]{\spacefactor3000\relax}%
\providecommand \BibitemShut  [1]{\csname bibitem#1\endcsname}%
\let\auto@bib@innerbib\@empty
\bibitem [{\citenamefont {Behler}(2016)}]{behler_perspective_2016}%
  \BibitemOpen
  \bibfield  {author} {\bibinfo {author} {\bibfnamefont {J.}~\bibnamefont
  {Behler}},\ }\href {https://doi.org/10.1063/1.4966192} {\bibfield  {journal}
  {\bibinfo  {journal} {J. Chem. Phys.}\ }\textbf {\bibinfo {volume} {145}},\
  \bibinfo {pages} {170901} (\bibinfo {year} {2016})}\BibitemShut {NoStop}%
\bibitem [{\citenamefont {Bart{\'o}k}\ \emph {et~al.}(2013)\citenamefont
  {Bart{\'o}k}, \citenamefont {Kondor},\ and\ \citenamefont
  {Cs{\'a}nyi}}]{bartok_representing_2013}%
  \BibitemOpen
  \bibfield  {author} {\bibinfo {author} {\bibfnamefont {A.~P.}\ \bibnamefont
  {Bart{\'o}k}}, \bibinfo {author} {\bibfnamefont {R.}~\bibnamefont {Kondor}},\
  and\ \bibinfo {author} {\bibfnamefont {G.}~\bibnamefont {Cs{\'a}nyi}},\
  }\bibfield  {journal} {\bibinfo  {journal} {Phys. Rev. B}\ }\textbf {\bibinfo
  {volume} {87}},\ \href {https://doi.org/10.1103/PhysRevB.87.184115}
  {10.1103/PhysRevB.87.184115} (\bibinfo {year} {2013})\BibitemShut {NoStop}%
\bibitem [{\citenamefont {Behler}\ and\ \citenamefont
  {Parrinello}(2007)}]{behler_generalized_2007}%
  \BibitemOpen
  \bibfield  {author} {\bibinfo {author} {\bibfnamefont {J.}~\bibnamefont
  {Behler}}\ and\ \bibinfo {author} {\bibfnamefont {M.}~\bibnamefont
  {Parrinello}},\ }\bibfield  {journal} {\bibinfo  {journal} {Phys. Rev.
  Lett.}\ }\textbf {\bibinfo {volume} {98}},\ \href
  {https://doi.org/10.1103/PhysRevLett.98.146401}
  {10.1103/PhysRevLett.98.146401} (\bibinfo {year} {2007})\BibitemShut
  {NoStop}%
\bibitem [{\citenamefont {S.~Smith}\ \emph {et~al.}(2017)\citenamefont
  {S.~Smith}, \citenamefont {Isayev},\ and\ \citenamefont
  {E.~Roitberg}}]{ssmith_ani-1_2017}%
  \BibitemOpen
  \bibfield  {author} {\bibinfo {author} {\bibfnamefont {J.}~\bibnamefont
  {S.~Smith}}, \bibinfo {author} {\bibfnamefont {O.}~\bibnamefont {Isayev}},\
  and\ \bibinfo {author} {\bibfnamefont {A.}~\bibnamefont {E.~Roitberg}},\
  }\href {https://doi.org/10.1039/C6SC05720A} {\bibfield  {journal} {\bibinfo
  {journal} {Chem. Sci.}\ }\textbf {\bibinfo {volume} {8}},\ \bibinfo {pages}
  {3192} (\bibinfo {year} {2017})}\BibitemShut {NoStop}%
\bibitem [{\citenamefont {Zhang}\ \emph {et~al.}(2018)\citenamefont {Zhang},
  \citenamefont {Han}, \citenamefont {Wang}, \citenamefont {Car},\ and\
  \citenamefont {E}}]{zhang_deep_2018}%
  \BibitemOpen
  \bibfield  {author} {\bibinfo {author} {\bibfnamefont {L.}~\bibnamefont
  {Zhang}}, \bibinfo {author} {\bibfnamefont {J.}~\bibnamefont {Han}}, \bibinfo
  {author} {\bibfnamefont {H.}~\bibnamefont {Wang}}, \bibinfo {author}
  {\bibfnamefont {R.}~\bibnamefont {Car}},\ and\ \bibinfo {author}
  {\bibfnamefont {W.}~\bibnamefont {E}},\ }\bibfield  {journal} {\bibinfo
  {journal} {Phys. Rev. Lett.}\ }\textbf {\bibinfo {volume} {120}},\ \href
  {https://doi.org/10.1103/PhysRevLett.120.143001}
  {10.1103/PhysRevLett.120.143001} (\bibinfo {year} {2018})\BibitemShut
  {NoStop}%
\bibitem [{\citenamefont {Sch{\"u}tt}\ \emph {et~al.}(2018)\citenamefont
  {Sch{\"u}tt}, \citenamefont {Sauceda}, \citenamefont {Kindermans},
  \citenamefont {Tkatchenko},\ and\ \citenamefont
  {M{\"u}ller}}]{schutt_schnet_2018}%
  \BibitemOpen
  \bibfield  {author} {\bibinfo {author} {\bibfnamefont {K.~T.}\ \bibnamefont
  {Sch{\"u}tt}}, \bibinfo {author} {\bibfnamefont {H.~E.}\ \bibnamefont
  {Sauceda}}, \bibinfo {author} {\bibfnamefont {P.-J.}\ \bibnamefont
  {Kindermans}}, \bibinfo {author} {\bibfnamefont {A.}~\bibnamefont
  {Tkatchenko}},\ and\ \bibinfo {author} {\bibfnamefont {K.-R.}\ \bibnamefont
  {M{\"u}ller}},\ }\href {https://doi.org/10.1063/1.5019779} {\bibfield
  {journal} {\bibinfo  {journal} {J. Chem. Phys.}\ }\textbf {\bibinfo {volume}
  {148}},\ \bibinfo {pages} {241722} (\bibinfo {year} {2018})}\BibitemShut
  {NoStop}%
\bibitem [{\citenamefont {Bart{\'o}k}\ \emph {et~al.}(2010)\citenamefont
  {Bart{\'o}k}, \citenamefont {Payne}, \citenamefont {Kondor},\ and\
  \citenamefont {Cs{\'a}nyi}}]{bartok_gaussian_2010}%
  \BibitemOpen
  \bibfield  {author} {\bibinfo {author} {\bibfnamefont {A.~P.}\ \bibnamefont
  {Bart{\'o}k}}, \bibinfo {author} {\bibfnamefont {M.~C.}\ \bibnamefont
  {Payne}}, \bibinfo {author} {\bibfnamefont {R.}~\bibnamefont {Kondor}},\ and\
  \bibinfo {author} {\bibfnamefont {G.}~\bibnamefont {Cs{\'a}nyi}},\ }\bibfield
   {journal} {\bibinfo  {journal} {Phys. Rev. Lett.}\ }\textbf {\bibinfo
  {volume} {104}},\ \href {https://doi.org/10.1103/PhysRevLett.104.136403}
  {10.1103/PhysRevLett.104.136403} (\bibinfo {year} {2010})\BibitemShut
  {NoStop}%
\bibitem [{\citenamefont {Glielmo}\ \emph {et~al.}(2018)\citenamefont
  {Glielmo}, \citenamefont {Zeni},\ and\ \citenamefont
  {De~Vita}}]{glielmo_efficient_2018}%
  \BibitemOpen
  \bibfield  {author} {\bibinfo {author} {\bibfnamefont {A.}~\bibnamefont
  {Glielmo}}, \bibinfo {author} {\bibfnamefont {C.}~\bibnamefont {Zeni}},\ and\
  \bibinfo {author} {\bibfnamefont {A.}~\bibnamefont {De~Vita}},\ }\bibfield
  {journal} {\bibinfo  {journal} {Phys. Rev. B}\ }\textbf {\bibinfo {volume}
  {97}},\ \href {https://doi.org/10.1103/PhysRevB.97.184307}
  {10.1103/PhysRevB.97.184307} (\bibinfo {year} {2018})\BibitemShut {NoStop}%
\bibitem [{\citenamefont {Vandermause}\ \emph {et~al.}(2020)\citenamefont
  {Vandermause}, \citenamefont {Torrisi}, \citenamefont {Batzner},
  \citenamefont {Xie}, \citenamefont {Sun}, \citenamefont {Kolpak},\ and\
  \citenamefont {Kozinsky}}]{vandermause_--fly_2020}%
  \BibitemOpen
  \bibfield  {author} {\bibinfo {author} {\bibfnamefont {J.}~\bibnamefont
  {Vandermause}}, \bibinfo {author} {\bibfnamefont {S.~B.}\ \bibnamefont
  {Torrisi}}, \bibinfo {author} {\bibfnamefont {S.}~\bibnamefont {Batzner}},
  \bibinfo {author} {\bibfnamefont {Y.}~\bibnamefont {Xie}}, \bibinfo {author}
  {\bibfnamefont {L.}~\bibnamefont {Sun}}, \bibinfo {author} {\bibfnamefont
  {A.~M.}\ \bibnamefont {Kolpak}},\ and\ \bibinfo {author} {\bibfnamefont
  {B.}~\bibnamefont {Kozinsky}},\ }\href
  {https://doi.org/10.1038/s41524-020-0283-z} {\bibfield  {journal} {\bibinfo
  {journal} {npj Comput Mater}\ }\textbf {\bibinfo {volume} {6}},\ \bibinfo
  {pages} {1} (\bibinfo {year} {2020})}\BibitemShut {NoStop}%
\bibitem [{\citenamefont {Thompson}\ \emph {et~al.}(2015)\citenamefont
  {Thompson}, \citenamefont {Swiler}, \citenamefont {Trott}, \citenamefont
  {Foiles},\ and\ \citenamefont {Tucker}}]{thompson_spectral_2015}%
  \BibitemOpen
  \bibfield  {author} {\bibinfo {author} {\bibfnamefont {A.~P.}\ \bibnamefont
  {Thompson}}, \bibinfo {author} {\bibfnamefont {L.~P.}\ \bibnamefont
  {Swiler}}, \bibinfo {author} {\bibfnamefont {C.~R.}\ \bibnamefont {Trott}},
  \bibinfo {author} {\bibfnamefont {S.~M.}\ \bibnamefont {Foiles}},\ and\
  \bibinfo {author} {\bibfnamefont {G.~J.}\ \bibnamefont {Tucker}},\ }\href
  {https://doi.org/10.1016/j.jcp.2014.12.018} {\bibfield  {journal} {\bibinfo
  {journal} {Journal of Computational Physics}\ }\textbf {\bibinfo {volume}
  {285}},\ \bibinfo {pages} {316} (\bibinfo {year} {2015})}\BibitemShut
  {NoStop}%
\bibitem [{\citenamefont {Shapeev}(2016)}]{shapeev_moment_2016}%
  \BibitemOpen
  \bibfield  {author} {\bibinfo {author} {\bibfnamefont {A.~V.}\ \bibnamefont
  {Shapeev}},\ }\href {https://doi.org/10.1137/15M1054183} {\bibfield
  {journal} {\bibinfo  {journal} {Multiscale Model. Simul.}\ }\textbf {\bibinfo
  {volume} {14}},\ \bibinfo {pages} {1153} (\bibinfo {year} {2016})},\ \Eprint
  {https://arxiv.org/abs/1512.06054} {arXiv:1512.06054} \BibitemShut {NoStop}%
\bibitem [{\citenamefont {Deringer}\ \emph {et~al.}(2019)\citenamefont
  {Deringer}, \citenamefont {Caro},\ and\ \citenamefont
  {Cs{\'a}nyi}}]{deringer_machine_2019}%
  \BibitemOpen
  \bibfield  {author} {\bibinfo {author} {\bibfnamefont {V.~L.}\ \bibnamefont
  {Deringer}}, \bibinfo {author} {\bibfnamefont {M.~A.}\ \bibnamefont {Caro}},\
  and\ \bibinfo {author} {\bibfnamefont {G.}~\bibnamefont {Cs{\'a}nyi}},\
  }\bibfield  {journal} {\bibinfo  {journal} {Adv. Mater.}\ }\href
  {https://doi.org/10.1002/adma.201902765} {10.1002/adma.201902765} (\bibinfo
  {year} {2019})\BibitemShut {NoStop}%
\bibitem [{\citenamefont {Mueller}\ \emph {et~al.}(2020)\citenamefont
  {Mueller}, \citenamefont {Hernandez},\ and\ \citenamefont
  {Wang}}]{mueller_machine_2020}%
  \BibitemOpen
  \bibfield  {author} {\bibinfo {author} {\bibfnamefont {T.}~\bibnamefont
  {Mueller}}, \bibinfo {author} {\bibfnamefont {A.}~\bibnamefont {Hernandez}},\
  and\ \bibinfo {author} {\bibfnamefont {C.}~\bibnamefont {Wang}},\ }\href
  {https://doi.org/10.1063/1.5126336} {\bibfield  {journal} {\bibinfo
  {journal} {The Journal of Chemical Physics}\ }\textbf {\bibinfo {volume}
  {152}},\ \bibinfo {pages} {050902} (\bibinfo {year} {2020})}\BibitemShut
  {NoStop}%
\bibitem [{\citenamefont {Chen}\ \emph {et~al.}(2017)\citenamefont {Chen},
  \citenamefont {Deng}, \citenamefont {Tran}, \citenamefont {Tang},
  \citenamefont {Chu},\ and\ \citenamefont {Ong}}]{chen_accurate_2017}%
  \BibitemOpen
  \bibfield  {author} {\bibinfo {author} {\bibfnamefont {C.}~\bibnamefont
  {Chen}}, \bibinfo {author} {\bibfnamefont {Z.}~\bibnamefont {Deng}}, \bibinfo
  {author} {\bibfnamefont {R.}~\bibnamefont {Tran}}, \bibinfo {author}
  {\bibfnamefont {H.}~\bibnamefont {Tang}}, \bibinfo {author} {\bibfnamefont
  {I.-H.}\ \bibnamefont {Chu}},\ and\ \bibinfo {author} {\bibfnamefont {S.~P.}\
  \bibnamefont {Ong}},\ }\href
  {https://doi.org/10.1103/PhysRevMaterials.1.043603} {\bibfield  {journal}
  {\bibinfo  {journal} {Phys. Rev. Materials}\ }\textbf {\bibinfo {volume}
  {1}},\ \bibinfo {pages} {043603} (\bibinfo {year} {2017})}\BibitemShut
  {NoStop}%
\bibitem [{\citenamefont {Wood}\ \emph {et~al.}(2019)\citenamefont {Wood},
  \citenamefont {Cusentino}, \citenamefont {Wirth},\ and\ \citenamefont
  {Thompson}}]{wood_data-driven_2019}%
  \BibitemOpen
  \bibfield  {author} {\bibinfo {author} {\bibfnamefont {M.~A.}\ \bibnamefont
  {Wood}}, \bibinfo {author} {\bibfnamefont {M.~A.}\ \bibnamefont {Cusentino}},
  \bibinfo {author} {\bibfnamefont {B.~D.}\ \bibnamefont {Wirth}},\ and\
  \bibinfo {author} {\bibfnamefont {A.~P.}\ \bibnamefont {Thompson}},\ }\href
  {https://doi.org/10.1103/PhysRevB.99.184305} {\bibfield  {journal} {\bibinfo
  {journal} {Phys. Rev. B}\ }\textbf {\bibinfo {volume} {99}},\ \bibinfo
  {pages} {184305} (\bibinfo {year} {2019})}\BibitemShut {NoStop}%
\bibitem [{\citenamefont {Zuo}\ \emph {et~al.}(2020)\citenamefont {Zuo},
  \citenamefont {Chen}, \citenamefont {Li}, \citenamefont {Deng}, \citenamefont
  {Chen}, \citenamefont {Behler}, \citenamefont {Cs{\'a}nyi}, \citenamefont
  {Shapeev}, \citenamefont {Thompson}, \citenamefont {Wood},\ and\
  \citenamefont {Ong}}]{zuo_performance_2020}%
  \BibitemOpen
  \bibfield  {author} {\bibinfo {author} {\bibfnamefont {Y.}~\bibnamefont
  {Zuo}}, \bibinfo {author} {\bibfnamefont {C.}~\bibnamefont {Chen}}, \bibinfo
  {author} {\bibfnamefont {X.}~\bibnamefont {Li}}, \bibinfo {author}
  {\bibfnamefont {Z.}~\bibnamefont {Deng}}, \bibinfo {author} {\bibfnamefont
  {Y.}~\bibnamefont {Chen}}, \bibinfo {author} {\bibfnamefont {J.}~\bibnamefont
  {Behler}}, \bibinfo {author} {\bibfnamefont {G.}~\bibnamefont {Cs{\'a}nyi}},
  \bibinfo {author} {\bibfnamefont {A.~V.}\ \bibnamefont {Shapeev}}, \bibinfo
  {author} {\bibfnamefont {A.~P.}\ \bibnamefont {Thompson}}, \bibinfo {author}
  {\bibfnamefont {M.~A.}\ \bibnamefont {Wood}},\ and\ \bibinfo {author}
  {\bibfnamefont {S.~P.}\ \bibnamefont {Ong}},\ }\href
  {https://doi.org/10.1021/acs.jpca.9b08723} {\bibfield  {journal} {\bibinfo
  {journal} {J. Phys. Chem. A}\ }\textbf {\bibinfo {volume} {124}},\ \bibinfo
  {pages} {731} (\bibinfo {year} {2020})}\BibitemShut {NoStop}%
\bibitem [{\citenamefont {Daw}\ and\ \citenamefont
  {Baskes}(1984)}]{daw_embedded-atom_1984}%
  \BibitemOpen
  \bibfield  {author} {\bibinfo {author} {\bibfnamefont {M.~S.}\ \bibnamefont
  {Daw}}\ and\ \bibinfo {author} {\bibfnamefont {M.~I.}\ \bibnamefont
  {Baskes}},\ }\href {https://doi.org/10.1103/PhysRevB.29.6443} {\bibfield
  {journal} {\bibinfo  {journal} {Phys. Rev. B}\ }\textbf {\bibinfo {volume}
  {29}},\ \bibinfo {pages} {6443} (\bibinfo {year} {1984})}\BibitemShut
  {NoStop}%
\bibitem [{\citenamefont {Ackland}\ and\ \citenamefont
  {Thetford}(1987)}]{ackland_improved_1987}%
  \BibitemOpen
  \bibfield  {author} {\bibinfo {author} {\bibfnamefont {G.~J.}\ \bibnamefont
  {Ackland}}\ and\ \bibinfo {author} {\bibfnamefont {R.}~\bibnamefont
  {Thetford}},\ }\href {https://doi.org/10.1080/01418618708204464} {\bibfield
  {journal} {\bibinfo  {journal} {Philos. Mag. A}\ }\textbf {\bibinfo {volume}
  {56}},\ \bibinfo {pages} {15} (\bibinfo {year} {1987})}\BibitemShut {NoStop}%
\bibitem [{\citenamefont {Derlet}\ \emph {et~al.}(2007)\citenamefont {Derlet},
  \citenamefont {{Nguyen-Manh}},\ and\ \citenamefont
  {Dudarev}}]{derlet_multiscale_2007}%
  \BibitemOpen
  \bibfield  {author} {\bibinfo {author} {\bibfnamefont {P.~M.}\ \bibnamefont
  {Derlet}}, \bibinfo {author} {\bibfnamefont {D.}~\bibnamefont
  {{Nguyen-Manh}}},\ and\ \bibinfo {author} {\bibfnamefont {S.~L.}\
  \bibnamefont {Dudarev}},\ }\href {https://doi.org/10.1103/PhysRevB.76.054107}
  {\bibfield  {journal} {\bibinfo  {journal} {Phys. Rev. B}\ }\textbf {\bibinfo
  {volume} {76}},\ \bibinfo {pages} {054107} (\bibinfo {year}
  {2007})}\BibitemShut {NoStop}%
\bibitem [{\citenamefont {Chen}\ \emph {et~al.}(2019)\citenamefont {Chen},
  \citenamefont {Fang}, \citenamefont {Liu}, \citenamefont {Hu}, \citenamefont
  {Gao}, \citenamefont {Gao},\ and\ \citenamefont
  {Deng}}]{chen_development_2019}%
  \BibitemOpen
  \bibfield  {author} {\bibinfo {author} {\bibfnamefont {Y.}~\bibnamefont
  {Chen}}, \bibinfo {author} {\bibfnamefont {J.}~\bibnamefont {Fang}}, \bibinfo
  {author} {\bibfnamefont {L.}~\bibnamefont {Liu}}, \bibinfo {author}
  {\bibfnamefont {W.}~\bibnamefont {Hu}}, \bibinfo {author} {\bibfnamefont
  {N.}~\bibnamefont {Gao}}, \bibinfo {author} {\bibfnamefont {F.}~\bibnamefont
  {Gao}},\ and\ \bibinfo {author} {\bibfnamefont {H.}~\bibnamefont {Deng}},\
  }\href {https://doi.org/10.1016/j.commatsci.2019.03.021} {\bibfield
  {journal} {\bibinfo  {journal} {Computational Materials Science}\ }\textbf
  {\bibinfo {volume} {163}},\ \bibinfo {pages} {91} (\bibinfo {year}
  {2019})}\BibitemShut {NoStop}%
\bibitem [{\citenamefont {Chen}\ \emph {et~al.}(2020)\citenamefont {Chen},
  \citenamefont {Liao}, \citenamefont {Gao}, \citenamefont {Hu}, \citenamefont
  {Gao},\ and\ \citenamefont {Deng}}]{chen_interatomic_2020}%
  \BibitemOpen
  \bibfield  {author} {\bibinfo {author} {\bibfnamefont {Y.}~\bibnamefont
  {Chen}}, \bibinfo {author} {\bibfnamefont {X.}~\bibnamefont {Liao}}, \bibinfo
  {author} {\bibfnamefont {N.}~\bibnamefont {Gao}}, \bibinfo {author}
  {\bibfnamefont {W.}~\bibnamefont {Hu}}, \bibinfo {author} {\bibfnamefont
  {F.}~\bibnamefont {Gao}},\ and\ \bibinfo {author} {\bibfnamefont
  {H.}~\bibnamefont {Deng}},\ }\href
  {https://doi.org/10.1016/j.jnucmat.2020.152020} {\bibfield  {journal}
  {\bibinfo  {journal} {Journal of Nuclear Materials}\ ,\ \bibinfo {pages}
  {152020}} (\bibinfo {year} {2020})}\BibitemShut {NoStop}%
\bibitem [{\citenamefont {Byggm{\"a}star}\ \emph {et~al.}(2019)\citenamefont
  {Byggm{\"a}star}, \citenamefont {Hamedani}, \citenamefont {Nordlund},\ and\
  \citenamefont {Djurabekova}}]{byggmastar_machine-learning_2019}%
  \BibitemOpen
  \bibfield  {author} {\bibinfo {author} {\bibfnamefont {J.}~\bibnamefont
  {Byggm{\"a}star}}, \bibinfo {author} {\bibfnamefont {A.}~\bibnamefont
  {Hamedani}}, \bibinfo {author} {\bibfnamefont {K.}~\bibnamefont {Nordlund}},\
  and\ \bibinfo {author} {\bibfnamefont {F.}~\bibnamefont {Djurabekova}},\
  }\href {https://doi.org/10.1103/PhysRevB.100.144105} {\bibfield  {journal}
  {\bibinfo  {journal} {Phys. Rev. B}\ }\textbf {\bibinfo {volume} {100}},\
  \bibinfo {pages} {144105} (\bibinfo {year} {2019})}\BibitemShut {NoStop}%
\bibitem [{\citenamefont {Bonny}\ \emph {et~al.}(2014)\citenamefont {Bonny},
  \citenamefont {Terentyev}, \citenamefont {Bakaev}, \citenamefont {Grigorev},\
  and\ \citenamefont {Van~Neck}}]{bonny_many-body_2014}%
  \BibitemOpen
  \bibfield  {author} {\bibinfo {author} {\bibfnamefont {G.}~\bibnamefont
  {Bonny}}, \bibinfo {author} {\bibfnamefont {D.}~\bibnamefont {Terentyev}},
  \bibinfo {author} {\bibfnamefont {A.}~\bibnamefont {Bakaev}}, \bibinfo
  {author} {\bibfnamefont {P.}~\bibnamefont {Grigorev}},\ and\ \bibinfo
  {author} {\bibfnamefont {D.}~\bibnamefont {Van~Neck}},\ }\href
  {https://doi.org/10.1088/0965-0393/22/5/053001} {\bibfield  {journal}
  {\bibinfo  {journal} {Model. Simul. Mater. Sci. Eng.}\ }\textbf {\bibinfo
  {volume} {22}},\ \bibinfo {pages} {053001} (\bibinfo {year}
  {2014})}\BibitemShut {NoStop}%
\bibitem [{\citenamefont {Rieth}\ \emph {et~al.}(2013)\citenamefont {Rieth},
  \citenamefont {Dudarev}, \citenamefont {{Gonzalez de Vicente}}, \citenamefont
  {Aktaa}, \citenamefont {Ahlgren}, \citenamefont {Antusch}, \citenamefont
  {Armstrong}, \citenamefont {Balden}, \citenamefont {Baluc}, \citenamefont
  {Barthe}, \citenamefont {Basuki}, \citenamefont {Battabyal}, \citenamefont
  {Becquart}, \citenamefont {Blagoeva}, \citenamefont {Boldyryeva},
  \citenamefont {Brinkmann}, \citenamefont {Celino}, \citenamefont {Ciupinski},
  \citenamefont {Correia}, \citenamefont {De~Backer}, \citenamefont {Domain},
  \citenamefont {Gaganidze}, \citenamefont {{Garc{\'i}a-Rosales}},
  \citenamefont {Gibson}, \citenamefont {Gilbert}, \citenamefont {Giusepponi},
  \citenamefont {Gludovatz}, \citenamefont {Greuner}, \citenamefont {Heinola},
  \citenamefont {H{\"o}schen}, \citenamefont {Hoffmann}, \citenamefont
  {Holstein}, \citenamefont {Koch}, \citenamefont {Krauss}, \citenamefont {Li},
  \citenamefont {Lindig}, \citenamefont {Linke}, \citenamefont {Linsmeier},
  \citenamefont {{L{\'o}pez-Ruiz}}, \citenamefont {Maier}, \citenamefont
  {Matejicek}, \citenamefont {Mishra}, \citenamefont {Muhammed}, \citenamefont
  {Mu{\~n}oz}, \citenamefont {Muzyk}, \citenamefont {Nordlund}, \citenamefont
  {{Nguyen-Manh}}, \citenamefont {Opschoor}, \citenamefont {Ord{\'a}s},
  \citenamefont {Palacios}, \citenamefont {Pintsuk}, \citenamefont {Pippan},
  \citenamefont {Reiser}, \citenamefont {Riesch}, \citenamefont {Roberts},
  \citenamefont {Romaner}, \citenamefont {Rosi{\'n}ski}, \citenamefont
  {Sanchez}, \citenamefont {Schulmeyer}, \citenamefont {Traxler}, \citenamefont
  {Ure{\~n}a}, \citenamefont {{van der Laan}}, \citenamefont {Veleva},
  \citenamefont {Wahlberg}, \citenamefont {Walter}, \citenamefont {Weber},
  \citenamefont {Weitkamp}, \citenamefont {Wurster}, \citenamefont {Yar},
  \citenamefont {You},\ and\ \citenamefont {Zivelonghi}}]{rieth_recent_2013}%
  \BibitemOpen
  \bibfield  {author} {\bibinfo {author} {\bibfnamefont {M.}~\bibnamefont
  {Rieth}}, \bibinfo {author} {\bibfnamefont {S.~L.}\ \bibnamefont {Dudarev}},
  \bibinfo {author} {\bibfnamefont {S.~M.}\ \bibnamefont {{Gonzalez de
  Vicente}}}, \bibinfo {author} {\bibfnamefont {J.}~\bibnamefont {Aktaa}},
  \bibinfo {author} {\bibfnamefont {T.}~\bibnamefont {Ahlgren}}, \bibinfo
  {author} {\bibfnamefont {S.}~\bibnamefont {Antusch}}, \bibinfo {author}
  {\bibfnamefont {D.~E.~J.}\ \bibnamefont {Armstrong}}, \bibinfo {author}
  {\bibfnamefont {M.}~\bibnamefont {Balden}}, \bibinfo {author} {\bibfnamefont
  {N.}~\bibnamefont {Baluc}}, \bibinfo {author} {\bibfnamefont {M.~F.}\
  \bibnamefont {Barthe}}, \bibinfo {author} {\bibfnamefont {W.~W.}\
  \bibnamefont {Basuki}}, \bibinfo {author} {\bibfnamefont {M.}~\bibnamefont
  {Battabyal}}, \bibinfo {author} {\bibfnamefont {C.~S.}\ \bibnamefont
  {Becquart}}, \bibinfo {author} {\bibfnamefont {D.}~\bibnamefont {Blagoeva}},
  \bibinfo {author} {\bibfnamefont {H.}~\bibnamefont {Boldyryeva}}, \bibinfo
  {author} {\bibfnamefont {J.}~\bibnamefont {Brinkmann}}, \bibinfo {author}
  {\bibfnamefont {M.}~\bibnamefont {Celino}}, \bibinfo {author} {\bibfnamefont
  {L.}~\bibnamefont {Ciupinski}}, \bibinfo {author} {\bibfnamefont {J.~B.}\
  \bibnamefont {Correia}}, \bibinfo {author} {\bibfnamefont {A.}~\bibnamefont
  {De~Backer}}, \bibinfo {author} {\bibfnamefont {C.}~\bibnamefont {Domain}},
  \bibinfo {author} {\bibfnamefont {E.}~\bibnamefont {Gaganidze}}, \bibinfo
  {author} {\bibfnamefont {C.}~\bibnamefont {{Garc{\'i}a-Rosales}}}, \bibinfo
  {author} {\bibfnamefont {J.}~\bibnamefont {Gibson}}, \bibinfo {author}
  {\bibfnamefont {M.~R.}\ \bibnamefont {Gilbert}}, \bibinfo {author}
  {\bibfnamefont {S.}~\bibnamefont {Giusepponi}}, \bibinfo {author}
  {\bibfnamefont {B.}~\bibnamefont {Gludovatz}}, \bibinfo {author}
  {\bibfnamefont {H.}~\bibnamefont {Greuner}}, \bibinfo {author} {\bibfnamefont
  {K.}~\bibnamefont {Heinola}}, \bibinfo {author} {\bibfnamefont
  {T.}~\bibnamefont {H{\"o}schen}}, \bibinfo {author} {\bibfnamefont
  {A.}~\bibnamefont {Hoffmann}}, \bibinfo {author} {\bibfnamefont
  {N.}~\bibnamefont {Holstein}}, \bibinfo {author} {\bibfnamefont
  {F.}~\bibnamefont {Koch}}, \bibinfo {author} {\bibfnamefont {W.}~\bibnamefont
  {Krauss}}, \bibinfo {author} {\bibfnamefont {H.}~\bibnamefont {Li}}, \bibinfo
  {author} {\bibfnamefont {S.}~\bibnamefont {Lindig}}, \bibinfo {author}
  {\bibfnamefont {J.}~\bibnamefont {Linke}}, \bibinfo {author} {\bibfnamefont
  {C.}~\bibnamefont {Linsmeier}}, \bibinfo {author} {\bibfnamefont
  {P.}~\bibnamefont {{L{\'o}pez-Ruiz}}}, \bibinfo {author} {\bibfnamefont
  {H.}~\bibnamefont {Maier}}, \bibinfo {author} {\bibfnamefont
  {J.}~\bibnamefont {Matejicek}}, \bibinfo {author} {\bibfnamefont {T.~P.}\
  \bibnamefont {Mishra}}, \bibinfo {author} {\bibfnamefont {M.}~\bibnamefont
  {Muhammed}}, \bibinfo {author} {\bibfnamefont {A.}~\bibnamefont {Mu{\~n}oz}},
  \bibinfo {author} {\bibfnamefont {M.}~\bibnamefont {Muzyk}}, \bibinfo
  {author} {\bibfnamefont {K.}~\bibnamefont {Nordlund}}, \bibinfo {author}
  {\bibfnamefont {D.}~\bibnamefont {{Nguyen-Manh}}}, \bibinfo {author}
  {\bibfnamefont {J.}~\bibnamefont {Opschoor}}, \bibinfo {author}
  {\bibfnamefont {N.}~\bibnamefont {Ord{\'a}s}}, \bibinfo {author}
  {\bibfnamefont {T.}~\bibnamefont {Palacios}}, \bibinfo {author}
  {\bibfnamefont {G.}~\bibnamefont {Pintsuk}}, \bibinfo {author} {\bibfnamefont
  {R.}~\bibnamefont {Pippan}}, \bibinfo {author} {\bibfnamefont
  {J.}~\bibnamefont {Reiser}}, \bibinfo {author} {\bibfnamefont
  {J.}~\bibnamefont {Riesch}}, \bibinfo {author} {\bibfnamefont {S.~G.}\
  \bibnamefont {Roberts}}, \bibinfo {author} {\bibfnamefont {L.}~\bibnamefont
  {Romaner}}, \bibinfo {author} {\bibfnamefont {M.}~\bibnamefont
  {Rosi{\'n}ski}}, \bibinfo {author} {\bibfnamefont {M.}~\bibnamefont
  {Sanchez}}, \bibinfo {author} {\bibfnamefont {W.}~\bibnamefont {Schulmeyer}},
  \bibinfo {author} {\bibfnamefont {H.}~\bibnamefont {Traxler}}, \bibinfo
  {author} {\bibfnamefont {A.}~\bibnamefont {Ure{\~n}a}}, \bibinfo {author}
  {\bibfnamefont {J.~G.}\ \bibnamefont {{van der Laan}}}, \bibinfo {author}
  {\bibfnamefont {L.}~\bibnamefont {Veleva}}, \bibinfo {author} {\bibfnamefont
  {S.}~\bibnamefont {Wahlberg}}, \bibinfo {author} {\bibfnamefont
  {M.}~\bibnamefont {Walter}}, \bibinfo {author} {\bibfnamefont
  {T.}~\bibnamefont {Weber}}, \bibinfo {author} {\bibfnamefont
  {T.}~\bibnamefont {Weitkamp}}, \bibinfo {author} {\bibfnamefont
  {S.}~\bibnamefont {Wurster}}, \bibinfo {author} {\bibfnamefont {M.~A.}\
  \bibnamefont {Yar}}, \bibinfo {author} {\bibfnamefont {J.~H.}\ \bibnamefont
  {You}},\ and\ \bibinfo {author} {\bibfnamefont {A.}~\bibnamefont
  {Zivelonghi}},\ }\href {https://doi.org/10.1016/j.jnucmat.2012.08.018}
  {\bibfield  {journal} {\bibinfo  {journal} {Journal of Nuclear Materials}\
  }\textbf {\bibinfo {volume} {432}},\ \bibinfo {pages} {482} (\bibinfo {year}
  {2013})}\BibitemShut {NoStop}%
\bibitem [{\citenamefont {Stork}\ \emph {et~al.}(2014)\citenamefont {Stork},
  \citenamefont {Agostini}, \citenamefont {Boutard}, \citenamefont
  {Buckthorpe}, \citenamefont {Diegele}, \citenamefont {Dudarev}, \citenamefont
  {English}, \citenamefont {Federici}, \citenamefont {Gilbert}, \citenamefont
  {Gonzalez}, \citenamefont {Ibarra}, \citenamefont {Linsmeier}, \citenamefont
  {Puma}, \citenamefont {Marbach}, \citenamefont {Packer}, \citenamefont {Raj},
  \citenamefont {Rieth}, \citenamefont {Tran}, \citenamefont {Ward},\ and\
  \citenamefont {Zinkle}}]{stork_materials_2014}%
  \BibitemOpen
  \bibfield  {author} {\bibinfo {author} {\bibfnamefont {D.}~\bibnamefont
  {Stork}}, \bibinfo {author} {\bibfnamefont {P.}~\bibnamefont {Agostini}},
  \bibinfo {author} {\bibfnamefont {J.-L.}\ \bibnamefont {Boutard}}, \bibinfo
  {author} {\bibfnamefont {D.}~\bibnamefont {Buckthorpe}}, \bibinfo {author}
  {\bibfnamefont {E.}~\bibnamefont {Diegele}}, \bibinfo {author} {\bibfnamefont
  {S.~L.}\ \bibnamefont {Dudarev}}, \bibinfo {author} {\bibfnamefont
  {C.}~\bibnamefont {English}}, \bibinfo {author} {\bibfnamefont
  {G.}~\bibnamefont {Federici}}, \bibinfo {author} {\bibfnamefont {M.~R.}\
  \bibnamefont {Gilbert}}, \bibinfo {author} {\bibfnamefont {S.}~\bibnamefont
  {Gonzalez}}, \bibinfo {author} {\bibfnamefont {A.}~\bibnamefont {Ibarra}},
  \bibinfo {author} {\bibfnamefont {C.}~\bibnamefont {Linsmeier}}, \bibinfo
  {author} {\bibfnamefont {A.~L.}\ \bibnamefont {Puma}}, \bibinfo {author}
  {\bibfnamefont {G.}~\bibnamefont {Marbach}}, \bibinfo {author} {\bibfnamefont
  {L.~W.}\ \bibnamefont {Packer}}, \bibinfo {author} {\bibfnamefont
  {B.}~\bibnamefont {Raj}}, \bibinfo {author} {\bibfnamefont {M.}~\bibnamefont
  {Rieth}}, \bibinfo {author} {\bibfnamefont {M.~Q.}\ \bibnamefont {Tran}},
  \bibinfo {author} {\bibfnamefont {D.~J.}\ \bibnamefont {Ward}},\ and\
  \bibinfo {author} {\bibfnamefont {S.~J.}\ \bibnamefont {Zinkle}},\ }\href
  {https://doi.org/10.1016/j.fusengdes.2013.11.007} {\bibfield  {journal}
  {\bibinfo  {journal} {Fusion Engineering and Design}\ }\bibinfo {series}
  {Proceedings of the 11th {{International Symposium}} on {{Fusion Nuclear
  Technology}}-11 ({{ISFNT}}-11) {{Barcelona}}, {{Spain}}, 15-20 {{September}},
  2013},\ \textbf {\bibinfo {volume} {89}},\ \bibinfo {pages} {1586} (\bibinfo
  {year} {2014})}\BibitemShut {NoStop}%
\bibitem [{\citenamefont {Litnovsky}\ \emph {et~al.}(2007)\citenamefont
  {Litnovsky}, \citenamefont {Wienhold}, \citenamefont {Philipps},
  \citenamefont {Sergienko}, \citenamefont {Schmitz}, \citenamefont
  {Kirschner}, \citenamefont {Kreter}, \citenamefont {Droste}, \citenamefont
  {Samm}, \citenamefont {Mertens}, \citenamefont {Donné}, \citenamefont
  {Rudakov}, \citenamefont {Allen}, \citenamefont {Boivin}, \citenamefont
  {McLean}, \citenamefont {Stangeby}, \citenamefont {West}, \citenamefont
  {Wong}, \citenamefont {Lipa}, \citenamefont {Schunke}, \citenamefont
  {Temmerman}, \citenamefont {Pitts}, \citenamefont {Costley}, \citenamefont
  {Voitsenya}, \citenamefont {Vukolov}, \citenamefont {Oelhafen}, \citenamefont
  {Rubel},\ and\ \citenamefont {Romanyuk}}]{litnovsky_diagnostic_2007}%
  \BibitemOpen
  \bibfield  {author} {\bibinfo {author} {\bibfnamefont {A.}~\bibnamefont
  {Litnovsky}}, \bibinfo {author} {\bibfnamefont {P.}~\bibnamefont {Wienhold}},
  \bibinfo {author} {\bibfnamefont {V.}~\bibnamefont {Philipps}}, \bibinfo
  {author} {\bibfnamefont {G.}~\bibnamefont {Sergienko}}, \bibinfo {author}
  {\bibfnamefont {O.}~\bibnamefont {Schmitz}}, \bibinfo {author} {\bibfnamefont
  {A.}~\bibnamefont {Kirschner}}, \bibinfo {author} {\bibfnamefont
  {A.}~\bibnamefont {Kreter}}, \bibinfo {author} {\bibfnamefont
  {S.}~\bibnamefont {Droste}}, \bibinfo {author} {\bibfnamefont
  {U.}~\bibnamefont {Samm}}, \bibinfo {author} {\bibfnamefont {P.}~\bibnamefont
  {Mertens}}, \bibinfo {author} {\bibfnamefont {A.}~\bibnamefont {Donné}},
  \bibinfo {author} {\bibfnamefont {D.}~\bibnamefont {Rudakov}}, \bibinfo
  {author} {\bibfnamefont {S.}~\bibnamefont {Allen}}, \bibinfo {author}
  {\bibfnamefont {R.}~\bibnamefont {Boivin}}, \bibinfo {author} {\bibfnamefont
  {A.}~\bibnamefont {McLean}}, \bibinfo {author} {\bibfnamefont
  {P.}~\bibnamefont {Stangeby}}, \bibinfo {author} {\bibfnamefont
  {W.}~\bibnamefont {West}}, \bibinfo {author} {\bibfnamefont {C.}~\bibnamefont
  {Wong}}, \bibinfo {author} {\bibfnamefont {M.}~\bibnamefont {Lipa}}, \bibinfo
  {author} {\bibfnamefont {B.}~\bibnamefont {Schunke}}, \bibinfo {author}
  {\bibfnamefont {G.~D.}\ \bibnamefont {Temmerman}}, \bibinfo {author}
  {\bibfnamefont {R.}~\bibnamefont {Pitts}}, \bibinfo {author} {\bibfnamefont
  {A.}~\bibnamefont {Costley}}, \bibinfo {author} {\bibfnamefont
  {V.}~\bibnamefont {Voitsenya}}, \bibinfo {author} {\bibfnamefont
  {K.}~\bibnamefont {Vukolov}}, \bibinfo {author} {\bibfnamefont
  {P.}~\bibnamefont {Oelhafen}}, \bibinfo {author} {\bibfnamefont
  {M.}~\bibnamefont {Rubel}},\ and\ \bibinfo {author} {\bibfnamefont
  {A.}~\bibnamefont {Romanyuk}},\ }\href
  {https://doi.org/https://doi.org/10.1016/j.jnucmat.2007.01.281} {\bibfield
  {journal} {\bibinfo  {journal} {Journal of Nuclear Materials}\ }\textbf
  {\bibinfo {volume} {363-365}},\ \bibinfo {pages} {1395 } (\bibinfo {year}
  {2007})},\ \bibinfo {note} {plasma-Surface Interactions-17}\BibitemShut
  {NoStop}%
\bibitem [{\citenamefont {Liu}\ \emph {et~al.}(2013)\citenamefont {Liu},
  \citenamefont {Zhang}, \citenamefont {Jiang}, \citenamefont {Ding},
  \citenamefont {Sun}, \citenamefont {Sun},\ and\ \citenamefont
  {Ma}}]{liu_nanostructured_2013}%
  \BibitemOpen
  \bibfield  {author} {\bibinfo {author} {\bibfnamefont {G.}~\bibnamefont
  {Liu}}, \bibinfo {author} {\bibfnamefont {G.~J.}\ \bibnamefont {Zhang}},
  \bibinfo {author} {\bibfnamefont {F.}~\bibnamefont {Jiang}}, \bibinfo
  {author} {\bibfnamefont {X.~D.}\ \bibnamefont {Ding}}, \bibinfo {author}
  {\bibfnamefont {Y.~J.}\ \bibnamefont {Sun}}, \bibinfo {author} {\bibfnamefont
  {J.}~\bibnamefont {Sun}},\ and\ \bibinfo {author} {\bibfnamefont
  {E.}~\bibnamefont {Ma}},\ }\href {https://doi.org/10.1038/nmat3544}
  {\bibfield  {journal} {\bibinfo  {journal} {Nature Mater}\ }\textbf {\bibinfo
  {volume} {12}},\ \bibinfo {pages} {344} (\bibinfo {year} {2013})}\BibitemShut
  {NoStop}%
\bibitem [{\citenamefont {Buckman}(2000)}]{buckman_new_2000}%
  \BibitemOpen
  \bibfield  {author} {\bibinfo {author} {\bibfnamefont {R.~W.}\ \bibnamefont
  {Buckman}},\ }\href {https://doi.org/10.1007/s11837-000-0100-6} {\bibfield
  {journal} {\bibinfo  {journal} {JOM}\ }\textbf {\bibinfo {volume} {52}},\
  \bibinfo {pages} {40} (\bibinfo {year} {2000})}\BibitemShut {NoStop}%
\bibitem [{\citenamefont {Sheftel}\ and\ \citenamefont
  {Bannykh}(1993)}]{sheftel_niobiumbase_1993}%
  \BibitemOpen
  \bibfield  {author} {\bibinfo {author} {\bibfnamefont {E.~N.}\ \bibnamefont
  {Sheftel}}\ and\ \bibinfo {author} {\bibfnamefont {O.~A.}\ \bibnamefont
  {Bannykh}},\ }\href {https://doi.org/10.1016/0263-4368(93)90038-H} {\bibfield
   {journal} {\bibinfo  {journal} {International Journal of Refractory Metals
  and Hard Materials}\ }\textbf {\bibinfo {volume} {12}},\ \bibinfo {pages}
  {303} (\bibinfo {year} {1993})}\BibitemShut {NoStop}%
\bibitem [{\citenamefont {Moskalyk}\ and\ \citenamefont
  {Alfantazi}(2003)}]{moskalyk_processing_2003}%
  \BibitemOpen
  \bibfield  {author} {\bibinfo {author} {\bibfnamefont {R.~R.}\ \bibnamefont
  {Moskalyk}}\ and\ \bibinfo {author} {\bibfnamefont {A.~M.}\ \bibnamefont
  {Alfantazi}},\ }\href {https://doi.org/10.1016/S0892-6875(03)00213-9}
  {\bibfield  {journal} {\bibinfo  {journal} {Minerals Engineering}\ }\textbf
  {\bibinfo {volume} {16}},\ \bibinfo {pages} {793} (\bibinfo {year}
  {2003})}\BibitemShut {NoStop}%
\bibitem [{\citenamefont {Bart{\'o}k}\ and\ \citenamefont
  {Cs{\'a}nyi}(2015)}]{bartok_gaussian_2015}%
  \BibitemOpen
  \bibfield  {author} {\bibinfo {author} {\bibfnamefont {A.~P.}\ \bibnamefont
  {Bart{\'o}k}}\ and\ \bibinfo {author} {\bibfnamefont {G.}~\bibnamefont
  {Cs{\'a}nyi}},\ }\href {https://doi.org/10.1002/qua.24927} {\bibfield
  {journal} {\bibinfo  {journal} {Int. J. Quantum Chem.}\ }\textbf {\bibinfo
  {volume} {115}},\ \bibinfo {pages} {1051} (\bibinfo {year}
  {2015})}\BibitemShut {NoStop}%
\bibitem [{\citenamefont {Nordlund}\ \emph {et~al.}(1997)\citenamefont
  {Nordlund}, \citenamefont {Runeberg},\ and\ \citenamefont
  {Sundholm}}]{nordlund_repulsive_1997}%
  \BibitemOpen
  \bibfield  {author} {\bibinfo {author} {\bibfnamefont {K.}~\bibnamefont
  {Nordlund}}, \bibinfo {author} {\bibfnamefont {N.}~\bibnamefont {Runeberg}},\
  and\ \bibinfo {author} {\bibfnamefont {D.}~\bibnamefont {Sundholm}},\ }\href
  {https://doi.org/10.1016/S0168-583X(97)00447-3} {\bibfield  {journal}
  {\bibinfo  {journal} {Nuclear Instruments and Methods in Physics Research
  Section B: Beam Interactions with Materials and Atoms}\ }\textbf {\bibinfo
  {volume} {132}},\ \bibinfo {pages} {45} (\bibinfo {year} {1997})}\BibitemShut
  {NoStop}%
\bibitem [{\citenamefont {Marinica}\ \emph {et~al.}(2012)\citenamefont
  {Marinica}, \citenamefont {Willaime},\ and\ \citenamefont
  {Crocombette}}]{marinica_irradiation-induced_2012}%
  \BibitemOpen
  \bibfield  {author} {\bibinfo {author} {\bibfnamefont {M.-C.}\ \bibnamefont
  {Marinica}}, \bibinfo {author} {\bibfnamefont {F.}~\bibnamefont {Willaime}},\
  and\ \bibinfo {author} {\bibfnamefont {J.-P.}\ \bibnamefont {Crocombette}},\
  }\href {https://doi.org/10.1103/PhysRevLett.108.025501} {\bibfield  {journal}
  {\bibinfo  {journal} {Phys. Rev. Lett.}\ }\textbf {\bibinfo {volume} {108}},\
  \bibinfo {pages} {025501} (\bibinfo {year} {2012})}\BibitemShut {NoStop}%
\bibitem [{\citenamefont {Kresse}\ and\ \citenamefont
  {Hafner}(1993)}]{kresse_ab_1993}%
  \BibitemOpen
  \bibfield  {author} {\bibinfo {author} {\bibfnamefont {G.}~\bibnamefont
  {Kresse}}\ and\ \bibinfo {author} {\bibfnamefont {J.}~\bibnamefont
  {Hafner}},\ }\href {https://doi.org/10.1103/PhysRevB.47.558} {\bibfield
  {journal} {\bibinfo  {journal} {Phys. Rev. B}\ }\textbf {\bibinfo {volume}
  {47}},\ \bibinfo {pages} {558} (\bibinfo {year} {1993})}\BibitemShut
  {NoStop}%
\bibitem [{\citenamefont {Kresse}\ and\ \citenamefont
  {Hafner}(1994)}]{kresse_ab_1994}%
  \BibitemOpen
  \bibfield  {author} {\bibinfo {author} {\bibfnamefont {G.}~\bibnamefont
  {Kresse}}\ and\ \bibinfo {author} {\bibfnamefont {J.}~\bibnamefont
  {Hafner}},\ }\href {https://doi.org/10.1103/PhysRevB.49.14251} {\bibfield
  {journal} {\bibinfo  {journal} {Phys. Rev. B}\ }\textbf {\bibinfo {volume}
  {49}},\ \bibinfo {pages} {14251} (\bibinfo {year} {1994})}\BibitemShut
  {NoStop}%
\bibitem [{\citenamefont {Kresse}\ and\ \citenamefont
  {Furthm{\"u}ller}(1996{\natexlab{a}})}]{kresse_efficiency_1996}%
  \BibitemOpen
  \bibfield  {author} {\bibinfo {author} {\bibfnamefont {G.}~\bibnamefont
  {Kresse}}\ and\ \bibinfo {author} {\bibfnamefont {J.}~\bibnamefont
  {Furthm{\"u}ller}},\ }\href {https://doi.org/10.1016/0927-0256(96)00008-0}
  {\bibfield  {journal} {\bibinfo  {journal} {Computational Materials Science}\
  }\textbf {\bibinfo {volume} {6}},\ \bibinfo {pages} {15} (\bibinfo {year}
  {1996}{\natexlab{a}})}\BibitemShut {NoStop}%
\bibitem [{\citenamefont {Kresse}\ and\ \citenamefont
  {Furthm{\"u}ller}(1996{\natexlab{b}})}]{kresse_efficient_1996}%
  \BibitemOpen
  \bibfield  {author} {\bibinfo {author} {\bibfnamefont {G.}~\bibnamefont
  {Kresse}}\ and\ \bibinfo {author} {\bibfnamefont {J.}~\bibnamefont
  {Furthm{\"u}ller}},\ }\href {https://doi.org/10.1103/PhysRevB.54.11169}
  {\bibfield  {journal} {\bibinfo  {journal} {Phys. Rev. B}\ }\textbf {\bibinfo
  {volume} {54}},\ \bibinfo {pages} {11169} (\bibinfo {year}
  {1996}{\natexlab{b}})}\BibitemShut {NoStop}%
\bibitem [{\citenamefont {Perdew}\ \emph {et~al.}(1996)\citenamefont {Perdew},
  \citenamefont {Burke},\ and\ \citenamefont
  {Ernzerhof}}]{perdew_generalized_1996}%
  \BibitemOpen
  \bibfield  {author} {\bibinfo {author} {\bibfnamefont {J.~P.}\ \bibnamefont
  {Perdew}}, \bibinfo {author} {\bibfnamefont {K.}~\bibnamefont {Burke}},\ and\
  \bibinfo {author} {\bibfnamefont {M.}~\bibnamefont {Ernzerhof}},\ }\href
  {https://doi.org/10.1103/PhysRevLett.77.3865} {\bibfield  {journal} {\bibinfo
   {journal} {Phys. Rev. Lett.}\ }\textbf {\bibinfo {volume} {77}},\ \bibinfo
  {pages} {3865} (\bibinfo {year} {1996})}\BibitemShut {NoStop}%
\bibitem [{\citenamefont {Bl{\"o}chl}(1994)}]{blochl_projector_1994}%
  \BibitemOpen
  \bibfield  {author} {\bibinfo {author} {\bibfnamefont {P.~E.}\ \bibnamefont
  {Bl{\"o}chl}},\ }\href {https://doi.org/10.1103/PhysRevB.50.17953} {\bibfield
   {journal} {\bibinfo  {journal} {Phys. Rev. B}\ }\textbf {\bibinfo {volume}
  {50}},\ \bibinfo {pages} {17953} (\bibinfo {year} {1994})}\BibitemShut
  {NoStop}%
\bibitem [{\citenamefont {Kresse}\ and\ \citenamefont
  {Joubert}(1999)}]{kresse_ultrasoft_1999}%
  \BibitemOpen
  \bibfield  {author} {\bibinfo {author} {\bibfnamefont {G.}~\bibnamefont
  {Kresse}}\ and\ \bibinfo {author} {\bibfnamefont {D.}~\bibnamefont
  {Joubert}},\ }\href {https://doi.org/10.1103/PhysRevB.59.1758} {\bibfield
  {journal} {\bibinfo  {journal} {Phys. Rev. B}\ }\textbf {\bibinfo {volume}
  {59}},\ \bibinfo {pages} {1758} (\bibinfo {year} {1999})}\BibitemShut
  {NoStop}%
\bibitem [{\citenamefont {Methfessel}\ and\ \citenamefont
  {Paxton}(1989)}]{methfessel_high-precision_1989}%
  \BibitemOpen
  \bibfield  {author} {\bibinfo {author} {\bibfnamefont {M.}~\bibnamefont
  {Methfessel}}\ and\ \bibinfo {author} {\bibfnamefont {A.~T.}\ \bibnamefont
  {Paxton}},\ }\href {https://doi.org/10.1103/PhysRevB.40.3616} {\bibfield
  {journal} {\bibinfo  {journal} {Phys. Rev. B}\ }\textbf {\bibinfo {volume}
  {40}},\ \bibinfo {pages} {3616} (\bibinfo {year} {1989})}\BibitemShut
  {NoStop}%
\bibitem [{\citenamefont {Monkhorst}\ and\ \citenamefont
  {Pack}(1976)}]{monkhorst_special_1976}%
  \BibitemOpen
  \bibfield  {author} {\bibinfo {author} {\bibfnamefont {H.~J.}\ \bibnamefont
  {Monkhorst}}\ and\ \bibinfo {author} {\bibfnamefont {J.~D.}\ \bibnamefont
  {Pack}},\ }\href {https://doi.org/10.1103/PhysRevB.13.5188} {\bibfield
  {journal} {\bibinfo  {journal} {Phys. Rev. B}\ }\textbf {\bibinfo {volume}
  {13}},\ \bibinfo {pages} {5188} (\bibinfo {year} {1976})}\BibitemShut
  {NoStop}%
\bibitem [{qui()}]{quip}%
  \BibitemOpen
  \href@noop {} {\bibinfo {title} {{QUIP} - {QUantum} mechanics and
  {Interatomic} {Potentials}}},\ \bibinfo {howpublished}
  {\url{https://github.com/libAtoms/QUIP}, \url{www.libatoms.org}}\BibitemShut
  {NoStop}%
\bibitem [{\citenamefont {Plimpton}(1995)}]{plimpton_fast_1995}%
  \BibitemOpen
  \bibfield  {author} {\bibinfo {author} {\bibfnamefont {S.}~\bibnamefont
  {Plimpton}},\ }\href {https://doi.org/10.1006/jcph.1995.1039} {\bibfield
  {journal} {\bibinfo  {journal} {Journal of Computational Physics}\ }\textbf
  {\bibinfo {volume} {117}},\ \bibinfo {pages} {1} (\bibinfo {year} {1995})},\
  \bibinfo {note} {http://lammps.sandia.gov}\BibitemShut {NoStop}%
\bibitem [{\citenamefont {Larsen}\ \emph {et~al.}(2017)\citenamefont {Larsen},
  \citenamefont {Mortensen}, \citenamefont {Blomqvist}, \citenamefont
  {Castelli}, \citenamefont {Christensen}, \citenamefont {{Marcin Du{\l}ak}},
  \citenamefont {Friis}, \citenamefont {Groves}, \citenamefont {Hammer},
  \citenamefont {Hargus}, \citenamefont {Hermes}, \citenamefont {Jennings},
  \citenamefont {Jensen}, \citenamefont {Kermode}, \citenamefont {Kitchin},
  \citenamefont {Kolsbjerg}, \citenamefont {Kubal}, \citenamefont {{Kristen
  Kaasbjerg}}, \citenamefont {Lysgaard}, \citenamefont {Maronsson},
  \citenamefont {Maxson}, \citenamefont {Olsen}, \citenamefont {Pastewka},
  \citenamefont {{Andrew Peterson}}, \citenamefont {Rostgaard}, \citenamefont
  {Schi{\o}tz}, \citenamefont {Sch{\"u}tt}, \citenamefont {Strange},
  \citenamefont {Thygesen}, \citenamefont {{Tejs Vegge}}, \citenamefont
  {Vilhelmsen}, \citenamefont {Walter}, \citenamefont {Zeng},\ and\
  \citenamefont {Jacobsen}}]{larsen_atomic_2017}%
  \BibitemOpen
  \bibfield  {author} {\bibinfo {author} {\bibfnamefont {A.~H.}\ \bibnamefont
  {Larsen}}, \bibinfo {author} {\bibfnamefont {J.~J.}\ \bibnamefont
  {Mortensen}}, \bibinfo {author} {\bibfnamefont {J.}~\bibnamefont
  {Blomqvist}}, \bibinfo {author} {\bibfnamefont {I.~E.}\ \bibnamefont
  {Castelli}}, \bibinfo {author} {\bibfnamefont {R.}~\bibnamefont
  {Christensen}}, \bibinfo {author} {\bibnamefont {{Marcin Du{\l}ak}}},
  \bibinfo {author} {\bibfnamefont {J.}~\bibnamefont {Friis}}, \bibinfo
  {author} {\bibfnamefont {M.~N.}\ \bibnamefont {Groves}}, \bibinfo {author}
  {\bibfnamefont {B.}~\bibnamefont {Hammer}}, \bibinfo {author} {\bibfnamefont
  {C.}~\bibnamefont {Hargus}}, \bibinfo {author} {\bibfnamefont {E.~D.}\
  \bibnamefont {Hermes}}, \bibinfo {author} {\bibfnamefont {P.~C.}\
  \bibnamefont {Jennings}}, \bibinfo {author} {\bibfnamefont {P.~B.}\
  \bibnamefont {Jensen}}, \bibinfo {author} {\bibfnamefont {J.}~\bibnamefont
  {Kermode}}, \bibinfo {author} {\bibfnamefont {J.~R.}\ \bibnamefont
  {Kitchin}}, \bibinfo {author} {\bibfnamefont {E.~L.}\ \bibnamefont
  {Kolsbjerg}}, \bibinfo {author} {\bibfnamefont {J.}~\bibnamefont {Kubal}},
  \bibinfo {author} {\bibnamefont {{Kristen Kaasbjerg}}}, \bibinfo {author}
  {\bibfnamefont {S.}~\bibnamefont {Lysgaard}}, \bibinfo {author}
  {\bibfnamefont {J.~B.}\ \bibnamefont {Maronsson}}, \bibinfo {author}
  {\bibfnamefont {T.}~\bibnamefont {Maxson}}, \bibinfo {author} {\bibfnamefont
  {T.}~\bibnamefont {Olsen}}, \bibinfo {author} {\bibfnamefont
  {L.}~\bibnamefont {Pastewka}}, \bibinfo {author} {\bibnamefont {{Andrew
  Peterson}}}, \bibinfo {author} {\bibfnamefont {C.}~\bibnamefont {Rostgaard}},
  \bibinfo {author} {\bibfnamefont {J.}~\bibnamefont {Schi{\o}tz}}, \bibinfo
  {author} {\bibfnamefont {O.}~\bibnamefont {Sch{\"u}tt}}, \bibinfo {author}
  {\bibfnamefont {M.}~\bibnamefont {Strange}}, \bibinfo {author} {\bibfnamefont
  {K.~S.}\ \bibnamefont {Thygesen}}, \bibinfo {author} {\bibnamefont {{Tejs
  Vegge}}}, \bibinfo {author} {\bibfnamefont {L.}~\bibnamefont {Vilhelmsen}},
  \bibinfo {author} {\bibfnamefont {M.}~\bibnamefont {Walter}}, \bibinfo
  {author} {\bibfnamefont {Z.}~\bibnamefont {Zeng}},\ and\ \bibinfo {author}
  {\bibfnamefont {K.~W.}\ \bibnamefont {Jacobsen}},\ }\href
  {https://doi.org/10.1088/1361-648X/aa680e} {\bibfield  {journal} {\bibinfo
  {journal} {J. Phys.: Condens. Matter}\ }\textbf {\bibinfo {volume} {29}},\
  \bibinfo {pages} {273002} (\bibinfo {year} {2017})}\BibitemShut {NoStop}%
\bibitem [{\citenamefont {Togo}\ and\ \citenamefont
  {Tanaka}(2015)}]{togo_first_2015}%
  \BibitemOpen
  \bibfield  {author} {\bibinfo {author} {\bibfnamefont {A.}~\bibnamefont
  {Togo}}\ and\ \bibinfo {author} {\bibfnamefont {I.}~\bibnamefont {Tanaka}},\
  }\href {https://doi.org/10.1016/j.scriptamat.2015.07.021} {\bibfield
  {journal} {\bibinfo  {journal} {Scripta Materialia}\ }\textbf {\bibinfo
  {volume} {108}},\ \bibinfo {pages} {1} (\bibinfo {year} {2015})}\BibitemShut
  {NoStop}%
\bibitem [{\citenamefont {Rumble}(2019)}]{rumble_crc_2019}%
  \BibitemOpen
  \bibinfo {editor} {\bibfnamefont {J.}~\bibnamefont {Rumble}},\ ed.,\
  \href@noop {} {\emph {\bibinfo {title} {{{CRC Handbook}} of {{Chemistry}} and
  {{Physics}}, 100th {{Edition}}}}},\ \bibinfo {edition} {100th}\ ed.\
  (\bibinfo  {publisher} {{CRC Press}},\ \bibinfo {address} {{Boca Raton,
  Fla.}},\ \bibinfo {year} {2019})\BibitemShut {NoStop}%
\bibitem [{\citenamefont {White}(1988)}]{white_heat_1988}%
  \BibitemOpen
  \bibfield  {author} {\bibinfo {author} {\bibfnamefont {G.~K.}\ \bibnamefont
  {White}},\ }\href {https://doi.org/10.1016/0378-4363(88)90251-3} {\bibfield
  {journal} {\bibinfo  {journal} {Physica B+C}\ }\textbf {\bibinfo {volume}
  {149}},\ \bibinfo {pages} {255} (\bibinfo {year} {1988})}\BibitemShut
  {NoStop}%
\bibitem [{\citenamefont {Luo}\ \emph {et~al.}(2007)\citenamefont {Luo},
  \citenamefont {Ahuja}, \citenamefont {Ding},\ and\ \citenamefont
  {Mao}}]{luo_unusual_2007}%
  \BibitemOpen
  \bibfield  {author} {\bibinfo {author} {\bibfnamefont {W.}~\bibnamefont
  {Luo}}, \bibinfo {author} {\bibfnamefont {R.}~\bibnamefont {Ahuja}}, \bibinfo
  {author} {\bibfnamefont {Y.}~\bibnamefont {Ding}},\ and\ \bibinfo {author}
  {\bibfnamefont {H.-k.}\ \bibnamefont {Mao}},\ }\href
  {https://doi.org/10.1073/pnas.0707377104} {\bibfield  {journal} {\bibinfo
  {journal} {Proc. Natl. Acad. Sci.}\ }\textbf {\bibinfo {volume} {104}},\
  \bibinfo {pages} {16428} (\bibinfo {year} {2007})}\BibitemShut {NoStop}%
\bibitem [{\citenamefont {Fellinger}\ \emph {et~al.}(2010)\citenamefont
  {Fellinger}, \citenamefont {Park},\ and\ \citenamefont
  {Wilkins}}]{fellinger_force-matched_2010}%
  \BibitemOpen
  \bibfield  {author} {\bibinfo {author} {\bibfnamefont {M.~R.}\ \bibnamefont
  {Fellinger}}, \bibinfo {author} {\bibfnamefont {H.}~\bibnamefont {Park}},\
  and\ \bibinfo {author} {\bibfnamefont {J.~W.}\ \bibnamefont {Wilkins}},\
  }\bibfield  {journal} {\bibinfo  {journal} {Phys. Rev. B}\ }\textbf {\bibinfo
  {volume} {81}},\ \href {https://doi.org/10.1103/PhysRevB.81.144119}
  {10.1103/PhysRevB.81.144119} (\bibinfo {year} {2010})\BibitemShut {NoStop}%
\bibitem [{\citenamefont {Park}\ \emph {et~al.}(2012)\citenamefont {Park},
  \citenamefont {Fellinger}, \citenamefont {Lenosky}, \citenamefont {Tipton},
  \citenamefont {Trinkle}, \citenamefont {Rudin}, \citenamefont {Woodward},
  \citenamefont {Wilkins},\ and\ \citenamefont {Hennig}}]{park_ab_2012}%
  \BibitemOpen
  \bibfield  {author} {\bibinfo {author} {\bibfnamefont {H.}~\bibnamefont
  {Park}}, \bibinfo {author} {\bibfnamefont {M.~R.}\ \bibnamefont {Fellinger}},
  \bibinfo {author} {\bibfnamefont {T.~J.}\ \bibnamefont {Lenosky}}, \bibinfo
  {author} {\bibfnamefont {W.~W.}\ \bibnamefont {Tipton}}, \bibinfo {author}
  {\bibfnamefont {D.~R.}\ \bibnamefont {Trinkle}}, \bibinfo {author}
  {\bibfnamefont {S.~P.}\ \bibnamefont {Rudin}}, \bibinfo {author}
  {\bibfnamefont {C.}~\bibnamefont {Woodward}}, \bibinfo {author}
  {\bibfnamefont {J.~W.}\ \bibnamefont {Wilkins}},\ and\ \bibinfo {author}
  {\bibfnamefont {R.~G.}\ \bibnamefont {Hennig}},\ }\bibfield  {journal}
  {\bibinfo  {journal} {Phys. Rev. B}\ }\textbf {\bibinfo {volume} {85}},\
  \href {https://doi.org/10.1103/PhysRevB.85.214121}
  {10.1103/PhysRevB.85.214121} (\bibinfo {year} {2012})\BibitemShut {NoStop}%
\bibitem [{\citenamefont {Taioli}\ \emph {et~al.}(2007)\citenamefont {Taioli},
  \citenamefont {Cazorla}, \citenamefont {Gillan},\ and\ \citenamefont
  {Alf{\`e}}}]{taioli_melting_2007}%
  \BibitemOpen
  \bibfield  {author} {\bibinfo {author} {\bibfnamefont {S.}~\bibnamefont
  {Taioli}}, \bibinfo {author} {\bibfnamefont {C.}~\bibnamefont {Cazorla}},
  \bibinfo {author} {\bibfnamefont {M.~J.}\ \bibnamefont {Gillan}},\ and\
  \bibinfo {author} {\bibfnamefont {D.}~\bibnamefont {Alf{\`e}}},\ }\bibfield
  {journal} {\bibinfo  {journal} {Phys. Rev. B}\ }\textbf {\bibinfo {volume}
  {75}},\ \href {https://doi.org/10.1103/PhysRevB.75.214103}
  {10.1103/PhysRevB.75.214103} (\bibinfo {year} {2007})\BibitemShut {NoStop}%
\bibitem [{\citenamefont {Einarsdotter}\ \emph {et~al.}(1997)\citenamefont
  {Einarsdotter}, \citenamefont {Sadigh}, \citenamefont {Grimvall},\ and\
  \citenamefont {Ozoli{\c n}{\v s}}}]{einarsdotter_phonon_1997}%
  \BibitemOpen
  \bibfield  {author} {\bibinfo {author} {\bibfnamefont {K.}~\bibnamefont
  {Einarsdotter}}, \bibinfo {author} {\bibfnamefont {B.}~\bibnamefont
  {Sadigh}}, \bibinfo {author} {\bibfnamefont {G.}~\bibnamefont {Grimvall}},\
  and\ \bibinfo {author} {\bibfnamefont {V.}~\bibnamefont {Ozoli{\c n}{\v
  s}}},\ }\href {https://doi.org/10.1103/PhysRevLett.79.2073} {\bibfield
  {journal} {\bibinfo  {journal} {Phys. Rev. Lett.}\ }\textbf {\bibinfo
  {volume} {79}},\ \bibinfo {pages} {2073} (\bibinfo {year}
  {1997})}\BibitemShut {NoStop}%
\bibitem [{\citenamefont {Bosak}\ \emph {et~al.}(2008)\citenamefont {Bosak},
  \citenamefont {Hoesch}, \citenamefont {Antonangeli}, \citenamefont {Farber},
  \citenamefont {Fischer},\ and\ \citenamefont
  {Krisch}}]{bosak_lattice_2008-1}%
  \BibitemOpen
  \bibfield  {author} {\bibinfo {author} {\bibfnamefont {A.}~\bibnamefont
  {Bosak}}, \bibinfo {author} {\bibfnamefont {M.}~\bibnamefont {Hoesch}},
  \bibinfo {author} {\bibfnamefont {D.}~\bibnamefont {Antonangeli}}, \bibinfo
  {author} {\bibfnamefont {D.~L.}\ \bibnamefont {Farber}}, \bibinfo {author}
  {\bibfnamefont {I.}~\bibnamefont {Fischer}},\ and\ \bibinfo {author}
  {\bibfnamefont {M.}~\bibnamefont {Krisch}},\ }\bibfield  {journal} {\bibinfo
  {journal} {Phys. Rev. B}\ }\textbf {\bibinfo {volume} {78}},\ \href
  {https://doi.org/10.1103/PhysRevB.78.020301} {10.1103/PhysRevB.78.020301}
  (\bibinfo {year} {2008})\BibitemShut {NoStop}%
\bibitem [{\citenamefont {Powell}\ \emph {et~al.}(1968)\citenamefont {Powell},
  \citenamefont {Martel},\ and\ \citenamefont {Woods}}]{powell_lattice_1968}%
  \BibitemOpen
  \bibfield  {author} {\bibinfo {author} {\bibfnamefont {B.~M.}\ \bibnamefont
  {Powell}}, \bibinfo {author} {\bibfnamefont {P.}~\bibnamefont {Martel}},\
  and\ \bibinfo {author} {\bibfnamefont {A.~D.~B.}\ \bibnamefont {Woods}},\
  }\href {https://doi.org/10.1103/PhysRev.171.727} {\bibfield  {journal}
  {\bibinfo  {journal} {Phys. Rev.}\ }\textbf {\bibinfo {volume} {171}},\
  \bibinfo {pages} {727} (\bibinfo {year} {1968})}\BibitemShut {NoStop}%
\bibitem [{\citenamefont {Woods}(1964)}]{woods_lattice_1964}%
  \BibitemOpen
  \bibfield  {author} {\bibinfo {author} {\bibfnamefont {A.~D.~B.}\
  \bibnamefont {Woods}},\ }\href {https://doi.org/10.1103/PhysRev.136.A781}
  {\bibfield  {journal} {\bibinfo  {journal} {Phys. Rev.}\ }\textbf {\bibinfo
  {volume} {136}},\ \bibinfo {pages} {A781} (\bibinfo {year}
  {1964})}\BibitemShut {NoStop}%
\bibitem [{\citenamefont {Larose}\ and\ \citenamefont
  {Brockhouse}(1976)}]{larose_lattice_1976}%
  \BibitemOpen
  \bibfield  {author} {\bibinfo {author} {\bibfnamefont {A.}~\bibnamefont
  {Larose}}\ and\ \bibinfo {author} {\bibfnamefont {B.~N.}\ \bibnamefont
  {Brockhouse}},\ }\href {https://doi.org/10.1139/p76-215} {\bibfield
  {journal} {\bibinfo  {journal} {Can. J. Phys.}\ }\textbf {\bibinfo {volume}
  {54}},\ \bibinfo {pages} {1819} (\bibinfo {year} {1976})}\BibitemShut
  {NoStop}%
\bibitem [{\citenamefont {Ma}\ and\ \citenamefont
  {Dudarev}(2019{\natexlab{a}})}]{ma_effect_2019}%
  \BibitemOpen
  \bibfield  {author} {\bibinfo {author} {\bibfnamefont {P.-W.}\ \bibnamefont
  {Ma}}\ and\ \bibinfo {author} {\bibfnamefont {S.~L.}\ \bibnamefont
  {Dudarev}},\ }\href {https://doi.org/10.1103/PhysRevMaterials.3.063601}
  {\bibfield  {journal} {\bibinfo  {journal} {Phys. Rev. Materials}\ }\textbf
  {\bibinfo {volume} {3}},\ \bibinfo {pages} {063601} (\bibinfo {year}
  {2019}{\natexlab{a}})}\BibitemShut {NoStop}%
\bibitem [{\citenamefont {Ullmaier}(1991)}]{ullmaier_atomic_1991}%
  \BibitemOpen
  \bibinfo {editor} {\bibfnamefont {H.}~\bibnamefont {Ullmaier}},\ ed.,\
  \href@noop {} {\emph {\bibinfo {title} {Atomic {{Defects}} in {{Metals}}}}},\
  Condensed {{Matter}}\ (\bibinfo  {publisher} {{Springer-Verlag}},\ \bibinfo
  {address} {{Berlin Heidelberg}},\ \bibinfo {year} {1991})\BibitemShut
  {NoStop}%
\bibitem [{\citenamefont {Ma}\ and\ \citenamefont
  {Dudarev}(2019{\natexlab{b}})}]{ma_universality_2019}%
  \BibitemOpen
  \bibfield  {author} {\bibinfo {author} {\bibfnamefont {P.-W.}\ \bibnamefont
  {Ma}}\ and\ \bibinfo {author} {\bibfnamefont {S.~L.}\ \bibnamefont
  {Dudarev}},\ }\bibfield  {journal} {\bibinfo  {journal} {Phys. Rev. Mater.}\
  }\textbf {\bibinfo {volume} {3}},\ \href
  {https://doi.org/10.1103/PhysRevMaterials.3.013605}
  {10.1103/PhysRevMaterials.3.013605} (\bibinfo {year}
  {2019}{\natexlab{b}})\BibitemShut {NoStop}%
\bibitem [{\citenamefont {Ma}\ and\ \citenamefont
  {Dudarev}(2019{\natexlab{c}})}]{ma_symmetry-broken_2019}%
  \BibitemOpen
  \bibfield  {author} {\bibinfo {author} {\bibfnamefont {P.-W.}\ \bibnamefont
  {Ma}}\ and\ \bibinfo {author} {\bibfnamefont {S.~L.}\ \bibnamefont
  {Dudarev}},\ }\bibfield  {journal} {\bibinfo  {journal} {Phys. Rev. Mater.}\
  }\textbf {\bibinfo {volume} {3}},\ \href
  {https://doi.org/10.1103/PhysRevMaterials.3.043606}
  {10.1103/PhysRevMaterials.3.043606} (\bibinfo {year}
  {2019}{\natexlab{c}})\BibitemShut {NoStop}%
\bibitem [{\citenamefont {Alexander}\ \emph {et~al.}(2016)\citenamefont
  {Alexander}, \citenamefont {Marinica}, \citenamefont {Proville},
  \citenamefont {Willaime}, \citenamefont {Arakawa}, \citenamefont {Gilbert},\
  and\ \citenamefont {Dudarev}}]{alexander_ab_2016}%
  \BibitemOpen
  \bibfield  {author} {\bibinfo {author} {\bibfnamefont {R.}~\bibnamefont
  {Alexander}}, \bibinfo {author} {\bibfnamefont {M.-C.}\ \bibnamefont
  {Marinica}}, \bibinfo {author} {\bibfnamefont {L.}~\bibnamefont {Proville}},
  \bibinfo {author} {\bibfnamefont {F.}~\bibnamefont {Willaime}}, \bibinfo
  {author} {\bibfnamefont {K.}~\bibnamefont {Arakawa}}, \bibinfo {author}
  {\bibfnamefont {M.~R.}\ \bibnamefont {Gilbert}},\ and\ \bibinfo {author}
  {\bibfnamefont {S.~L.}\ \bibnamefont {Dudarev}},\ }\href
  {https://doi.org/10.1103/PhysRevB.94.024103} {\bibfield  {journal} {\bibinfo
  {journal} {Phys. Rev. B}\ }\textbf {\bibinfo {volume} {94}},\ \bibinfo
  {pages} {024103} (\bibinfo {year} {2016})}\BibitemShut {NoStop}%
\bibitem [{\citenamefont {Han}\ \emph {et~al.}(2003)\citenamefont {Han},
  \citenamefont {{Zepeda-Ruiz}}, \citenamefont {Ackland}, \citenamefont {Car},\
  and\ \citenamefont {Srolovitz}}]{han_interatomic_2003}%
  \BibitemOpen
  \bibfield  {author} {\bibinfo {author} {\bibfnamefont {S.}~\bibnamefont
  {Han}}, \bibinfo {author} {\bibfnamefont {L.~A.}\ \bibnamefont
  {{Zepeda-Ruiz}}}, \bibinfo {author} {\bibfnamefont {G.~J.}\ \bibnamefont
  {Ackland}}, \bibinfo {author} {\bibfnamefont {R.}~\bibnamefont {Car}},\ and\
  \bibinfo {author} {\bibfnamefont {D.~J.}\ \bibnamefont {Srolovitz}},\ }\href
  {https://doi.org/10.1063/1.1555275} {\bibfield  {journal} {\bibinfo
  {journal} {J. Appl. Phys.}\ }\textbf {\bibinfo {volume} {93}},\ \bibinfo
  {pages} {3328} (\bibinfo {year} {2003})}\BibitemShut {NoStop}%
\bibitem [{sup()}]{supplemental}%
  \BibitemOpen
  \href@noop {} {\bibinfo {title} {See {Supplemental} {Material} for
  potential files, training data, and comparison with other interatomic
  potentials~\cite{ackland_improved_1987,derlet_multiscale_2007,han_interatomic_2003,fellinger_force-matched_2010,park_ab_2012,chen_accurate_2017,li_embedded-atom-method_2003,ravelo_shock-induced_2013,purja_pun_angular-dependent_2015,marinica_interatomic_2013}.}}\BibitemShut
  {Stop}%
\bibitem [{\citenamefont {Mathon}(1988)}]{mathon_magnetism_1988}%
  \BibitemOpen
  \bibfield  {author} {\bibinfo {author} {\bibfnamefont {J.}~\bibnamefont
  {Mathon}},\ }\href {https://doi.org/10.1088/0034-4885/51/1/001} {\bibfield
  {journal} {\bibinfo  {journal} {Reports on Progress in Physics}\ }\textbf
  {\bibinfo {volume} {51}},\ \bibinfo {pages} {1} (\bibinfo {year}
  {1988})}\BibitemShut {NoStop}%
\bibitem [{\citenamefont {Ziegler}\ \emph {et~al.}(1985)\citenamefont
  {Ziegler}, \citenamefont {Biersack},\ and\ \citenamefont
  {Littmarck}}]{ziegler_stopping_1985}%
  \BibitemOpen
  \bibfield  {author} {\bibinfo {author} {\bibfnamefont {J.~F.}\ \bibnamefont
  {Ziegler}}, \bibinfo {author} {\bibfnamefont {J.~P.}\ \bibnamefont
  {Biersack}},\ and\ \bibinfo {author} {\bibfnamefont {U.}~\bibnamefont
  {Littmarck}},\ }in\ \href@noop {} {\emph {\bibinfo {booktitle} {Treatise on
  {{Heavy}}-{{Ion Science}}}}}\ (\bibinfo  {publisher} {{Pergamon}},\ \bibinfo
  {address} {{New York}},\ \bibinfo {year} {1985})\ pp.\ \bibinfo {pages}
  {93--129}\BibitemShut {NoStop}%
\bibitem [{\citenamefont {Miller}\ and\ \citenamefont
  {Chaplin}(1974)}]{miller_defect_1974}%
  \BibitemOpen
  \bibfield  {author} {\bibinfo {author} {\bibfnamefont {M.~G.}\ \bibnamefont
  {Miller}}\ and\ \bibinfo {author} {\bibfnamefont {R.~L.}\ \bibnamefont
  {Chaplin}},\ }\href {https://doi.org/10.1080/00337577408232154} {\bibfield
  {journal} {\bibinfo  {journal} {Radiat. Eff.}\ }\textbf {\bibinfo {volume}
  {22}},\ \bibinfo {pages} {107} (\bibinfo {year} {1974})}\BibitemShut
  {NoStop}%
\bibitem [{\citenamefont {Jung}\ and\ \citenamefont
  {Lucki}(1975)}]{jung_damage_1975}%
  \BibitemOpen
  \bibfield  {author} {\bibinfo {author} {\bibfnamefont {P.}~\bibnamefont
  {Jung}}\ and\ \bibinfo {author} {\bibfnamefont {G.}~\bibnamefont {Lucki}},\
  }\href {https://doi.org/10.1080/00337577508237426} {\bibfield  {journal}
  {\bibinfo  {journal} {Radiat. Eff.}\ }\textbf {\bibinfo {volume} {26}},\
  \bibinfo {pages} {99} (\bibinfo {year} {1975})}\BibitemShut {NoStop}%
\bibitem [{\citenamefont {Maury}\ \emph {et~al.}(1975)\citenamefont {Maury},
  \citenamefont {Vajda}, \citenamefont {Biget}, \citenamefont {Lucasson},\ and\
  \citenamefont {Lucasson}}]{maury_anisotropy_1975}%
  \BibitemOpen
  \bibfield  {author} {\bibinfo {author} {\bibfnamefont {F.}~\bibnamefont
  {Maury}}, \bibinfo {author} {\bibfnamefont {P.}~\bibnamefont {Vajda}},
  \bibinfo {author} {\bibfnamefont {M.}~\bibnamefont {Biget}}, \bibinfo
  {author} {\bibfnamefont {A.}~\bibnamefont {Lucasson}},\ and\ \bibinfo
  {author} {\bibfnamefont {P.}~\bibnamefont {Lucasson}},\ }\href
  {https://doi.org/10.1080/00337577508235387} {\bibfield  {journal} {\bibinfo
  {journal} {Radiat. Eff.}\ }\textbf {\bibinfo {volume} {25}},\ \bibinfo
  {pages} {175} (\bibinfo {year} {1975})}\BibitemShut {NoStop}%
\bibitem [{\citenamefont {Biget}\ \emph {et~al.}(1974)\citenamefont {Biget},
  \citenamefont {Vajda}, \citenamefont {Lucasson},\ and\ \citenamefont
  {Lucasson}}]{biget_study_1974}%
  \BibitemOpen
  \bibfield  {author} {\bibinfo {author} {\bibfnamefont {M.}~\bibnamefont
  {Biget}}, \bibinfo {author} {\bibfnamefont {P.}~\bibnamefont {Vajda}},
  \bibinfo {author} {\bibfnamefont {A.}~\bibnamefont {Lucasson}},\ and\
  \bibinfo {author} {\bibfnamefont {P.}~\bibnamefont {Lucasson}},\ }\href
  {https://doi.org/10.1080/00337577408232410} {\bibfield  {journal} {\bibinfo
  {journal} {Radiat. Eff.}\ }\textbf {\bibinfo {volume} {21}},\ \bibinfo
  {pages} {229} (\bibinfo {year} {1974})}\BibitemShut {NoStop}%
\bibitem [{\citenamefont {Biget}\ \emph {et~al.}(1979)\citenamefont {Biget},
  \citenamefont {Maury}, \citenamefont {Vajda}, \citenamefont {Lucasson},\ and\
  \citenamefont {Lucasson}}]{biget_near-threshold_1979}%
  \BibitemOpen
  \bibfield  {author} {\bibinfo {author} {\bibfnamefont {M.}~\bibnamefont
  {Biget}}, \bibinfo {author} {\bibfnamefont {F.}~\bibnamefont {Maury}},
  \bibinfo {author} {\bibfnamefont {P.}~\bibnamefont {Vajda}}, \bibinfo
  {author} {\bibfnamefont {A.}~\bibnamefont {Lucasson}},\ and\ \bibinfo
  {author} {\bibfnamefont {P.}~\bibnamefont {Lucasson}},\ }\href
  {https://doi.org/10.1103/PhysRevB.19.820} {\bibfield  {journal} {\bibinfo
  {journal} {Phys. Rev. B}\ }\textbf {\bibinfo {volume} {19}},\ \bibinfo
  {pages} {820} (\bibinfo {year} {1979})}\BibitemShut {NoStop}%
\bibitem [{\citenamefont {Maury}\ \emph {et~al.}(1978)\citenamefont {Maury},
  \citenamefont {Biget}, \citenamefont {Vajda}, \citenamefont {Lucasson},\ and\
  \citenamefont {Lucasson}}]{maury_frenkel_1978}%
  \BibitemOpen
  \bibfield  {author} {\bibinfo {author} {\bibfnamefont {F.}~\bibnamefont
  {Maury}}, \bibinfo {author} {\bibfnamefont {M.}~\bibnamefont {Biget}},
  \bibinfo {author} {\bibfnamefont {P.}~\bibnamefont {Vajda}}, \bibinfo
  {author} {\bibfnamefont {A.}~\bibnamefont {Lucasson}},\ and\ \bibinfo
  {author} {\bibfnamefont {P.}~\bibnamefont {Lucasson}},\ }\href
  {https://doi.org/10.1080/00337577808233209} {\bibfield  {journal} {\bibinfo
  {journal} {Radiat. Eff.}\ }\textbf {\bibinfo {volume} {38}},\ \bibinfo
  {pages} {53} (\bibinfo {year} {1978})}\BibitemShut {NoStop}%
\bibitem [{\citenamefont {Nordlund}\ \emph {et~al.}(2006)\citenamefont
  {Nordlund}, \citenamefont {Wallenius},\ and\ \citenamefont
  {Malerba}}]{nordlund_molecular_2006}%
  \BibitemOpen
  \bibfield  {author} {\bibinfo {author} {\bibfnamefont {K.}~\bibnamefont
  {Nordlund}}, \bibinfo {author} {\bibfnamefont {J.}~\bibnamefont
  {Wallenius}},\ and\ \bibinfo {author} {\bibfnamefont {L.}~\bibnamefont
  {Malerba}},\ }\href {https://doi.org/10.1016/j.nimb.2006.01.003} {\bibfield
  {journal} {\bibinfo  {journal} {Nuclear Instruments and Methods in Physics
  Research Section B: Beam Interactions with Materials and Atoms}\ }\textbf
  {\bibinfo {volume} {246}},\ \bibinfo {pages} {322} (\bibinfo {year}
  {2006})}\BibitemShut {NoStop}%
\bibitem [{\citenamefont {Vinet}\ \emph {et~al.}(1993)\citenamefont {Vinet},
  \citenamefont {Garandet},\ and\ \citenamefont
  {Cortella}}]{vinet_surface_1993}%
  \BibitemOpen
  \bibfield  {author} {\bibinfo {author} {\bibfnamefont {B.}~\bibnamefont
  {Vinet}}, \bibinfo {author} {\bibfnamefont {J.~P.}\ \bibnamefont
  {Garandet}},\ and\ \bibinfo {author} {\bibfnamefont {L.}~\bibnamefont
  {Cortella}},\ }\href {https://doi.org/10.1063/1.352891} {\bibfield  {journal}
  {\bibinfo  {journal} {J. Appl. Phys.}\ }\textbf {\bibinfo {volume} {73}},\
  \bibinfo {pages} {3830} (\bibinfo {year} {1993})}\BibitemShut {NoStop}%
\bibitem [{\citenamefont {Jinnouchi}\ \emph {et~al.}(2019)\citenamefont
  {Jinnouchi}, \citenamefont {Karsai},\ and\ \citenamefont
  {Kresse}}]{jinnouchi_on-the-fly_2019}%
  \BibitemOpen
  \bibfield  {author} {\bibinfo {author} {\bibfnamefont {R.}~\bibnamefont
  {Jinnouchi}}, \bibinfo {author} {\bibfnamefont {F.}~\bibnamefont {Karsai}},\
  and\ \bibinfo {author} {\bibfnamefont {G.}~\bibnamefont {Kresse}},\ }\href
  {https://doi.org/10.1103/PhysRevB.100.014105} {\bibfield  {journal} {\bibinfo
   {journal} {Phys. Rev. B}\ }\textbf {\bibinfo {volume} {100}},\ \bibinfo
  {pages} {014105} (\bibinfo {year} {2019})}\BibitemShut {NoStop}%
\bibitem [{\citenamefont {Cazorla}\ \emph {et~al.}(2007)\citenamefont
  {Cazorla}, \citenamefont {Gillan}, \citenamefont {Taioli},\ and\
  \citenamefont {Alf{\`e}}}]{cazorla_ab_2007}%
  \BibitemOpen
  \bibfield  {author} {\bibinfo {author} {\bibfnamefont {C.}~\bibnamefont
  {Cazorla}}, \bibinfo {author} {\bibfnamefont {M.~J.}\ \bibnamefont {Gillan}},
  \bibinfo {author} {\bibfnamefont {S.}~\bibnamefont {Taioli}},\ and\ \bibinfo
  {author} {\bibfnamefont {D.}~\bibnamefont {Alf{\`e}}},\ }\href
  {https://doi.org/10.1063/1.2735324} {\bibfield  {journal} {\bibinfo
  {journal} {J. Chem. Phys.}\ }\textbf {\bibinfo {volume} {126}},\ \bibinfo
  {pages} {194502} (\bibinfo {year} {2007})}\BibitemShut {NoStop}%
\bibitem [{\citenamefont {Belonoshko}\ \emph {et~al.}(2008)\citenamefont
  {Belonoshko}, \citenamefont {Burakovsky}, \citenamefont {Chen}, \citenamefont
  {Johansson}, \citenamefont {Mikhaylushkin}, \citenamefont {Preston},
  \citenamefont {Simak},\ and\ \citenamefont
  {Swift}}]{belonoshko_molybdenum_2008}%
  \BibitemOpen
  \bibfield  {author} {\bibinfo {author} {\bibfnamefont {A.~B.}\ \bibnamefont
  {Belonoshko}}, \bibinfo {author} {\bibfnamefont {L.}~\bibnamefont
  {Burakovsky}}, \bibinfo {author} {\bibfnamefont {S.~P.}\ \bibnamefont
  {Chen}}, \bibinfo {author} {\bibfnamefont {B.}~\bibnamefont {Johansson}},
  \bibinfo {author} {\bibfnamefont {A.~S.}\ \bibnamefont {Mikhaylushkin}},
  \bibinfo {author} {\bibfnamefont {D.~L.}\ \bibnamefont {Preston}}, \bibinfo
  {author} {\bibfnamefont {S.~I.}\ \bibnamefont {Simak}},\ and\ \bibinfo
  {author} {\bibfnamefont {D.~C.}\ \bibnamefont {Swift}},\ }\bibfield
  {journal} {\bibinfo  {journal} {Phys. Rev. Lett.}\ }\textbf {\bibinfo
  {volume} {100}},\ \href {https://doi.org/10.1103/PhysRevLett.100.135701}
  {10.1103/PhysRevLett.100.135701} (\bibinfo {year} {2008})\BibitemShut
  {NoStop}%
\bibitem [{\citenamefont {Burakovsky}\ \emph {et~al.}(2010)\citenamefont
  {Burakovsky}, \citenamefont {Chen}, \citenamefont {Preston}, \citenamefont
  {Belonoshko}, \citenamefont {Rosengren}, \citenamefont {Mikhaylushkin},
  \citenamefont {Simak},\ and\ \citenamefont
  {Moriarty}}]{burakovsky_high-pressurehigh-temperature_2010}%
  \BibitemOpen
  \bibfield  {author} {\bibinfo {author} {\bibfnamefont {L.}~\bibnamefont
  {Burakovsky}}, \bibinfo {author} {\bibfnamefont {S.~P.}\ \bibnamefont
  {Chen}}, \bibinfo {author} {\bibfnamefont {D.~L.}\ \bibnamefont {Preston}},
  \bibinfo {author} {\bibfnamefont {A.~B.}\ \bibnamefont {Belonoshko}},
  \bibinfo {author} {\bibfnamefont {A.}~\bibnamefont {Rosengren}}, \bibinfo
  {author} {\bibfnamefont {A.~S.}\ \bibnamefont {Mikhaylushkin}}, \bibinfo
  {author} {\bibfnamefont {S.~I.}\ \bibnamefont {Simak}},\ and\ \bibinfo
  {author} {\bibfnamefont {J.~A.}\ \bibnamefont {Moriarty}},\ }\bibfield
  {journal} {\bibinfo  {journal} {Phys. Rev. Lett.}\ }\textbf {\bibinfo
  {volume} {104}},\ \href {https://doi.org/10.1103/PhysRevLett.104.255702}
  {10.1103/PhysRevLett.104.255702} (\bibinfo {year} {2010})\BibitemShut
  {NoStop}%
\bibitem [{\citenamefont {Zhang}\ \emph {et~al.}(2008)\citenamefont {Zhang},
  \citenamefont {Liu}, \citenamefont {Gu}, \citenamefont {Cai},\ and\
  \citenamefont {Jing}}]{zhang_temperature_2008}%
  \BibitemOpen
  \bibfield  {author} {\bibinfo {author} {\bibfnamefont {X.-l.}\ \bibnamefont
  {Zhang}}, \bibinfo {author} {\bibfnamefont {Z.-l.}\ \bibnamefont {Liu}},
  \bibinfo {author} {\bibfnamefont {Y.-j.}\ \bibnamefont {Gu}}, \bibinfo
  {author} {\bibfnamefont {L.-c.}\ \bibnamefont {Cai}},\ and\ \bibinfo {author}
  {\bibfnamefont {F.-q.}\ \bibnamefont {Jing}},\ }\href
  {https://doi.org/10.1016/j.physb.2008.04.012} {\bibfield  {journal} {\bibinfo
   {journal} {Physica B: Condensed Matter}\ }\textbf {\bibinfo {volume}
  {403}},\ \bibinfo {pages} {3261} (\bibinfo {year} {2008})}\BibitemShut
  {NoStop}%
\bibitem [{\citenamefont {Wang}\ \emph {et~al.}(2015)\citenamefont {Wang},
  \citenamefont {Coppari}, \citenamefont {Smith}, \citenamefont {Eggert},
  \citenamefont {Lazicki}, \citenamefont {Fratanduono}, \citenamefont {Rygg},
  \citenamefont {Boehly}, \citenamefont {Collins},\ and\ \citenamefont
  {Duffy}}]{wang_x-ray_2015}%
  \BibitemOpen
  \bibfield  {author} {\bibinfo {author} {\bibfnamefont {J.}~\bibnamefont
  {Wang}}, \bibinfo {author} {\bibfnamefont {F.}~\bibnamefont {Coppari}},
  \bibinfo {author} {\bibfnamefont {R.~F.}\ \bibnamefont {Smith}}, \bibinfo
  {author} {\bibfnamefont {J.~H.}\ \bibnamefont {Eggert}}, \bibinfo {author}
  {\bibfnamefont {A.~E.}\ \bibnamefont {Lazicki}}, \bibinfo {author}
  {\bibfnamefont {D.~E.}\ \bibnamefont {Fratanduono}}, \bibinfo {author}
  {\bibfnamefont {J.~R.}\ \bibnamefont {Rygg}}, \bibinfo {author}
  {\bibfnamefont {T.~R.}\ \bibnamefont {Boehly}}, \bibinfo {author}
  {\bibfnamefont {G.~W.}\ \bibnamefont {Collins}},\ and\ \bibinfo {author}
  {\bibfnamefont {T.~S.}\ \bibnamefont {Duffy}},\ }\href
  {https://doi.org/10.1103/PhysRevB.92.174114} {\bibfield  {journal} {\bibinfo
  {journal} {Phys. Rev. B}\ }\textbf {\bibinfo {volume} {92}},\ \bibinfo
  {pages} {174114} (\bibinfo {year} {2015})}\BibitemShut {NoStop}%
\bibitem [{\citenamefont {Hrubiak}\ \emph {et~al.}(2017)\citenamefont
  {Hrubiak}, \citenamefont {Meng},\ and\ \citenamefont
  {Shen}}]{hrubiak_microstructures_2017}%
  \BibitemOpen
  \bibfield  {author} {\bibinfo {author} {\bibfnamefont {R.}~\bibnamefont
  {Hrubiak}}, \bibinfo {author} {\bibfnamefont {Y.}~\bibnamefont {Meng}},\ and\
  \bibinfo {author} {\bibfnamefont {G.}~\bibnamefont {Shen}},\ }\href
  {https://doi.org/10.1038/ncomms14562} {\bibfield  {journal} {\bibinfo
  {journal} {Nat Commun}\ }\textbf {\bibinfo {volume} {8}},\ \bibinfo {pages}
  {1} (\bibinfo {year} {2017})}\BibitemShut {NoStop}%
\bibitem [{\citenamefont {Dewaele}\ \emph {et~al.}(2010)\citenamefont
  {Dewaele}, \citenamefont {Mezouar}, \citenamefont {Guignot},\ and\
  \citenamefont {Loubeyre}}]{dewaele_high_2010}%
  \BibitemOpen
  \bibfield  {author} {\bibinfo {author} {\bibfnamefont {A.}~\bibnamefont
  {Dewaele}}, \bibinfo {author} {\bibfnamefont {M.}~\bibnamefont {Mezouar}},
  \bibinfo {author} {\bibfnamefont {N.}~\bibnamefont {Guignot}},\ and\ \bibinfo
  {author} {\bibfnamefont {P.}~\bibnamefont {Loubeyre}},\ }\bibfield  {journal}
  {\bibinfo  {journal} {Phys. Rev. Lett.}\ }\textbf {\bibinfo {volume} {104}},\
  \href {https://doi.org/10.1103/PhysRevLett.104.255701}
  {10.1103/PhysRevLett.104.255701} (\bibinfo {year} {2010})\BibitemShut
  {NoStop}%
\bibitem [{\citenamefont {Dai}\ \emph {et~al.}(2009)\citenamefont {Dai},
  \citenamefont {Hu},\ and\ \citenamefont {Tan}}]{dai_hugoniot_2009}%
  \BibitemOpen
  \bibfield  {author} {\bibinfo {author} {\bibfnamefont {C.}~\bibnamefont
  {Dai}}, \bibinfo {author} {\bibfnamefont {J.}~\bibnamefont {Hu}},\ and\
  \bibinfo {author} {\bibfnamefont {H.}~\bibnamefont {Tan}},\ }\href
  {https://doi.org/10.1063/1.3204941} {\bibfield  {journal} {\bibinfo
  {journal} {J. Appl. Phys.}\ }\textbf {\bibinfo {volume} {106}},\ \bibinfo
  {pages} {043519} (\bibinfo {year} {2009})}\BibitemShut {NoStop}%
\bibitem [{\citenamefont {Errandonea}\ \emph {et~al.}(2001)\citenamefont
  {Errandonea}, \citenamefont {Schwager}, \citenamefont {Ditz}, \citenamefont
  {Gessmann}, \citenamefont {Boehler},\ and\ \citenamefont
  {Ross}}]{errandonea_systematics_2001}%
  \BibitemOpen
  \bibfield  {author} {\bibinfo {author} {\bibfnamefont {D.}~\bibnamefont
  {Errandonea}}, \bibinfo {author} {\bibfnamefont {B.}~\bibnamefont
  {Schwager}}, \bibinfo {author} {\bibfnamefont {R.}~\bibnamefont {Ditz}},
  \bibinfo {author} {\bibfnamefont {C.}~\bibnamefont {Gessmann}}, \bibinfo
  {author} {\bibfnamefont {R.}~\bibnamefont {Boehler}},\ and\ \bibinfo {author}
  {\bibfnamefont {M.}~\bibnamefont {Ross}},\ }\href
  {https://doi.org/10.1103/PhysRevB.63.132104} {\bibfield  {journal} {\bibinfo
  {journal} {Phys. Rev. B}\ }\textbf {\bibinfo {volume} {63}},\ \bibinfo
  {pages} {132104} (\bibinfo {year} {2001})}\BibitemShut {NoStop}%
\bibitem [{\citenamefont {Errandonea}\ \emph {et~al.}(2019)\citenamefont
  {Errandonea}, \citenamefont {MacLeod}, \citenamefont {Burakovsky},
  \citenamefont {{Santamaria-Perez}}, \citenamefont {Proctor}, \citenamefont
  {Cynn},\ and\ \citenamefont {Mezouar}}]{errandonea_melting_2019}%
  \BibitemOpen
  \bibfield  {author} {\bibinfo {author} {\bibfnamefont {D.}~\bibnamefont
  {Errandonea}}, \bibinfo {author} {\bibfnamefont {S.~G.}\ \bibnamefont
  {MacLeod}}, \bibinfo {author} {\bibfnamefont {L.}~\bibnamefont {Burakovsky}},
  \bibinfo {author} {\bibfnamefont {D.}~\bibnamefont {{Santamaria-Perez}}},
  \bibinfo {author} {\bibfnamefont {J.~E.}\ \bibnamefont {Proctor}}, \bibinfo
  {author} {\bibfnamefont {H.}~\bibnamefont {Cynn}},\ and\ \bibinfo {author}
  {\bibfnamefont {M.}~\bibnamefont {Mezouar}},\ }\bibfield  {journal} {\bibinfo
   {journal} {Phys. Rev. B}\ }\textbf {\bibinfo {volume} {100}},\ \href
  {https://doi.org/10.1103/PhysRevB.100.094111} {10.1103/PhysRevB.100.094111}
  (\bibinfo {year} {2019})\BibitemShut {NoStop}%
\bibitem [{\citenamefont {Alfè}\ \emph {et~al.}(2002)\citenamefont {Alfè},
  \citenamefont {Gillan},\ and\ \citenamefont
  {Price}}]{alfe_complementary_2002}%
  \BibitemOpen
  \bibfield  {author} {\bibinfo {author} {\bibfnamefont {D.}~\bibnamefont
  {Alfè}}, \bibinfo {author} {\bibfnamefont {M.~J.}\ \bibnamefont {Gillan}},\
  and\ \bibinfo {author} {\bibfnamefont {G.~D.}\ \bibnamefont {Price}},\ }\href
  {https://doi.org/10.1063/1.1460865} {\bibfield  {journal} {\bibinfo
  {journal} {The Journal of Chemical Physics}\ }\textbf {\bibinfo {volume}
  {116}},\ \bibinfo {pages} {6170} (\bibinfo {year} {2002})}\BibitemShut
  {NoStop}%
\bibitem [{\citenamefont {Belonoshko}\ \emph {et~al.}(2006)\citenamefont
  {Belonoshko}, \citenamefont {Skorodumova}, \citenamefont {Rosengren},\ and\
  \citenamefont {Johansson}}]{belonoshko_melting_2006}%
  \BibitemOpen
  \bibfield  {author} {\bibinfo {author} {\bibfnamefont {A.~B.}\ \bibnamefont
  {Belonoshko}}, \bibinfo {author} {\bibfnamefont {N.~V.}\ \bibnamefont
  {Skorodumova}}, \bibinfo {author} {\bibfnamefont {A.}~\bibnamefont
  {Rosengren}},\ and\ \bibinfo {author} {\bibfnamefont {B.}~\bibnamefont
  {Johansson}},\ }\bibfield  {journal} {\bibinfo  {journal} {Phys. Rev. B}\
  }\textbf {\bibinfo {volume} {73}},\ \href
  {https://doi.org/10.1103/PhysRevB.73.012201} {10.1103/PhysRevB.73.012201}
  (\bibinfo {year} {2006})\BibitemShut {NoStop}%
\bibitem [{\citenamefont {Simon}\ and\ \citenamefont
  {Glatzel}(1929)}]{simon_bemerkungen_1929}%
  \BibitemOpen
  \bibfield  {author} {\bibinfo {author} {\bibfnamefont {F.}~\bibnamefont
  {Simon}}\ and\ \bibinfo {author} {\bibfnamefont {G.}~\bibnamefont
  {Glatzel}},\ }\href {https://doi.org/10.1002/zaac.19291780123} {\bibfield
  {journal} {\bibinfo  {journal} {Zeitschrift für anorganische und allgemeine
  Chemie}\ }\textbf {\bibinfo {volume} {178}},\ \bibinfo {pages} {309}
  (\bibinfo {year} {1929})}\BibitemShut {NoStop}%
\bibitem [{\citenamefont {Zhang}\ \emph
  {et~al.}(2019{\natexlab{a}})\citenamefont {Zhang}, \citenamefont {Wang},
  \citenamefont {Song}, \citenamefont {Duan},\ and\ \citenamefont
  {Liu}}]{zhang_melting_2019}%
  \BibitemOpen
  \bibfield  {author} {\bibinfo {author} {\bibfnamefont {T.}~\bibnamefont
  {Zhang}}, \bibinfo {author} {\bibfnamefont {S.}~\bibnamefont {Wang}},
  \bibinfo {author} {\bibfnamefont {H.}~\bibnamefont {Song}}, \bibinfo {author}
  {\bibfnamefont {S.}~\bibnamefont {Duan}},\ and\ \bibinfo {author}
  {\bibfnamefont {H.}~\bibnamefont {Liu}},\ }\href
  {https://doi.org/10.1063/1.5124520} {\bibfield  {journal} {\bibinfo
  {journal} {J. Appl. Phys.}\ }\textbf {\bibinfo {volume} {126}},\ \bibinfo
  {pages} {205901} (\bibinfo {year} {2019}{\natexlab{a}})}\BibitemShut
  {NoStop}%
\bibitem [{\citenamefont {Hixson}\ \emph {et~al.}(1989)\citenamefont {Hixson},
  \citenamefont {Boness}, \citenamefont {Shaner},\ and\ \citenamefont
  {Moriarty}}]{hixson_acoustic_1989}%
  \BibitemOpen
  \bibfield  {author} {\bibinfo {author} {\bibfnamefont {R.~S.}\ \bibnamefont
  {Hixson}}, \bibinfo {author} {\bibfnamefont {D.~A.}\ \bibnamefont {Boness}},
  \bibinfo {author} {\bibfnamefont {J.~W.}\ \bibnamefont {Shaner}},\ and\
  \bibinfo {author} {\bibfnamefont {J.~A.}\ \bibnamefont {Moriarty}},\ }\href
  {https://doi.org/10.1103/PhysRevLett.62.637} {\bibfield  {journal} {\bibinfo
  {journal} {Phys. Rev. Lett.}\ }\textbf {\bibinfo {volume} {62}},\ \bibinfo
  {pages} {637} (\bibinfo {year} {1989})}\BibitemShut {NoStop}%
\bibitem [{\citenamefont {Cazorla}\ \emph {et~al.}(2012)\citenamefont
  {Cazorla}, \citenamefont {Alf{\`e}},\ and\ \citenamefont
  {Gillan}}]{cazorla_constraints_2012}%
  \BibitemOpen
  \bibfield  {author} {\bibinfo {author} {\bibfnamefont {C.}~\bibnamefont
  {Cazorla}}, \bibinfo {author} {\bibfnamefont {D.}~\bibnamefont {Alf{\`e}}},\
  and\ \bibinfo {author} {\bibfnamefont {M.~J.}\ \bibnamefont {Gillan}},\
  }\bibfield  {journal} {\bibinfo  {journal} {Phys. Rev. B}\ }\textbf {\bibinfo
  {volume} {85}},\ \href {https://doi.org/10.1103/PhysRevB.85.064113}
  {10.1103/PhysRevB.85.064113} (\bibinfo {year} {2012})\BibitemShut {NoStop}%
\bibitem [{\citenamefont {Haskins}\ \emph {et~al.}(2012)\citenamefont
  {Haskins}, \citenamefont {Moriarty},\ and\ \citenamefont
  {Hood}}]{haskins_polymorphism_2012}%
  \BibitemOpen
  \bibfield  {author} {\bibinfo {author} {\bibfnamefont {J.~B.}\ \bibnamefont
  {Haskins}}, \bibinfo {author} {\bibfnamefont {J.~A.}\ \bibnamefont
  {Moriarty}},\ and\ \bibinfo {author} {\bibfnamefont {R.~Q.}\ \bibnamefont
  {Hood}},\ }\href {https://doi.org/10.1103/PhysRevB.86.224104} {\bibfield
  {journal} {\bibinfo  {journal} {Phys. Rev. B}\ }\textbf {\bibinfo {volume}
  {86}},\ \bibinfo {pages} {224104} (\bibinfo {year} {2012})}\BibitemShut
  {NoStop}%
\bibitem [{\citenamefont {Bj{\"o}rkas}\ and\ \citenamefont
  {Nordlund}(2009)}]{bjorkas_assessment_2009}%
  \BibitemOpen
  \bibfield  {author} {\bibinfo {author} {\bibfnamefont {C.}~\bibnamefont
  {Bj{\"o}rkas}}\ and\ \bibinfo {author} {\bibfnamefont {K.}~\bibnamefont
  {Nordlund}},\ }\href {https://doi.org/10.1016/j.nimb.2009.03.080} {\bibfield
  {journal} {\bibinfo  {journal} {Nuclear Instruments and Methods in Physics
  Research Section B: Beam Interactions with Materials and Atoms}\ }\textbf
  {\bibinfo {volume} {267}},\ \bibinfo {pages} {1830} (\bibinfo {year}
  {2009})}\BibitemShut {NoStop}%
\bibitem [{\citenamefont {Byggm{\"a}star}\ and\ \citenamefont
  {Granberg}(2020)}]{byggmastar_dynamical_2020}%
  \BibitemOpen
  \bibfield  {author} {\bibinfo {author} {\bibfnamefont {J.}~\bibnamefont
  {Byggm{\"a}star}}\ and\ \bibinfo {author} {\bibfnamefont {F.}~\bibnamefont
  {Granberg}},\ }\href {https://doi.org/10.1016/j.jnucmat.2019.151893}
  {\bibfield  {journal} {\bibinfo  {journal} {Journal of Nuclear Materials}\
  }\textbf {\bibinfo {volume} {528}},\ \bibinfo {pages} {151893} (\bibinfo
  {year} {2020})}\BibitemShut {NoStop}%
\bibitem [{\citenamefont {Ercolessi}\ and\ \citenamefont
  {Adams}(1994)}]{ercolessi_interatomic_1994}%
  \BibitemOpen
  \bibfield  {author} {\bibinfo {author} {\bibfnamefont {F.}~\bibnamefont
  {Ercolessi}}\ and\ \bibinfo {author} {\bibfnamefont {J.~B.}\ \bibnamefont
  {Adams}},\ }\href {https://doi.org/10.1209/0295-5075/26/8/005} {\bibfield
  {journal} {\bibinfo  {journal} {Europhys. Lett. EPL}\ }\textbf {\bibinfo
  {volume} {26}},\ \bibinfo {pages} {583} (\bibinfo {year} {1994})}\BibitemShut
  {NoStop}%
\bibitem [{\citenamefont {Mendelev}\ \emph {et~al.}(2003)\citenamefont
  {Mendelev}, \citenamefont {Han}, \citenamefont {Srolovitz}, \citenamefont
  {Ackland}, \citenamefont {Sun},\ and\ \citenamefont
  {Asta}}]{mendelev_development_2003}%
  \BibitemOpen
  \bibfield  {author} {\bibinfo {author} {\bibfnamefont {M.~I.}\ \bibnamefont
  {Mendelev}}, \bibinfo {author} {\bibfnamefont {S.}~\bibnamefont {Han}},
  \bibinfo {author} {\bibfnamefont {D.~J.}\ \bibnamefont {Srolovitz}}, \bibinfo
  {author} {\bibfnamefont {G.~J.}\ \bibnamefont {Ackland}}, \bibinfo {author}
  {\bibfnamefont {D.~Y.}\ \bibnamefont {Sun}},\ and\ \bibinfo {author}
  {\bibfnamefont {M.}~\bibnamefont {Asta}},\ }\href
  {https://doi.org/10.1080/14786430310001613264} {\bibfield  {journal}
  {\bibinfo  {journal} {Philos. Mag.}\ }\textbf {\bibinfo {volume} {83}},\
  \bibinfo {pages} {3977} (\bibinfo {year} {2003})}\BibitemShut {NoStop}%
\bibitem [{\citenamefont {Bernstein}\ \emph {et~al.}(2019)\citenamefont
  {Bernstein}, \citenamefont {Cs{\'a}nyi},\ and\ \citenamefont
  {Deringer}}]{bernstein_novo_2019}%
  \BibitemOpen
  \bibfield  {author} {\bibinfo {author} {\bibfnamefont {N.}~\bibnamefont
  {Bernstein}}, \bibinfo {author} {\bibfnamefont {G.}~\bibnamefont
  {Cs{\'a}nyi}},\ and\ \bibinfo {author} {\bibfnamefont {V.~L.}\ \bibnamefont
  {Deringer}},\ }\href {https://doi.org/10.1038/s41524-019-0236-6} {\bibfield
  {journal} {\bibinfo  {journal} {npj Comput Mater}\ }\textbf {\bibinfo
  {volume} {5}},\ \bibinfo {pages} {1} (\bibinfo {year} {2019})}\BibitemShut
  {NoStop}%
\bibitem [{\citenamefont {Deringer}\ \emph {et~al.}(2018)\citenamefont
  {Deringer}, \citenamefont {Pickard},\ and\ \citenamefont
  {Cs{\'a}nyi}}]{deringer_data-driven_2018}%
  \BibitemOpen
  \bibfield  {author} {\bibinfo {author} {\bibfnamefont {V.~L.}\ \bibnamefont
  {Deringer}}, \bibinfo {author} {\bibfnamefont {C.~J.}\ \bibnamefont
  {Pickard}},\ and\ \bibinfo {author} {\bibfnamefont {G.}~\bibnamefont
  {Cs{\'a}nyi}},\ }\bibfield  {journal} {\bibinfo  {journal} {Phys. Rev.
  Lett.}\ }\textbf {\bibinfo {volume} {120}},\ \href
  {https://doi.org/10.1103/PhysRevLett.120.156001}
  {10.1103/PhysRevLett.120.156001} (\bibinfo {year} {2018})\BibitemShut
  {NoStop}%
\bibitem [{\citenamefont {Zhang}\ \emph
  {et~al.}(2019{\natexlab{b}})\citenamefont {Zhang}, \citenamefont {Lin},
  \citenamefont {Wang}, \citenamefont {Car},\ and\ \citenamefont
  {E}}]{zhang_active_2019}%
  \BibitemOpen
  \bibfield  {author} {\bibinfo {author} {\bibfnamefont {L.}~\bibnamefont
  {Zhang}}, \bibinfo {author} {\bibfnamefont {D.-Y.}\ \bibnamefont {Lin}},
  \bibinfo {author} {\bibfnamefont {H.}~\bibnamefont {Wang}}, \bibinfo {author}
  {\bibfnamefont {R.}~\bibnamefont {Car}},\ and\ \bibinfo {author}
  {\bibfnamefont {W.}~\bibnamefont {E}},\ }\href
  {https://doi.org/10.1103/PhysRevMaterials.3.023804} {\bibfield  {journal}
  {\bibinfo  {journal} {Phys. Rev. Materials}\ }\textbf {\bibinfo {volume}
  {3}},\ \bibinfo {pages} {023804} (\bibinfo {year}
  {2019}{\natexlab{b}})}\BibitemShut {NoStop}%
\bibitem [{git()}]{gitlab_gap}%
  \BibitemOpen
  \href@noop {} {}\bibinfo {howpublished}
  {\url{https://gitlab.com/acclab/gap-data}}\BibitemShut {NoStop}%
\bibitem [{\citenamefont {Li}\ \emph {et~al.}(2003)\citenamefont {Li},
  \citenamefont {Siegel}, \citenamefont {Adams},\ and\ \citenamefont
  {Liu}}]{li_embedded-atom-method_2003}%
  \BibitemOpen
  \bibfield  {author} {\bibinfo {author} {\bibfnamefont {Y.}~\bibnamefont
  {Li}}, \bibinfo {author} {\bibfnamefont {D.~J.}\ \bibnamefont {Siegel}},
  \bibinfo {author} {\bibfnamefont {J.~B.}\ \bibnamefont {Adams}},\ and\
  \bibinfo {author} {\bibfnamefont {X.-Y.}\ \bibnamefont {Liu}},\ }\bibfield
  {journal} {\bibinfo  {journal} {Phys. Rev. B}\ }\textbf {\bibinfo {volume}
  {67}},\ \href {https://doi.org/10.1103/PhysRevB.67.125101}
  {10.1103/PhysRevB.67.125101} (\bibinfo {year} {2003})\BibitemShut {NoStop}%
\bibitem [{\citenamefont {Ravelo}\ \emph {et~al.}(2013)\citenamefont {Ravelo},
  \citenamefont {Germann}, \citenamefont {Guerrero}, \citenamefont {An},\ and\
  \citenamefont {Holian}}]{ravelo_shock-induced_2013}%
  \BibitemOpen
  \bibfield  {author} {\bibinfo {author} {\bibfnamefont {R.}~\bibnamefont
  {Ravelo}}, \bibinfo {author} {\bibfnamefont {T.~C.}\ \bibnamefont {Germann}},
  \bibinfo {author} {\bibfnamefont {O.}~\bibnamefont {Guerrero}}, \bibinfo
  {author} {\bibfnamefont {Q.}~\bibnamefont {An}},\ and\ \bibinfo {author}
  {\bibfnamefont {B.~L.}\ \bibnamefont {Holian}},\ }\bibfield  {journal}
  {\bibinfo  {journal} {Phys. Rev. B}\ }\textbf {\bibinfo {volume} {88}},\
  \href {https://doi.org/10.1103/PhysRevB.88.134101}
  {10.1103/PhysRevB.88.134101} (\bibinfo {year} {2013})\BibitemShut {NoStop}%
\bibitem [{\citenamefont {Purja~Pun}\ \emph {et~al.}(2015)\citenamefont
  {Purja~Pun}, \citenamefont {Darling}, \citenamefont {Kecskes},\ and\
  \citenamefont {Mishin}}]{purja_pun_angular-dependent_2015}%
  \BibitemOpen
  \bibfield  {author} {\bibinfo {author} {\bibfnamefont {G.~P.}\ \bibnamefont
  {Purja~Pun}}, \bibinfo {author} {\bibfnamefont {K.~A.}\ \bibnamefont
  {Darling}}, \bibinfo {author} {\bibfnamefont {L.~J.}\ \bibnamefont
  {Kecskes}},\ and\ \bibinfo {author} {\bibfnamefont {Y.}~\bibnamefont
  {Mishin}},\ }\href {https://doi.org/10.1016/j.actamat.2015.08.052} {\bibfield
   {journal} {\bibinfo  {journal} {Acta Materialia}\ }\textbf {\bibinfo
  {volume} {100}},\ \bibinfo {pages} {377} (\bibinfo {year}
  {2015})}\BibitemShut {NoStop}%
\bibitem [{\citenamefont {Marinica}\ \emph {et~al.}(2013)\citenamefont
  {Marinica}, \citenamefont {Ventelon}, \citenamefont {Gilbert}, \citenamefont
  {Proville}, \citenamefont {Dudarev}, \citenamefont {Marian}, \citenamefont
  {Bencteux},\ and\ \citenamefont {Willaime}}]{marinica_interatomic_2013}%
  \BibitemOpen
  \bibfield  {author} {\bibinfo {author} {\bibfnamefont {M.-C.}\ \bibnamefont
  {Marinica}}, \bibinfo {author} {\bibfnamefont {L.}~\bibnamefont {Ventelon}},
  \bibinfo {author} {\bibfnamefont {M.~R.}\ \bibnamefont {Gilbert}}, \bibinfo
  {author} {\bibfnamefont {L.}~\bibnamefont {Proville}}, \bibinfo {author}
  {\bibfnamefont {S.~L.}\ \bibnamefont {Dudarev}}, \bibinfo {author}
  {\bibfnamefont {J.}~\bibnamefont {Marian}}, \bibinfo {author} {\bibfnamefont
  {G.}~\bibnamefont {Bencteux}},\ and\ \bibinfo {author} {\bibfnamefont
  {F.}~\bibnamefont {Willaime}},\ }\href
  {https://doi.org/10.1088/0953-8984/25/39/395502} {\bibfield  {journal}
  {\bibinfo  {journal} {J. Phys.: Condens. Matter}\ }\textbf {\bibinfo {volume}
  {25}},\ \bibinfo {pages} {395502} (\bibinfo {year} {2013})}\BibitemShut
  {NoStop}%
\end{thebibliography}%


\begin{thebibliography}{11}%
\makeatletter
\providecommand \@ifxundefined [1]{%
 \@ifx{#1\undefined}
}%
\providecommand \@ifnum [1]{%
 \ifnum #1\expandafter \@firstoftwo
 \else \expandafter \@secondoftwo
 \fi
}%
\providecommand \@ifx [1]{%
 \ifx #1\expandafter \@firstoftwo
 \else \expandafter \@secondoftwo
 \fi
}%
\providecommand \natexlab [1]{#1}%
\providecommand \enquote  [1]{``#1''}%
\providecommand \bibnamefont  [1]{#1}%
\providecommand \bibfnamefont [1]{#1}%
\providecommand \citenamefont [1]{#1}%
\providecommand \href@noop [0]{\@secondoftwo}%
\providecommand \href [0]{\begingroup \@sanitize@url \@href}%
\providecommand \@href[1]{\@@startlink{#1}\@@href}%
\providecommand \@@href[1]{\endgroup#1\@@endlink}%
\providecommand \@sanitize@url [0]{\catcode `\\12\catcode `\$12\catcode
  `\&12\catcode `\#12\catcode `\^12\catcode `\_12\catcode `\%12\relax}%
\providecommand \@@startlink[1]{}%
\providecommand \@@endlink[0]{}%
\providecommand \url  [0]{\begingroup\@sanitize@url \@url }%
\providecommand \@url [1]{\endgroup\@href {#1}{\urlprefix }}%
\providecommand \urlprefix  [0]{URL }%
\providecommand \Eprint [0]{\href }%
\providecommand \doibase [0]{https://doi.org/}%
\providecommand \selectlanguage [0]{\@gobble}%
\providecommand \bibinfo  [0]{\@secondoftwo}%
\providecommand \bibfield  [0]{\@secondoftwo}%
\providecommand \translation [1]{[#1]}%
\providecommand \BibitemOpen [0]{}%
\providecommand \bibitemStop [0]{}%
\providecommand \bibitemNoStop [0]{.\EOS\space}%
\providecommand \EOS [0]{\spacefactor3000\relax}%
\providecommand \BibitemShut  [1]{\csname bibitem#1\endcsname}%
\let\auto@bib@innerbib\@empty
\bibitem [{\citenamefont {Ackland}\ and\ \citenamefont
  {Thetford}(1987)}]{ackland_improved_1987}%
  \BibitemOpen
  \bibfield  {author} {\bibinfo {author} {\bibfnamefont {G.~J.}\ \bibnamefont
  {Ackland}}\ and\ \bibinfo {author} {\bibfnamefont {R.}~\bibnamefont
  {Thetford}},\ }\href {https://doi.org/10.1080/01418618708204464} {\bibfield
  {journal} {\bibinfo  {journal} {Philos. Mag. A}\ }\textbf {\bibinfo {volume}
  {56}},\ \bibinfo {pages} {15} (\bibinfo {year} {1987})}\BibitemShut {NoStop}%
\bibitem [{\citenamefont {Derlet}\ \emph {et~al.}(2007)\citenamefont {Derlet},
  \citenamefont {{Nguyen-Manh}},\ and\ \citenamefont
  {Dudarev}}]{derlet_multiscale_2007}%
  \BibitemOpen
  \bibfield  {author} {\bibinfo {author} {\bibfnamefont {P.~M.}\ \bibnamefont
  {Derlet}}, \bibinfo {author} {\bibfnamefont {D.}~\bibnamefont
  {{Nguyen-Manh}}},\ and\ \bibinfo {author} {\bibfnamefont {S.~L.}\
  \bibnamefont {Dudarev}},\ }\href {https://doi.org/10.1103/PhysRevB.76.054107}
  {\bibfield  {journal} {\bibinfo  {journal} {Phys. Rev. B}\ }\textbf {\bibinfo
  {volume} {76}},\ \bibinfo {pages} {054107} (\bibinfo {year}
  {2007})}\BibitemShut {NoStop}%
\bibitem [{\citenamefont {Han}\ \emph {et~al.}(2003)\citenamefont {Han},
  \citenamefont {{Zepeda-Ruiz}}, \citenamefont {Ackland}, \citenamefont {Car},\
  and\ \citenamefont {Srolovitz}}]{han_interatomic_2003}%
  \BibitemOpen
  \bibfield  {author} {\bibinfo {author} {\bibfnamefont {S.}~\bibnamefont
  {Han}}, \bibinfo {author} {\bibfnamefont {L.~A.}\ \bibnamefont
  {{Zepeda-Ruiz}}}, \bibinfo {author} {\bibfnamefont {G.~J.}\ \bibnamefont
  {Ackland}}, \bibinfo {author} {\bibfnamefont {R.}~\bibnamefont {Car}},\ and\
  \bibinfo {author} {\bibfnamefont {D.~J.}\ \bibnamefont {Srolovitz}},\ }\href
  {https://doi.org/10.1063/1.1555275} {\bibfield  {journal} {\bibinfo
  {journal} {J. Appl. Phys.}\ }\textbf {\bibinfo {volume} {93}},\ \bibinfo
  {pages} {3328} (\bibinfo {year} {2003})}\BibitemShut {NoStop}%
\bibitem [{\citenamefont {Fellinger}\ \emph {et~al.}(2010)\citenamefont
  {Fellinger}, \citenamefont {Park},\ and\ \citenamefont
  {Wilkins}}]{fellinger_force-matched_2010}%
  \BibitemOpen
  \bibfield  {author} {\bibinfo {author} {\bibfnamefont {M.~R.}\ \bibnamefont
  {Fellinger}}, \bibinfo {author} {\bibfnamefont {H.}~\bibnamefont {Park}},\
  and\ \bibinfo {author} {\bibfnamefont {J.~W.}\ \bibnamefont {Wilkins}},\
  }\bibfield  {journal} {\bibinfo  {journal} {Phys. Rev. B}\ }\textbf {\bibinfo
  {volume} {81}},\ \href {https://doi.org/10.1103/PhysRevB.81.144119}
  {10.1103/PhysRevB.81.144119} (\bibinfo {year} {2010})\BibitemShut {NoStop}%
\bibitem [{\citenamefont {Park}\ \emph {et~al.}(2012)\citenamefont {Park},
  \citenamefont {Fellinger}, \citenamefont {Lenosky}, \citenamefont {Tipton},
  \citenamefont {Trinkle}, \citenamefont {Rudin}, \citenamefont {Woodward},
  \citenamefont {Wilkins},\ and\ \citenamefont {Hennig}}]{park_ab_2012}%
  \BibitemOpen
  \bibfield  {author} {\bibinfo {author} {\bibfnamefont {H.}~\bibnamefont
  {Park}}, \bibinfo {author} {\bibfnamefont {M.~R.}\ \bibnamefont {Fellinger}},
  \bibinfo {author} {\bibfnamefont {T.~J.}\ \bibnamefont {Lenosky}}, \bibinfo
  {author} {\bibfnamefont {W.~W.}\ \bibnamefont {Tipton}}, \bibinfo {author}
  {\bibfnamefont {D.~R.}\ \bibnamefont {Trinkle}}, \bibinfo {author}
  {\bibfnamefont {S.~P.}\ \bibnamefont {Rudin}}, \bibinfo {author}
  {\bibfnamefont {C.}~\bibnamefont {Woodward}}, \bibinfo {author}
  {\bibfnamefont {J.~W.}\ \bibnamefont {Wilkins}},\ and\ \bibinfo {author}
  {\bibfnamefont {R.~G.}\ \bibnamefont {Hennig}},\ }\bibfield  {journal}
  {\bibinfo  {journal} {Phys. Rev. B}\ }\textbf {\bibinfo {volume} {85}},\
  \href {https://doi.org/10.1103/PhysRevB.85.214121}
  {10.1103/PhysRevB.85.214121} (\bibinfo {year} {2012})\BibitemShut {NoStop}%
\bibitem [{\citenamefont {Thompson}\ \emph {et~al.}(2015)\citenamefont
  {Thompson}, \citenamefont {Swiler}, \citenamefont {Trott}, \citenamefont
  {Foiles},\ and\ \citenamefont {Tucker}}]{thompson_spectral_2015}%
  \BibitemOpen
  \bibfield  {author} {\bibinfo {author} {\bibfnamefont {A.~P.}\ \bibnamefont
  {Thompson}}, \bibinfo {author} {\bibfnamefont {L.~P.}\ \bibnamefont
  {Swiler}}, \bibinfo {author} {\bibfnamefont {C.~R.}\ \bibnamefont {Trott}},
  \bibinfo {author} {\bibfnamefont {S.~M.}\ \bibnamefont {Foiles}},\ and\
  \bibinfo {author} {\bibfnamefont {G.~J.}\ \bibnamefont {Tucker}},\ }\href
  {https://doi.org/10.1016/j.jcp.2014.12.018} {\bibfield  {journal} {\bibinfo
  {journal} {Journal of Computational Physics}\ }\textbf {\bibinfo {volume}
  {285}},\ \bibinfo {pages} {316} (\bibinfo {year} {2015})}\BibitemShut
  {NoStop}%
\bibitem [{\citenamefont {Chen}\ \emph {et~al.}(2017)\citenamefont {Chen},
  \citenamefont {Deng}, \citenamefont {Tran}, \citenamefont {Tang},
  \citenamefont {Chu},\ and\ \citenamefont {Ong}}]{chen_accurate_2017}%
  \BibitemOpen
  \bibfield  {author} {\bibinfo {author} {\bibfnamefont {C.}~\bibnamefont
  {Chen}}, \bibinfo {author} {\bibfnamefont {Z.}~\bibnamefont {Deng}}, \bibinfo
  {author} {\bibfnamefont {R.}~\bibnamefont {Tran}}, \bibinfo {author}
  {\bibfnamefont {H.}~\bibnamefont {Tang}}, \bibinfo {author} {\bibfnamefont
  {I.-H.}\ \bibnamefont {Chu}},\ and\ \bibinfo {author} {\bibfnamefont {S.~P.}\
  \bibnamefont {Ong}},\ }\href
  {https://doi.org/10.1103/PhysRevMaterials.1.043603} {\bibfield  {journal}
  {\bibinfo  {journal} {Phys. Rev. Materials}\ }\textbf {\bibinfo {volume}
  {1}},\ \bibinfo {pages} {043603} (\bibinfo {year} {2017})}\BibitemShut
  {NoStop}%
\bibitem [{\citenamefont {Li}\ \emph {et~al.}(2003)\citenamefont {Li},
  \citenamefont {Siegel}, \citenamefont {Adams},\ and\ \citenamefont
  {Liu}}]{li_embedded-atom-method_2003}%
  \BibitemOpen
  \bibfield  {author} {\bibinfo {author} {\bibfnamefont {Y.}~\bibnamefont
  {Li}}, \bibinfo {author} {\bibfnamefont {D.~J.}\ \bibnamefont {Siegel}},
  \bibinfo {author} {\bibfnamefont {J.~B.}\ \bibnamefont {Adams}},\ and\
  \bibinfo {author} {\bibfnamefont {X.-Y.}\ \bibnamefont {Liu}},\ }\bibfield
  {journal} {\bibinfo  {journal} {Phys. Rev. B}\ }\textbf {\bibinfo {volume}
  {67}},\ \href {https://doi.org/10.1103/PhysRevB.67.125101}
  {10.1103/PhysRevB.67.125101} (\bibinfo {year} {2003})\BibitemShut {NoStop}%
\bibitem [{\citenamefont {Ravelo}\ \emph {et~al.}(2013)\citenamefont {Ravelo},
  \citenamefont {Germann}, \citenamefont {Guerrero}, \citenamefont {An},\ and\
  \citenamefont {Holian}}]{ravelo_shock-induced_2013}%
  \BibitemOpen
  \bibfield  {author} {\bibinfo {author} {\bibfnamefont {R.}~\bibnamefont
  {Ravelo}}, \bibinfo {author} {\bibfnamefont {T.~C.}\ \bibnamefont {Germann}},
  \bibinfo {author} {\bibfnamefont {O.}~\bibnamefont {Guerrero}}, \bibinfo
  {author} {\bibfnamefont {Q.}~\bibnamefont {An}},\ and\ \bibinfo {author}
  {\bibfnamefont {B.~L.}\ \bibnamefont {Holian}},\ }\bibfield  {journal}
  {\bibinfo  {journal} {Phys. Rev. B}\ }\textbf {\bibinfo {volume} {88}},\
  \href {https://doi.org/10.1103/PhysRevB.88.134101}
  {10.1103/PhysRevB.88.134101} (\bibinfo {year} {2013})\BibitemShut {NoStop}%
\bibitem [{\citenamefont {Purja~Pun}\ \emph {et~al.}(2015)\citenamefont
  {Purja~Pun}, \citenamefont {Darling}, \citenamefont {Kecskes},\ and\
  \citenamefont {Mishin}}]{purja_pun_angular-dependent_2015}%
  \BibitemOpen
  \bibfield  {author} {\bibinfo {author} {\bibfnamefont {G.~P.}\ \bibnamefont
  {Purja~Pun}}, \bibinfo {author} {\bibfnamefont {K.~A.}\ \bibnamefont
  {Darling}}, \bibinfo {author} {\bibfnamefont {L.~J.}\ \bibnamefont
  {Kecskes}},\ and\ \bibinfo {author} {\bibfnamefont {Y.}~\bibnamefont
  {Mishin}},\ }\href {https://doi.org/10.1016/j.actamat.2015.08.052} {\bibfield
   {journal} {\bibinfo  {journal} {Acta Materialia}\ }\textbf {\bibinfo
  {volume} {100}},\ \bibinfo {pages} {377} (\bibinfo {year}
  {2015})}\BibitemShut {NoStop}%
\bibitem [{\citenamefont {Marinica}\ \emph {et~al.}(2013)\citenamefont
  {Marinica}, \citenamefont {Ventelon}, \citenamefont {Gilbert}, \citenamefont
  {Proville}, \citenamefont {Dudarev}, \citenamefont {Marian}, \citenamefont
  {Bencteux},\ and\ \citenamefont {Willaime}}]{marinica_interatomic_2013}%
  \BibitemOpen
  \bibfield  {author} {\bibinfo {author} {\bibfnamefont {M.-C.}\ \bibnamefont
  {Marinica}}, \bibinfo {author} {\bibfnamefont {L.}~\bibnamefont {Ventelon}},
  \bibinfo {author} {\bibfnamefont {M.~R.}\ \bibnamefont {Gilbert}}, \bibinfo
  {author} {\bibfnamefont {L.}~\bibnamefont {Proville}}, \bibinfo {author}
  {\bibfnamefont {S.~L.}\ \bibnamefont {Dudarev}}, \bibinfo {author}
  {\bibfnamefont {J.}~\bibnamefont {Marian}}, \bibinfo {author} {\bibfnamefont
  {G.}~\bibnamefont {Bencteux}},\ and\ \bibinfo {author} {\bibfnamefont
  {F.}~\bibnamefont {Willaime}},\ }\href
  {https://doi.org/10.1088/0953-8984/25/39/395502} {\bibfield  {journal}
  {\bibinfo  {journal} {J. Phys.: Condens. Matter}\ }\textbf {\bibinfo {volume}
  {25}},\ \bibinfo {pages} {395502} (\bibinfo {year} {2013})}\BibitemShut
  {NoStop}%
\end{thebibliography}%

\end{document}


\title{Supplemental material for: Gaussian approximation potentials for body-centered cubic transition metals}

\author{J. Byggmästar}
\thanks{Corresponding author}
\email{jesper.byggmastar@helsinki.fi}
\affiliation{Department of Physics, P.O. Box 43, FI-00014 University of Helsinki, Finland}
\author{K. Nordlund}
\affiliation{Department of Physics, P.O. Box 43, FI-00014 University of Helsinki, Finland}
\author{F. Djurabekova}
\affiliation{Department of Physics, P.O. Box 43, FI-00014 University of Helsinki, Finland}
\affiliation{Helsinki Institute of Physics, Helsinki, Finland}

\date{\today}

\maketitle

\section{Comparison with other interatomic potentials}
\label{sec:intro}

\begin{figure*}
    \centering
        \begin{subfigure}{0.48\linewidth}
        \centering
        \includegraphics[width=\linewidth]{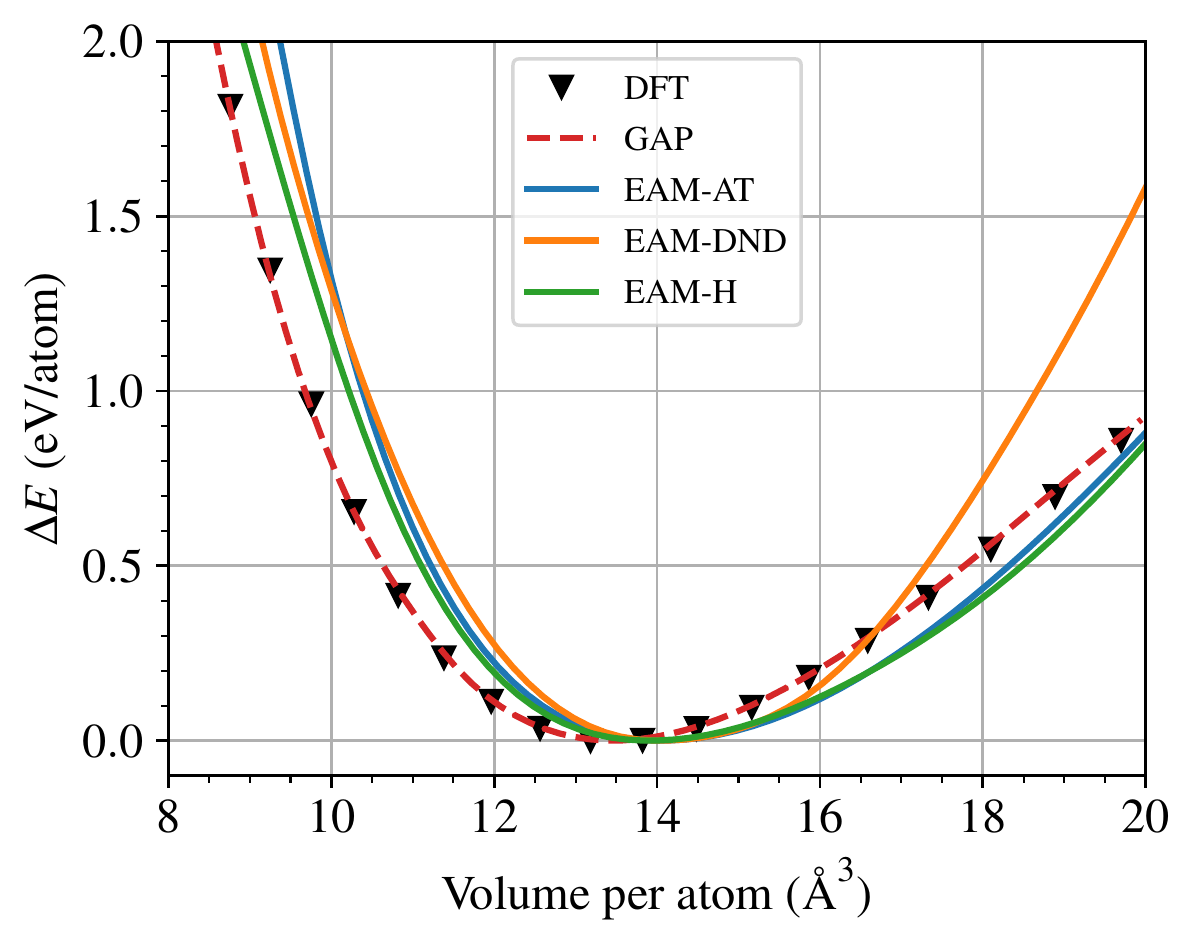}
        \caption{V.}
    \end{subfigure}
    \begin{subfigure}{0.48\linewidth}
        \centering
        \includegraphics[width=\linewidth]{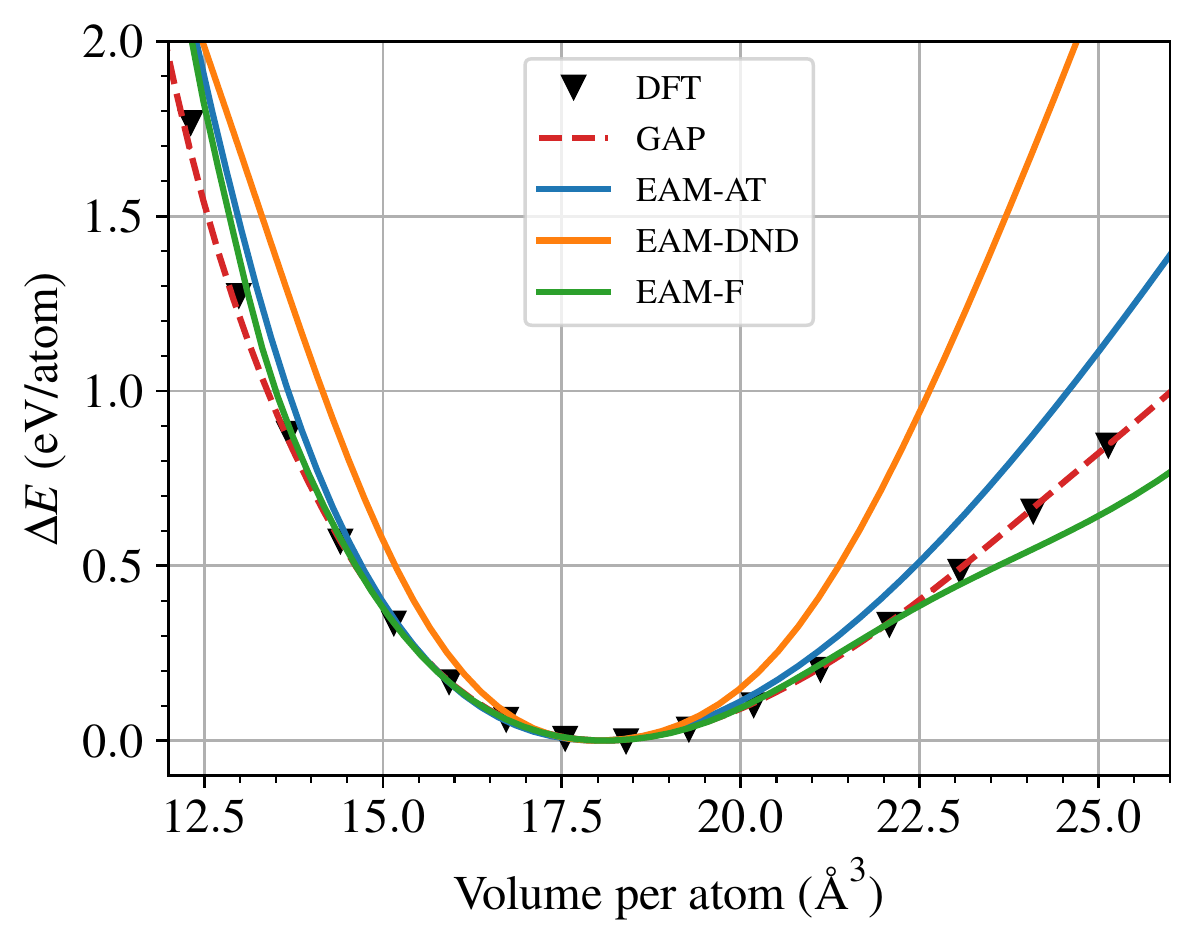}
        \caption{Nb.}
    \end{subfigure}
    \begin{subfigure}{0.48\linewidth}
        \centering
        \includegraphics[width=\linewidth]{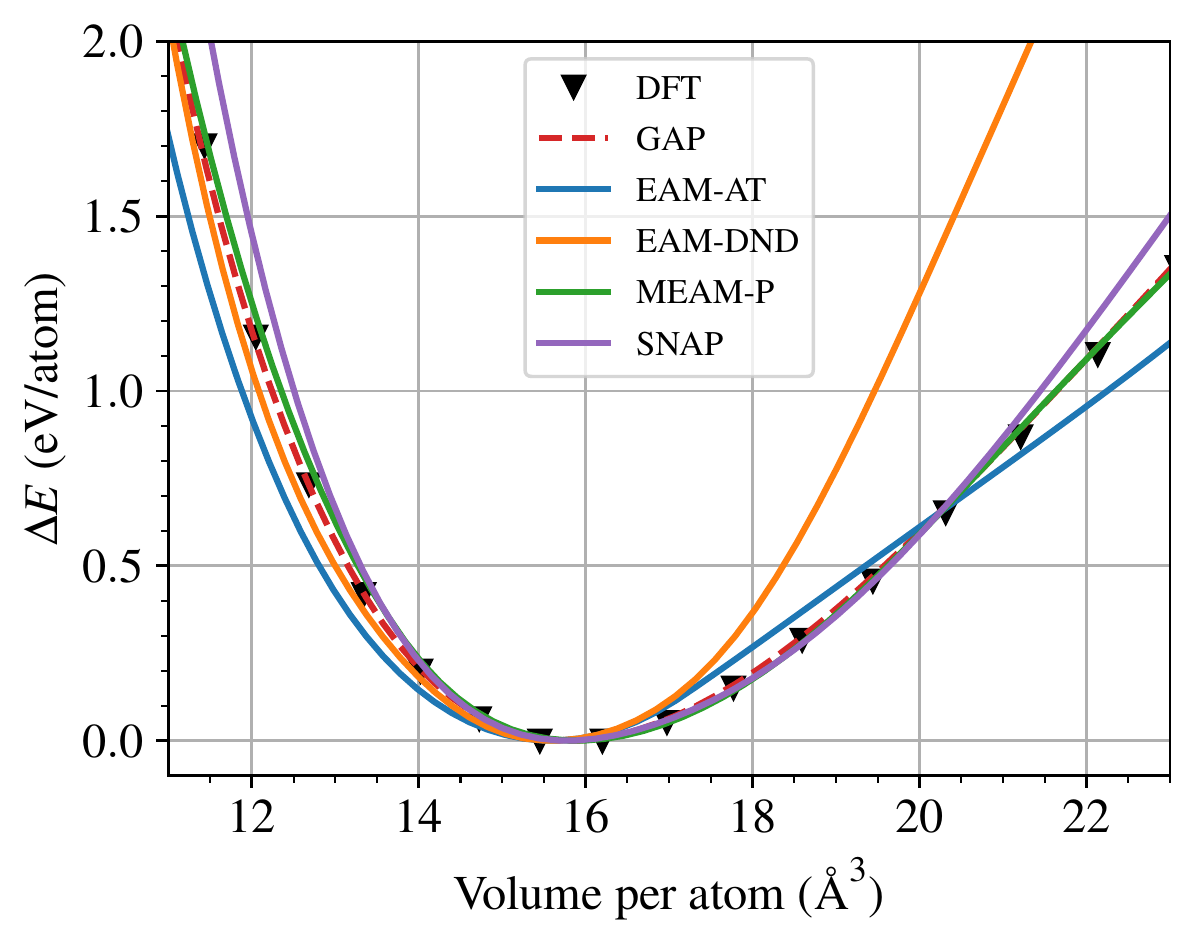}
        \caption{Mo.}
    \end{subfigure}
    \begin{subfigure}{0.48\linewidth}
        \centering
        \includegraphics[width=\linewidth]{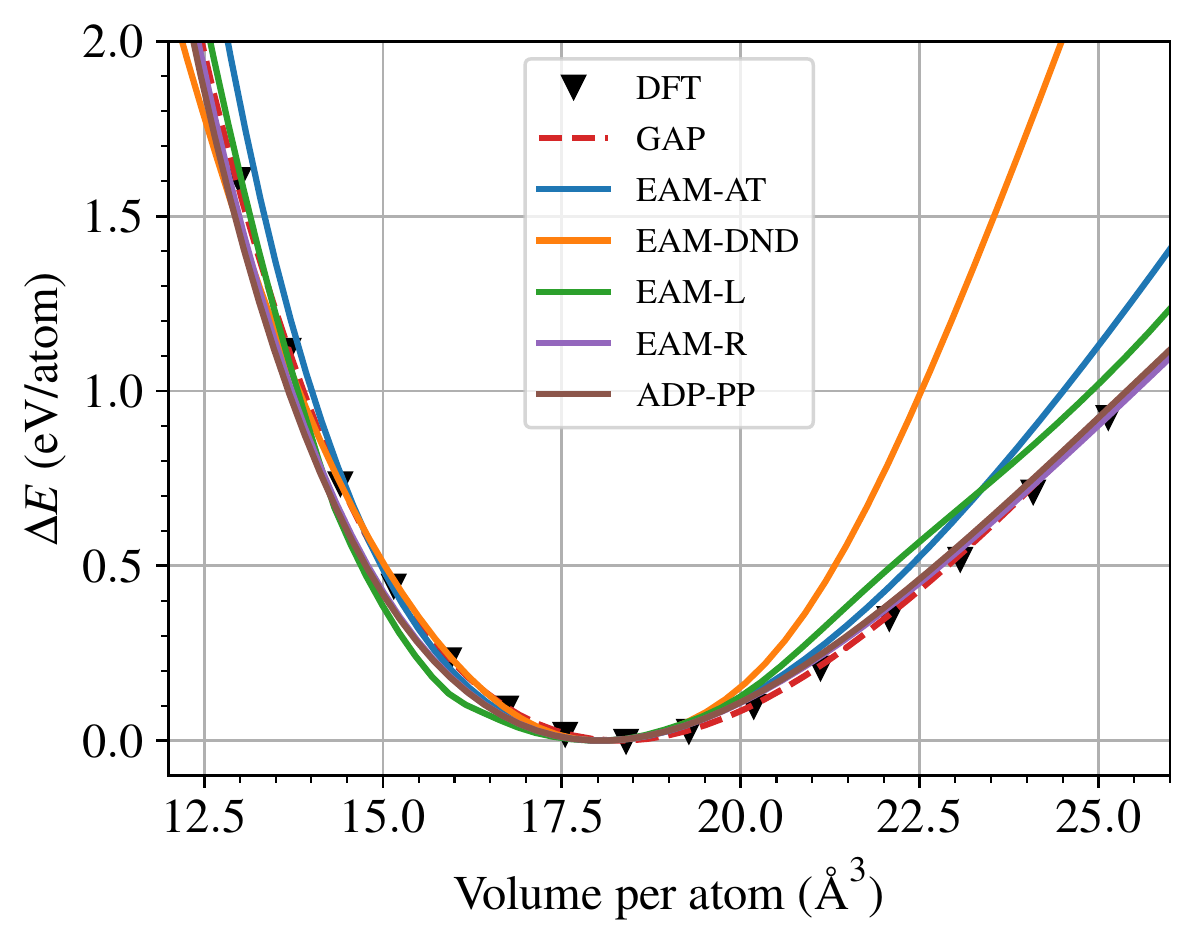}
        \caption{Ta.}
    \end{subfigure}
    \begin{subfigure}{0.48\linewidth}
        \centering
        \includegraphics[width=\linewidth]{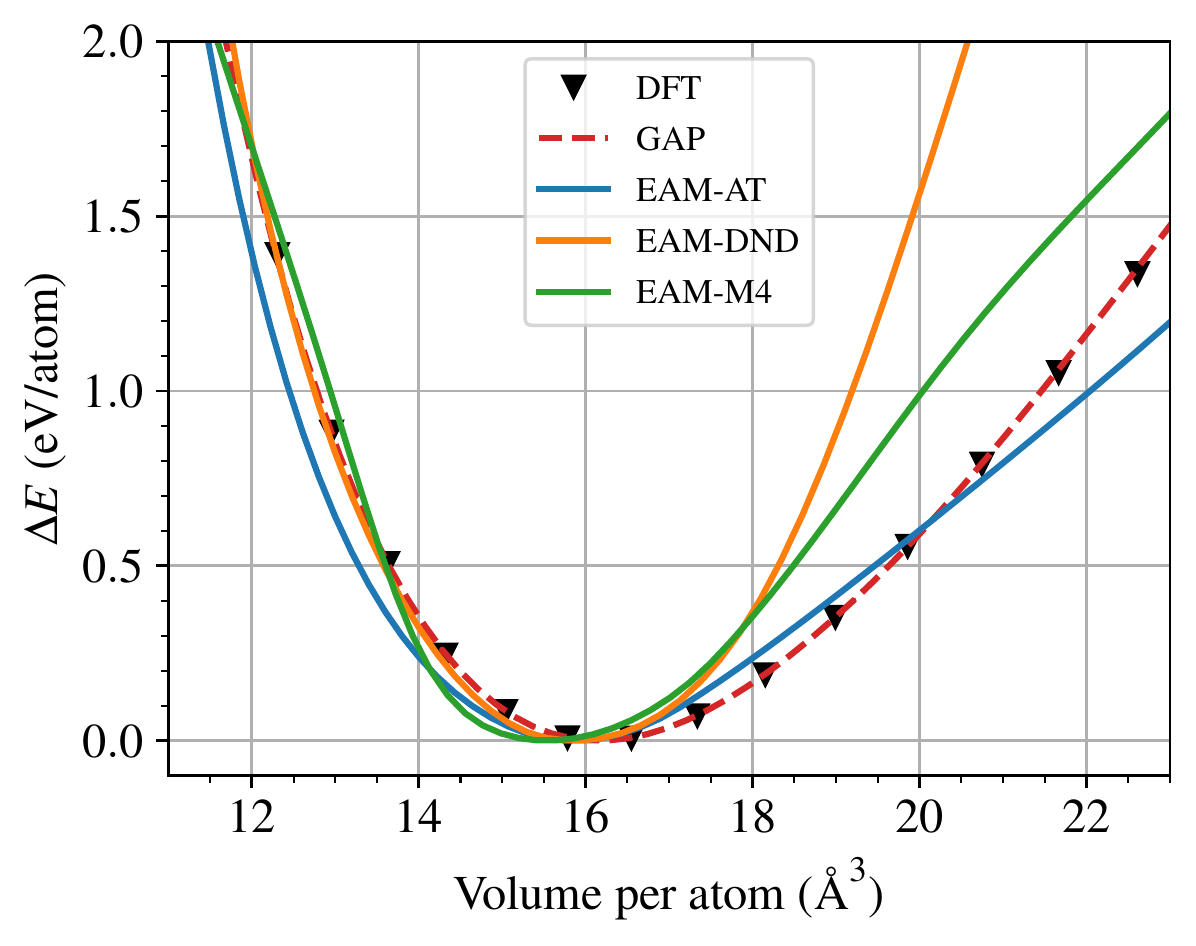}
        \caption{W.}
    \end{subfigure}
    \caption{Energy--volume curves of bcc for each element, compared between various interatomic potentials.}
    \label{fig:ev}
\end{figure*}

\begin{figure*}
    \centering
        \begin{subfigure}{0.48\linewidth}
        \centering
        \includegraphics[width=\linewidth]{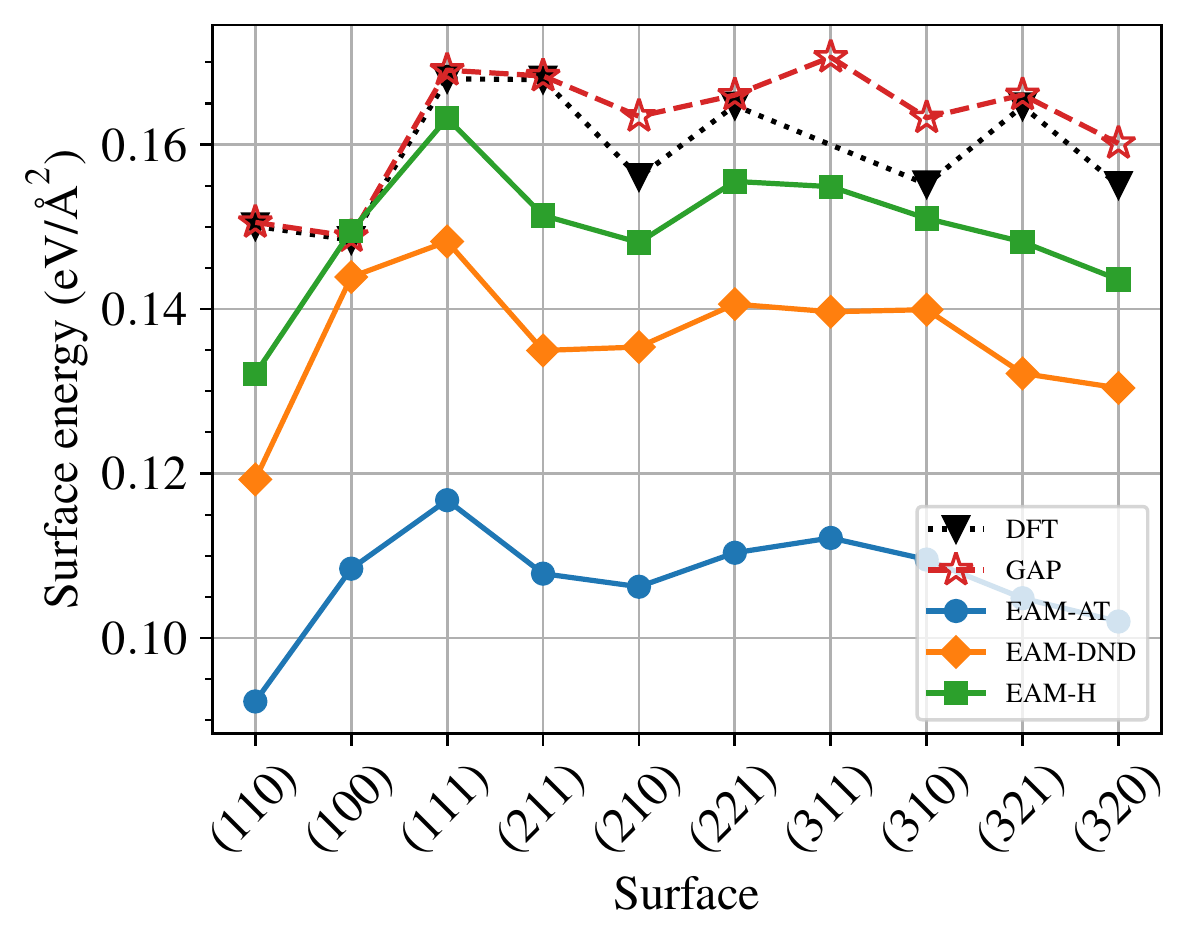}
        \caption{V.}
    \end{subfigure}
    \begin{subfigure}{0.48\linewidth}
        \centering
        \includegraphics[width=\linewidth]{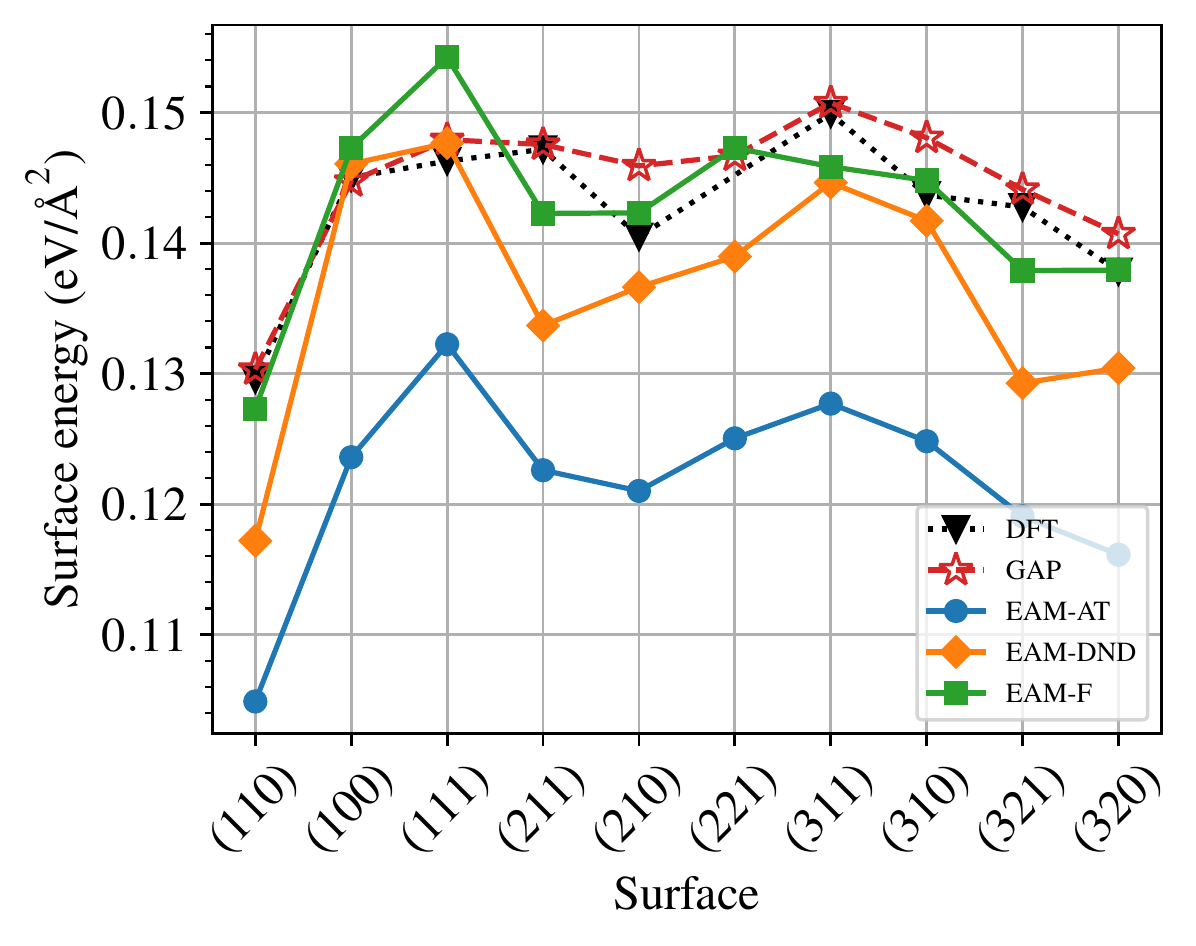}
        \caption{Nb.}
    \end{subfigure}
    \begin{subfigure}{0.48\linewidth}
        \centering
        \includegraphics[width=\linewidth]{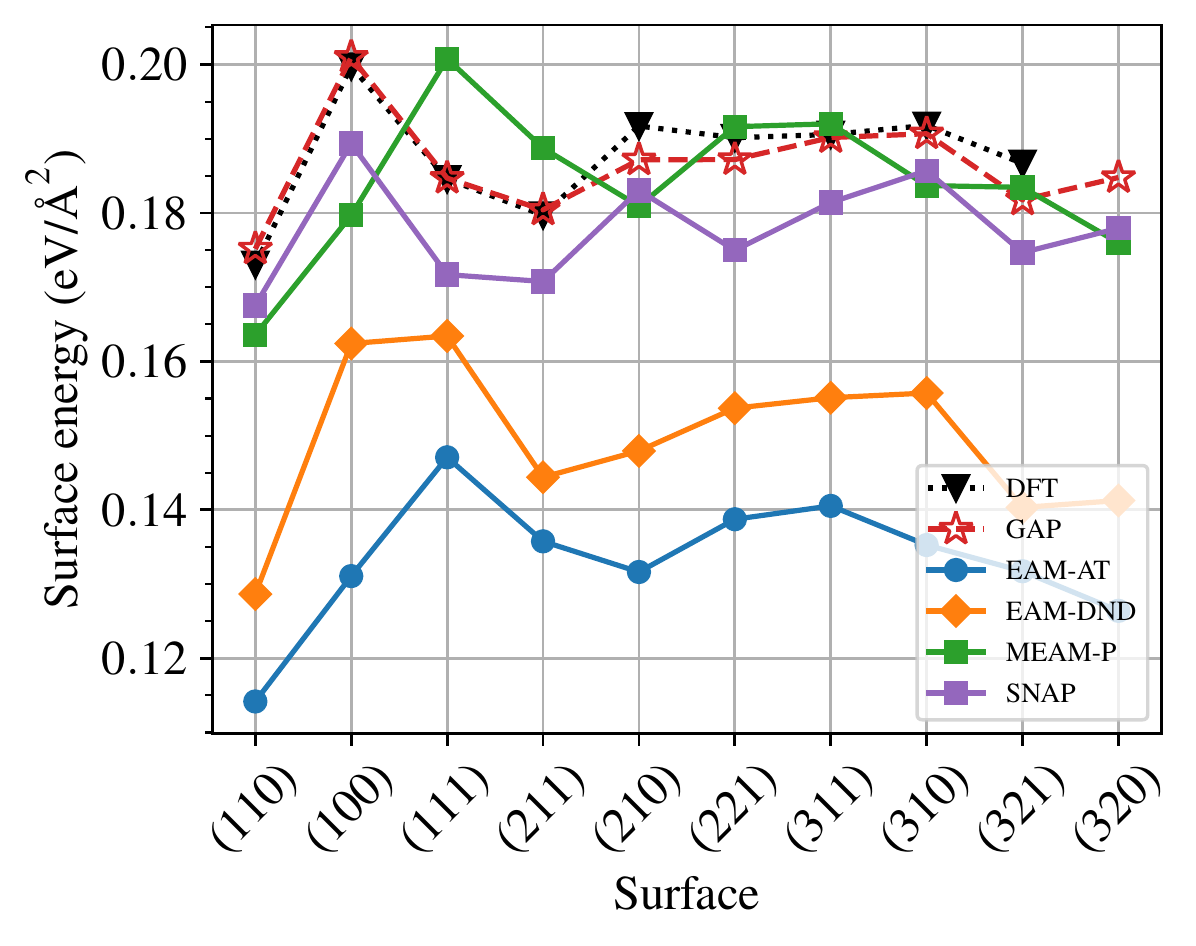}
        \caption{Mo.}
    \end{subfigure}
    \begin{subfigure}{0.48\linewidth}
        \centering
        \includegraphics[width=\linewidth]{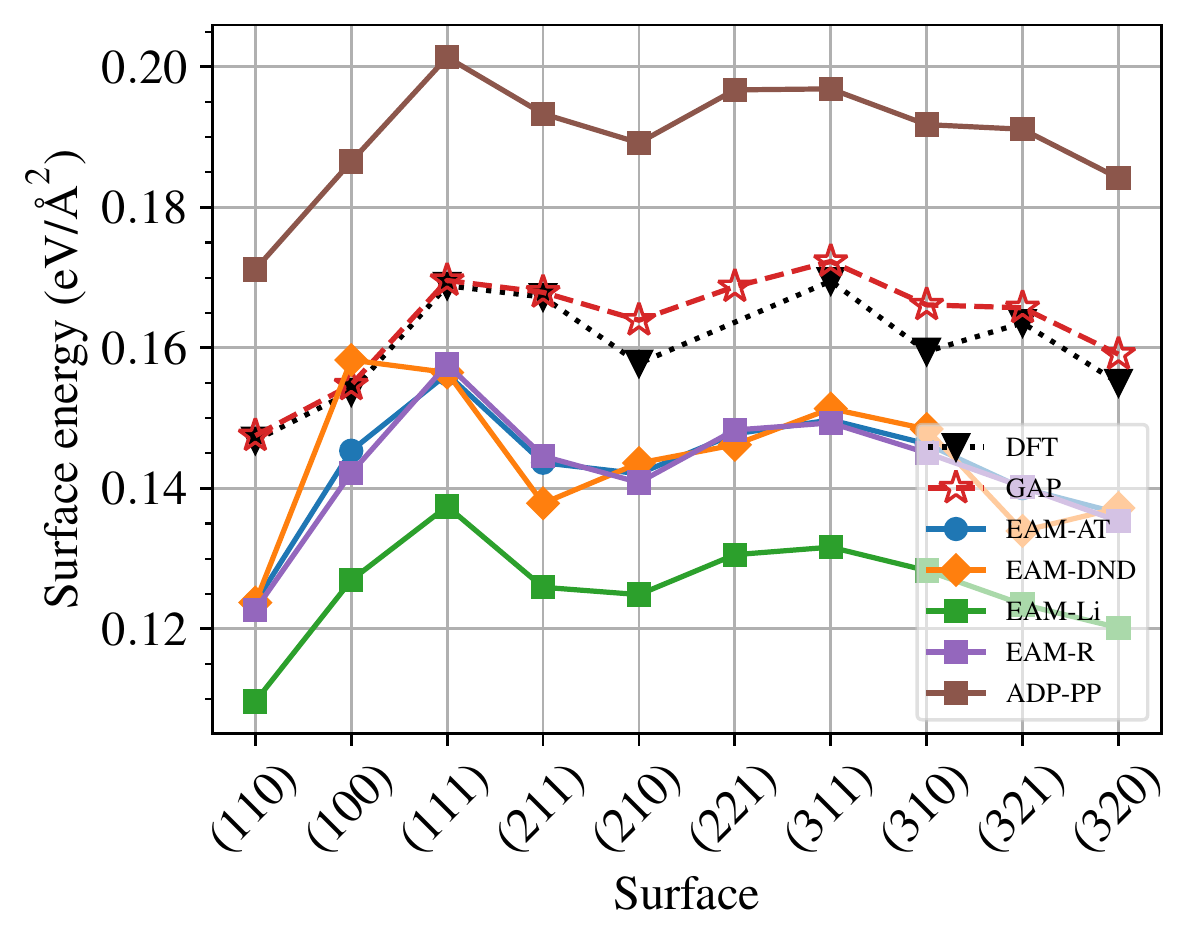}
        \caption{Ta.}
    \end{subfigure}
    \begin{subfigure}{0.48\linewidth}
        \centering
        \includegraphics[width=\linewidth]{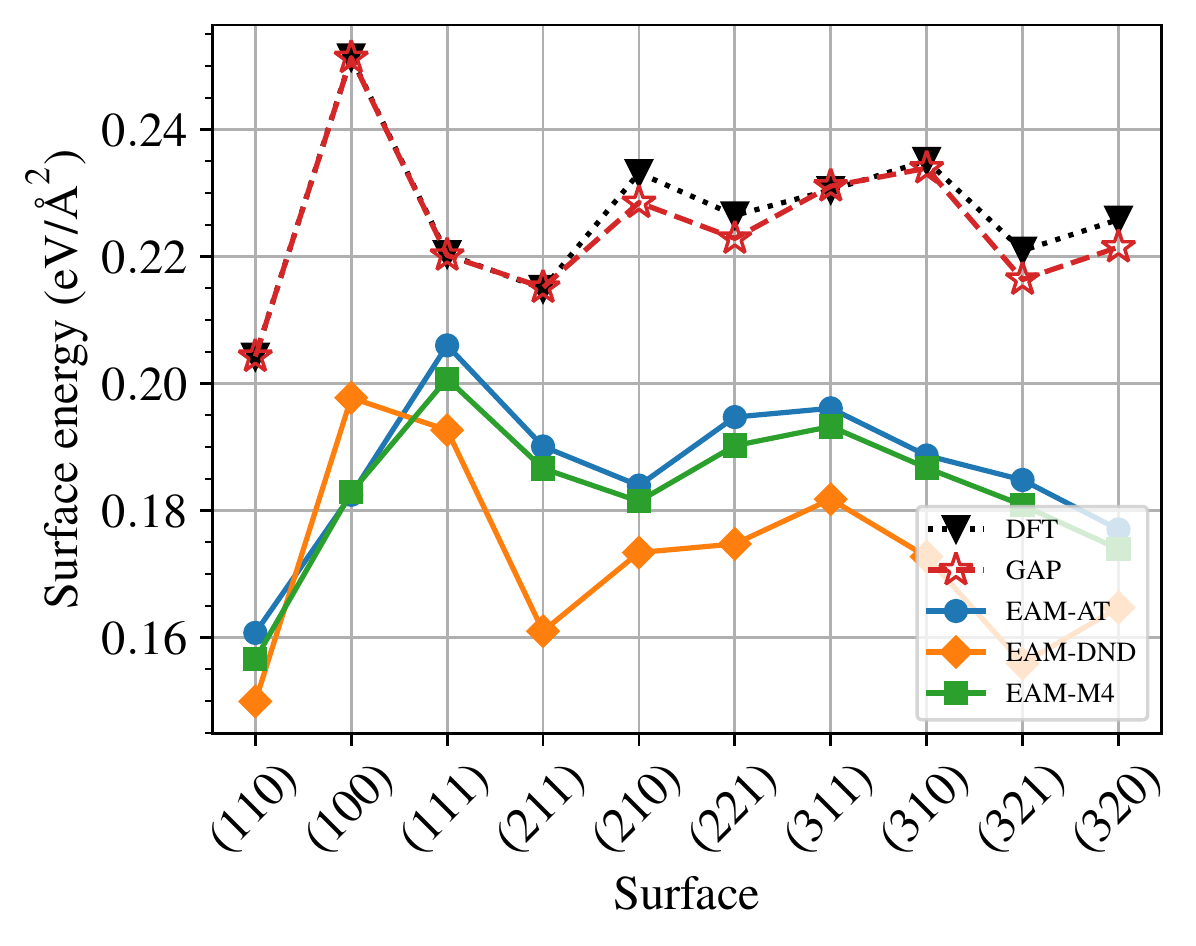}
        \caption{W.}
    \end{subfigure}
    \caption{Surface energies of each element, compared between various interatomic potentials.}
    \label{fig:esurf}
\end{figure*}

\begin{figure*}
    \centering
        \begin{subfigure}{0.48\linewidth}
        \centering
        \includegraphics[width=\linewidth]{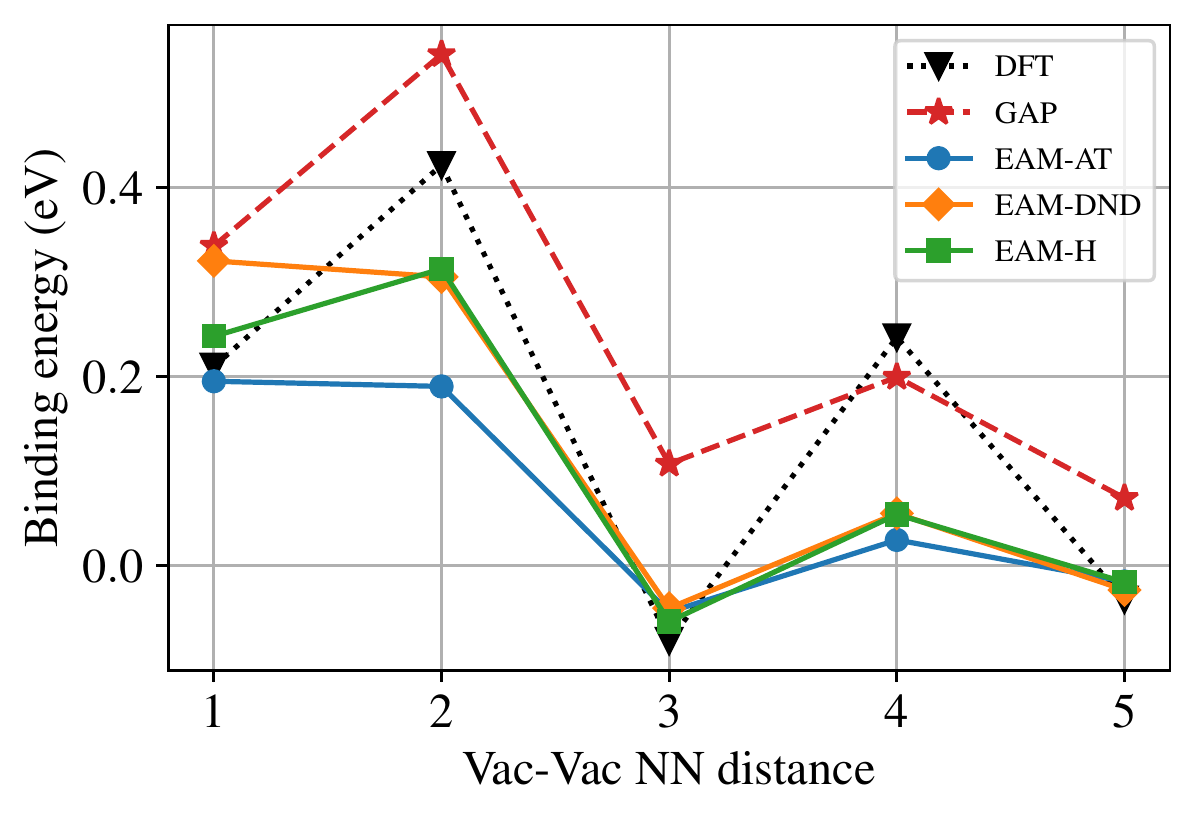}
        \caption{V.}
    \end{subfigure}
    \begin{subfigure}{0.48\linewidth}
        \centering
        \includegraphics[width=\linewidth]{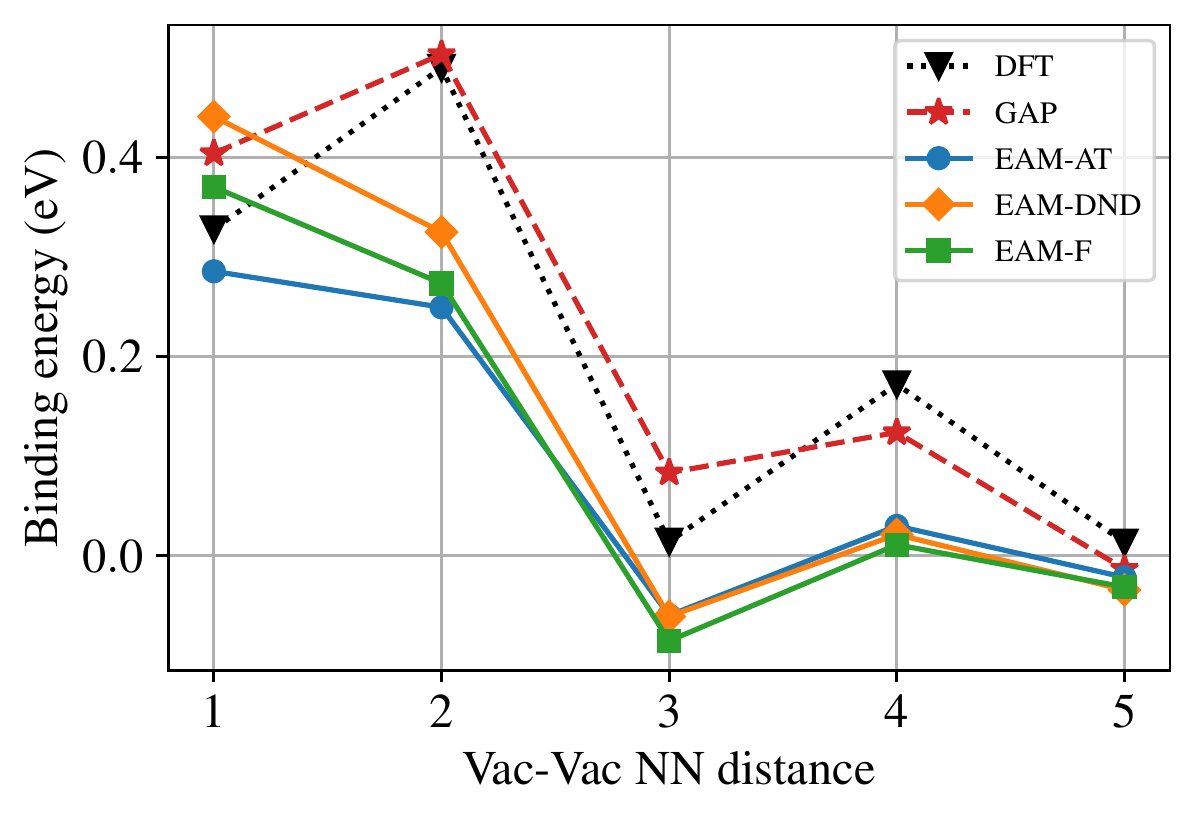}
        \caption{Nb.}
    \end{subfigure}
    \begin{subfigure}{0.48\linewidth}
        \centering
        \includegraphics[width=\linewidth]{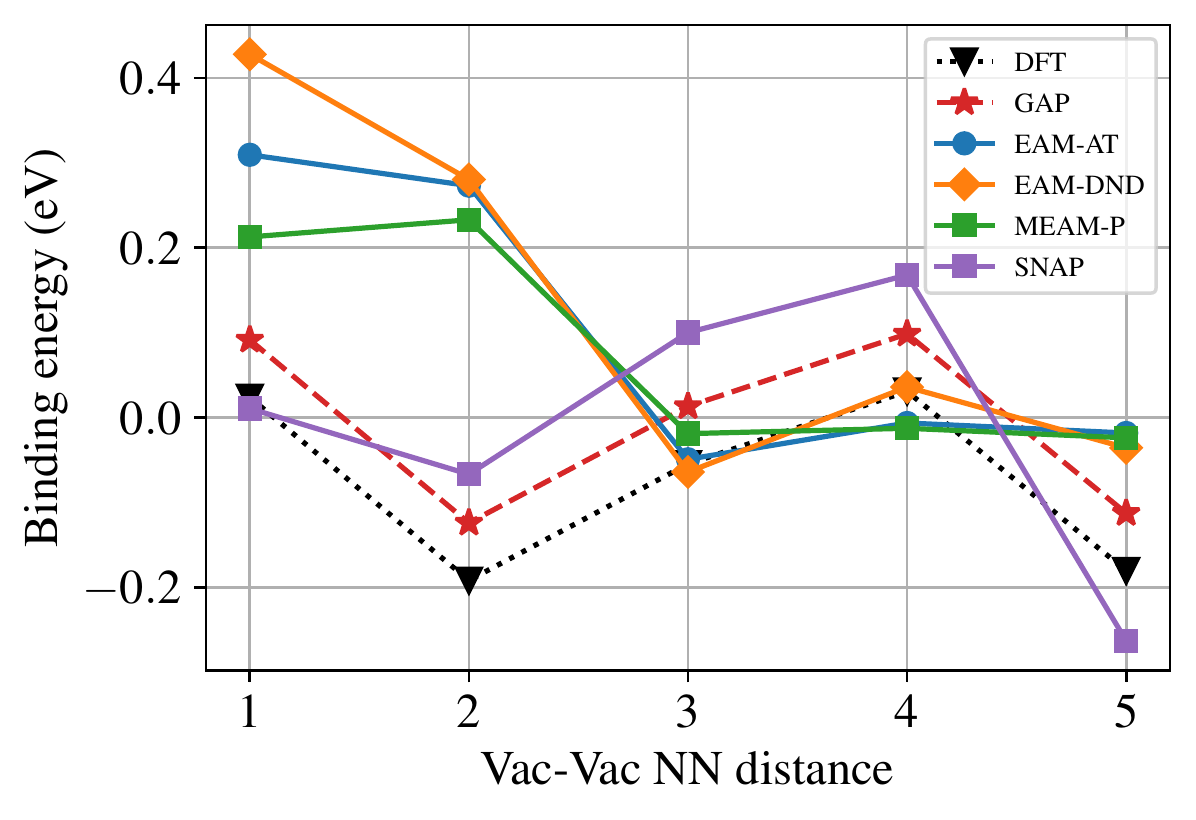}
        \caption{Mo.}
    \end{subfigure}
    \begin{subfigure}{0.48\linewidth}
        \centering
        \includegraphics[width=\linewidth]{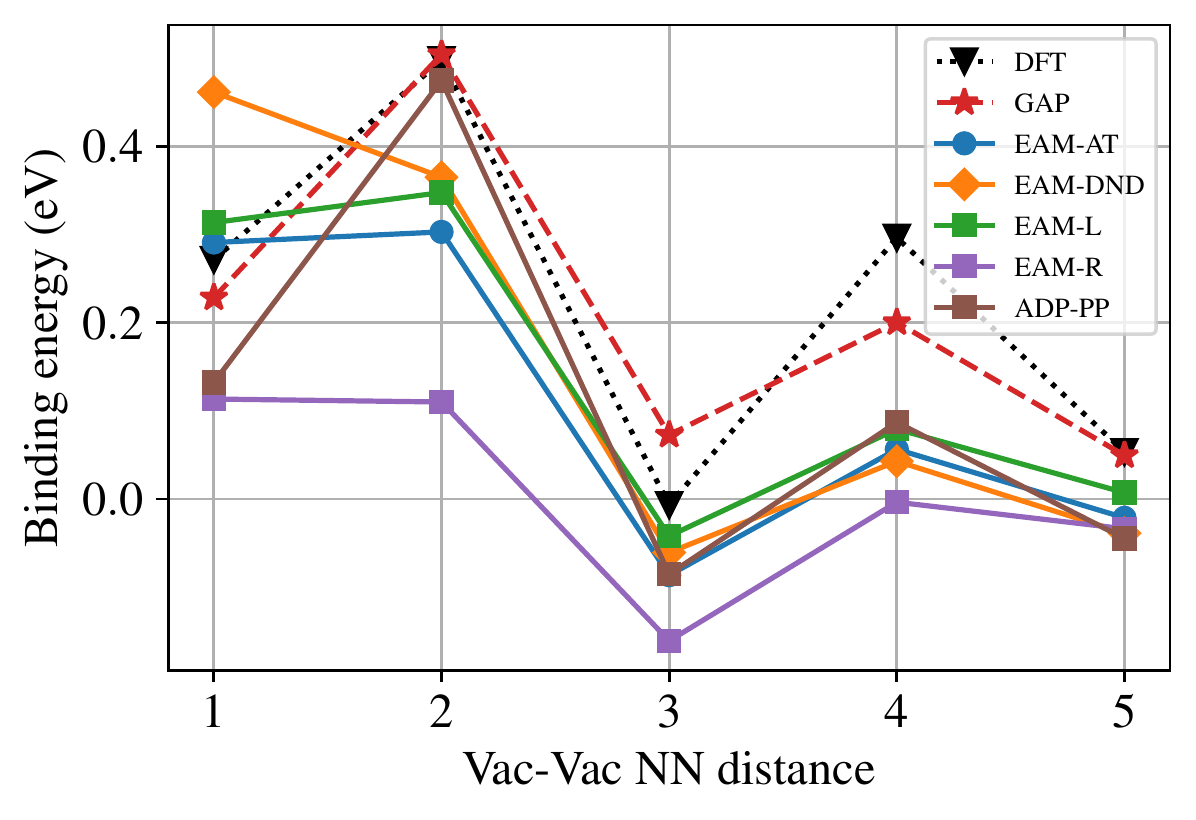}
        \caption{Ta.}
    \end{subfigure}
    \begin{subfigure}{0.48\linewidth}
        \centering
        \includegraphics[width=\linewidth]{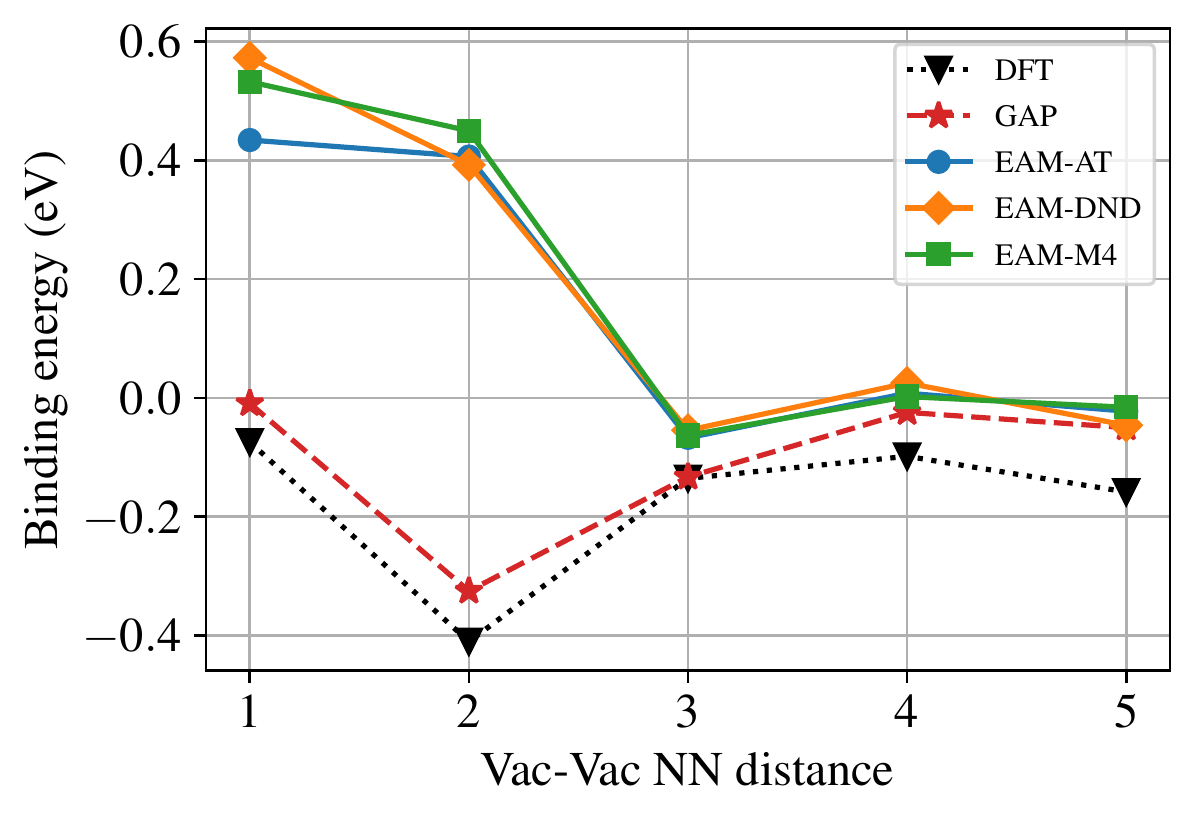}
        \caption{W.}
    \end{subfigure}
    \caption{Binding energies of divacancies in each element, compared between various interatomic potentials.}
    \label{fig:divac}
\end{figure*}

Here we compare a number of other interatomic potentials with the GAP, DFT, and experimental data discussed in the main text. The potentials are listed in Tab.~\ref{tab:pots}. Most of the potentials are chosen simply because the potential files are easily available for download from the NIST repository (\url{https://www.ctcms.nist.gov/potentials/}). All properties listed in Tabs.~\ref{tab:V}--\ref{tab:W} and Figs.~\ref{fig:ev}--\ref{fig:divac} are calculated in this work and in the same way as described in the main text for the GAP results.

Tabs.~\ref{tab:V}--\ref{tab:W} list basic bulk properties along with formation energies and relaxation volumes of vacancies and self-interstitial atoms. The listed properties are: cohesive energy of bcc ($E_\mathrm{coh}$), lattice constant of bcc ($a$), elastic constants ($C_{ij}$), minimum surface energy ($E_\mathrm{surf.}$), formation energies ($E_\mathrm{f}$), vacancy migration energy ($E_\mathrm{mig.}^\mathrm{vac.}$), relaxation volumes ($\Omega_\mathrm{rel.}$), linear thermal expansion coefficient at 300 K ($\alpha_L$), and melting temperature $T_\mathrm{melt}$. Note that the thermal expansion coefficients listed for the potentials are computed from MD simulations, and the DFT values are the quasi-harmonic approximation results from the main article (which may not be fully consistent as discussed in the main text).

Fig.~\ref{fig:ev} shows energy--volume curves of the bcc phase for each element.

Fig.~\ref{fig:esurf} shows surface energies of the 10 lowest-index surfaces for each element.

Fig.~\ref{fig:divac} shows binding energies of divacancies at different nearest-neighbour (NN) separations in each element.

\begin{table*}
    \centering
    \caption{Interatomic potentials used in this supplementary document.}
    \label{tab:pots}
    \begin{tabular}{llll}
        \toprule
        Element & Abbreviation & Potential type & Reference \\
        \midrule
         V, Nb, Mo, Ta, W & EAM-AT & Embedded atom method & Ackland \& Thetford~\cite{ackland_improved_1987} \\
         V, Nb, Mo, Ta, W & EAM-DND & Embedded atom method & Derlet et al.~\cite{derlet_multiscale_2007} \\
         V & EAM-H & Embedded atom method & Han et al.~\cite{han_interatomic_2003} \\
         Nb & EAM-F & Embedded atom method & Fellinger et al.~\cite{fellinger_force-matched_2010} \\
         Mo & MEAM-P & Modified embedded atom method & Park et al.~\cite{park_ab_2012} \\
         Mo & SNAP & SNAP~\cite{thompson_spectral_2015} (machine learning, linear regression) & Chen et al.~\cite{chen_accurate_2017} \\
         Ta & EAM-L & Embedded atom method & Li et al.~\cite{li_embedded-atom-method_2003} \\
         Ta & EAM-R & Embedded atom method & Ravelo et al.~\cite{ravelo_shock-induced_2013} \\
         Ta & ADP-PP & Angular-dependent potential & Purja Pun et al.~\cite{purja_pun_angular-dependent_2015} \\
         W & EAM-M4 & Embedded atom method & Marinica et al.~\cite{marinica_interatomic_2013} (v. 4) \\
        \bottomrule 
    \end{tabular}
\end{table*}

\begin{table*}
 \centering
 \caption{V. See the main text for references of experimental and DFT data.}
 \label{tab:V}
  \begin{tabular}{lllllll}
   \toprule
    & EAM-AT & EAM-DND & EAM-H & GAP & DFT & Expt. \\
   \midrule
    $E_\mathrm{coh}$ (eV/atom) & $-5.31$ & $-5.31$ & $-5.30$ & $-5.384$ & $-5.384$ & $-5.329$ \\
    $a$ (Å) & 3.04 & 3.04 & 3.03 & 2.997 & 2.997 & 3.024 \\
    $C_{11}$ (GPa) & 228 & 229 & 228 & 271 & 269 & 229 \\
    $C_{12}$ (GPa) & 119 & 119 & 119 & 145 & 146 & 119 \\
    $C_{44}$ (GPa) & 43 & 43 & 43 & 23 & 22 & 43 \\
    \\
    $E_\mathrm{surf.}$ (meV/Å$^2$) & 92 & 119 & 132 & 149 & 148 \\
    \\
    $E_\mathrm{f}^\mathrm{vac.}$ (eV) & 1.85 & 2.52 & 2.63 & 2.56 & 2.49 & 2.1--2.2 \\
    $E_\mathrm{mig.}^\mathrm{vac.}$ (eV) & 0.70 & 0.96 & 0.75 & 0.39 & 0.65 & 0.5 \\
    $\Omega_\mathrm{rel.}$ (at. vol.) & $-0.26$ & $-0.12$ & $-0.12$ & $-0.45$ & $-0.47$ \\
    \\
    $E_\mathrm{f}^{\hkl<111>}$ (eV) & 4.48 & 3.34 & 3.27 & 2.80 & 2.41*, 2.75* \\
    $E_\mathrm{f}^{\hkl<110>}$ (eV) & 4.12 & 3.60 & 3.71 & 3.06 & 2.68* \\
    $E_\mathrm{f}^{\hkl<100>}$ (eV) & 4.91 & 3.89 & 3.54 & 3.30 & 2.83* \\
    $E_\mathrm{f}^\mathrm{octa}$ (eV) & 4.75 & 3.94 & 4.25 & 3.37 & 2.90* \\
    $E_\mathrm{f}^\mathrm{tetra}$ (eV) & 4.74 & 3.78 & 3.63 & 3.32 & 2.90* \\
    $\Omega_\mathrm{rel.}^{\hkl<111>}$ (at. vol.) & 1.79 & 1.10 & 1.36 & 1.40 & 1.47 \\
    \\
    $\alpha_L$ ($10^{-6}$ K$^{-1}$) & 10.0 & $-1.0$ & 13.3 & 10.0 & 10.8 & 8.4 \\
    $T_\mathrm{melt}$ (K) & $2480 \pm 20$ & $2240 \pm 20$ & $2900 \pm 20$ & $2130 \pm 10$ & & 2183 \\
   \bottomrule
   *see discussion in main text
 \end{tabular}
\end{table*}

\begin{table*}
 \centering
 \caption{Nb. See the main text for references of experimental and DFT data.}
 \label{tab:Nb}
  \begin{tabular}{lllllll}
   \toprule
    & EAM-AT & EAM-DND & EAM-F & GAP & DFT & Expt. \\
   \midrule
    $E_\mathrm{coh}$ (eV/atom) & $-7.57$ & $-7.57$ & $-7.09$ & $-7.004$ & $-7.003$ & $-7.523$ \\
    $a$ (Å) & 3.301 & 3.301 & 3.308 & 3.308 & 3.307 & 3.300 \\
    $C_{11}$ (GPa) & 247 & 247 & 233 & 243 & 237 & 247 \\
    $C_{12}$ (GPa) & 133 & 134 & 124 & 137 & 138 & 135 \\
    $C_{44}$ (GPa) & 28 & 28 & 32 & 13 & 11 & 29 \\
    \\
    $E_\mathrm{surf.}$ (meV/Å$^2$) & 105 & 117 & 127 & 130 & 130 \\
    \\
    $E_\mathrm{f}^\mathrm{vac.}$ (eV) & 2.51 & 2.99 & 3.10 & 2.85 & 2.77 & 2.6--3.07 \\
    $E_\mathrm{mig.}^\mathrm{vac.}$ (eV) & 0.86 & 1.17 & 0.78 & 0.42 & 0.65 & 0.55 \\
    $\Omega_\mathrm{rel.}$ (at. vol.) & $-0.30$ & $-0.22$ & $-0.37$ & $-0.39$ & $-0.40$ \\
    \\
    $E_\mathrm{f}^{\hkl<111>}$ (eV) & 4.77 & 5.15 & 3.97 & 4.01 & 3.95 \\
    $E_\mathrm{f}^{\hkl<110>}$ (eV) & 4.45 & 5.62 & 3.84 & 4.20 & 4.20 \\
    $E_\mathrm{f}^{\hkl<100>}$ (eV) & 4.80 & 5.88 & 4.49 & 4.63 & 4.50 \\
    $E_\mathrm{f}^\mathrm{octa}$ (eV) & 4.86 & 5.94 & 4.36 & 4.76 & 4.62 \\
    $E_\mathrm{f}^\mathrm{tetra}$ (eV) & 4.89 & 5.65 & 4.49 & 4.66 & 4.42 \\
    $\Omega_\mathrm{rel.}^{\hkl<111>}$ (at. vol.) & 1.35 & 1.08 & 1.53 & 1.48 & 1.55 \\
    \\
    $\alpha_L$ ($10^{-6}$ K$^{-1}$) & 6.7 & 0.7 & 9.2 & 8.5 & 8.0 & 7.3 \\
    $T_\mathrm{melt}$ (K) & $3080 \pm 20$ & $3100 \pm 20$ & $2660 \pm 20$ & $2550 \pm 10$ & & 2750 \\
   \bottomrule
 \end{tabular}
\end{table*}

\begin{table*}
 \centering
 \caption{Mo. See the main text for references of experimental and DFT data.}
 \label{tab:Mo}
  \begin{tabular}{llllllll}
   \toprule
    & EAM-AT & EAM-DND & MEAM-P & SNAP & GAP & DFT & Expt. \\
   \midrule
    $E_\mathrm{coh}$ (eV/atom) & $-6.82$ & $-6.82$ & $-6.82$ & $-5.13$ & $-6.288$ & $-6.288$ & $-6.821$ \\
    $a$ (Å) & 3.147 & 3.147 & 3.167 & 3.160 & 3.163 & 3.163 & 3.147 \\
    $C_{11}$ (GPa) & 465 & 466 & 423 & 473 & 472 & 468 & 464 \\
    $C_{12}$ (GPa) & 162 & 162 & 143 & 152 & 163 & 155 & 159 \\
    $C_{44}$ (GPa) & 109 & 109 & 95 & 107 & 105 & 100 & 109 \\
    \\
    $E_\mathrm{surf.}$ (meV/Å$^2$) & 114 & 129 & 164 & 168 & 175 & 173 \\
    \\
    $E_\mathrm{f}^\mathrm{vac.}$ (eV) & 2.55 & 2.97 & 2.96 & 2.55 & 2.84 & 2.83 & 3.0--3.24 \\
    $E_\mathrm{mig.}^\mathrm{vac.}$ (eV) & 1.28 & 1.68 & 1.63 & 1.44 & 1.28 & 1.24 & 1.35--1.62 \\
    $\Omega_\mathrm{rel.}$ (at. vol.) & $-0.15$ & $-0.10$ & $-0.42$ & $-0.36$ & $-0.33$ & $-0.38$ \\
    \\
    $E_\mathrm{f}^{\hkl<11\xi>}$ (eV) & -- & -- & 7.56 & -- & 7.47 & 7.40 \\
    $E_\mathrm{f}^{\hkl<111>}$ (eV) & 7.19 & 7.31 & 7.63 & 8.15 & 7.56 & 7.48 \\
    $E_\mathrm{f}^{\hkl<110>}$ (eV) & 6.95 & 7.52 & 7.66 & 8.29 & 7.61 & 7.58 \\
    $E_\mathrm{f}^{\hkl<100>}$ (eV) & 7.18 & 8.74 & 7.79 & 7.57 & 8.99 & 8.89 \\
    $E_\mathrm{f}^\mathrm{octa}$ (eV) & 7.56 & 8.91 & 8.04 & 7.26 & 9.00 & 8.92 \\
    $E_\mathrm{f}^\mathrm{tetra}$ (eV) & 7.35 & 8.32 & 8.18 & -- & 8.44 & 8.36 \\
    $\Omega_\mathrm{rel.}^{\hkl<111>}$ (at. vol.) & 1.12 & 1.26 & 1.57 & 1.87 & 1.58 & 1.54 \\
    \\
    $\alpha_L$ ($10^{-6}$ K$^{-1}$) & 2.8 & $-3.6$ & 5.7 & 7.6 & 6.3 & 5.8 & 4.8 \\
    $T_\mathrm{melt}$ (K) & $3080 \pm 20$ & $3280 \pm 20$ & $3200 \pm 20$ & $2580 \pm 20$ & $2750 \pm 10$ & & 2895 \\
   \bottomrule
 \end{tabular}
\end{table*}

\begin{table*}
 \centering
 \caption{Ta. See the main text for references of experimental and DFT data.}
 \label{tab:Ta}
  \begin{tabular}{lllllllll}
   \toprule
    & EAM-AT & EAM-DND & EAM-L & EAM-R & ADP-PP & GAP & DFT & Expt. \\
   \midrule
    $E_\mathrm{coh}$ (eV/atom) & $-8.1$ & $-8.1$ & $-8.09$ & $-8.10$ & $-8.10$ & $-8.114$ & $-8.113$ & $-8.105$ \\
    $a$ (Å) & 3.306 & 3.306 & 3.303 & 3.304 & 3.304 & 3.321 & 3.319 & 3.303 \\
    $C_{11}$ (GPa) & 266 & 267 & 247 & 263 & 269 & 267 & 266 & 260 \\
    $C_{12}$ (GPa) & 161 & 162 & 144 & 161 & 158 & 161 & 161 & 154 \\
    $C_{44}$ (GPa) & 82 & 83 & 87 & 82 & 90 & 77 & 77 & 83 \\
    \\
    $E_\mathrm{surf.}$ (meV/Å$^2$) & 124 & 124 & 110 & 123 & 171 & 148 & 147 \\
    \\
    $E_\mathrm{f}^\mathrm{vac.}$ (eV) & 2.90 & 3.14 & 2.75 & 2.39 & 3.06 & 3.09 & 2.95 & 2.2--3.1 \\
    $E_\mathrm{mig.}^\mathrm{vac.}$ (eV) & 1.07 & 1.41 & 1.11 & 1.01 & 0.84 & 0.62 & 0.76 & 0.7 \\
    $\Omega_\mathrm{rel.}$ (at. vol.) & $-0.26$ & $-0.18$ & $-0.35$ & $-0.46$ & $-0.30$ & $-0.47$ & $-0.44$ \\
    \\
    $E_\mathrm{f}^{\hkl<111>}$ (eV) & 7.10 & 5.75 & 6.53 & 4.45 & 5.61 & 4.84 & 4.77 \\
    $E_\mathrm{f}^{\hkl<110>}$ (eV) & 6.78 & 6.39 & 6.08 & 4.56 & 5.67 & 5.47 & 5.48 \\
    $E_\mathrm{f}^{\hkl<100>}$ (eV) & 8.03 & 6.94 & 8.03 & 5.32 & 5.85 & 5.99 & 5.89 \\
    $E_\mathrm{f}^\mathrm{octa}$ (eV) & 7.91 & 6.72 & 7.53 & 5.30 & 6.01 & 6.06 & 5.95 \\
    $E_\mathrm{f}^\mathrm{tetra}$ (eV) & 7.60 & 6.59 & 7.02 & 5.09 & 5.75 & 5.98 & 5.77 \\
    $\Omega_\mathrm{rel.}^{\hkl<111>}$ (at. vol.) & 1.54 & 0.92 & 1.02 & 0.94 & 1.17 & 1.35 & 1.52 \\
    \\
    $\alpha_L$ ($10^{-6}$ K$^{-1}$) & 6.2 & $-3.1$ & 0.6 & 2.6 & 6.3 & 7.7 & 8.6 & 6.3 \\
    $T_\mathrm{melt}$ (K) & $3860 \pm 20$ & $3420 \pm 20$ & $3260 \pm 20$ & $3060 \pm 20$ & $3300 \pm 20$ & $3010 \pm 10$ & & 3290 \\
   \bottomrule
 \end{tabular}
\end{table*}

\begin{table*}
 \centering
 \caption{W. See the main text for references of experimental and DFT data.}
 \label{tab:W}
  \begin{tabular}{lllllll}
   \toprule
    & EAM-AT & EAM-DND & EAM-M4 & GAP & DFT & Expt. \\
   \midrule
    $E_\mathrm{coh}$ (eV/atom) & $-8.9$ & $-8.9$ & $-8.9$ & $-8.39$ & $-8.386$ & $-8.803$ \\
    $a$ (Å) & 3.165 & 3.165 & 3.143 & 3.185 & 3.185 & 3.165 \\
    $C_{11}$ (GPa) & 522 & 524 & 523 & 524 & 521 & 522 \\
    $C_{12}$ (GPa) & 204 & 205 & 202 & 200 & 195 & 204 \\
    $C_{44}$ (GPa) & 161 & 161 & 161 & 148 & 147 & 161 \\
    \\
    $E_\mathrm{surf.}$ (meV/Å$^2$) & 161 & 150 & 157 & 204 & 204 \\
    \\
    $E_\mathrm{f}^\mathrm{vac.}$ (eV) & 3.63 & 3.56 & 3.82 & 3.32 & 3.36 & 3.51--4.1 \\
    $E_\mathrm{mig.}^\mathrm{vac.}$ (eV) & 1.44 & 2.07 & 1.85 & 1.71 & 1.73 & 1.70--2.02 \\
    $\Omega_\mathrm{rel.}$ (at. vol.) & $-0.14$ & $-0.11$ & $-0.04$ & $-0.31$ & $-0.36$ \\
    \\
    $E_\mathrm{f}^{\hkl<11\xi>}$ (eV) & -- & -- & -- & 10.32 & 10.25 \\
    $E_\mathrm{f}^{\hkl<111>}$ (eV) & 8.86 & 9.43 & 10.38 & 10.35 & 10.29 \\
    $E_\mathrm{f}^{\hkl<110>}$ (eV) & 9.53 & 9.76 & 10.82 & 10.58 & 10.58 \\
    $E_\mathrm{f}^{\hkl<100>}$ (eV) & 9.76 & 11.36 & 12.77 & 12.23 & 12.20 \\
    $E_\mathrm{f}^\mathrm{octa}$ (eV) & 9.94 & 9.92 & 12.56 & 12.33 & 12.27 \\
    $E_\mathrm{f}^\mathrm{tetra}$ (eV) & 9.77 & 10.90 & 11.89 & 11.77 & 11.72 \\
    $\Omega_\mathrm{rel.}^{\hkl<111>}$ (at. vol.) & 1.39 & 1.25 & 1.23 & 1.85 & 1.71 \\
    \\
    $\alpha_L$ ($10^{-6}$ K$^{-1}$) & 3.8 & $-3.5$ & 8.1 & 5.2 & 4.9 & 4.5 \\
    $T_\mathrm{melt}$ (K) & $4100 \pm 20$ & $3780 \pm 20$ & $4620 \pm 20$ & $3540 \pm 10$ & & 3687 \\
   \bottomrule
 \end{tabular}
\end{table*}

\clearpage
\bibliography{mybib}